\documentclass{cernyrep}
\usepackage{texnames}
\usepackage[T1]{fontenc}
\usepackage[bookmarks, colorlinks=true, linktoc=page, pdftex, linkcolor=black, citecolor=black, urlcolor=blue]{hyperref}
\usepackage{hepnames}
\usepackage{esint}
\usepackage{physics}
\usepackage{xcolor}

\usepackage{varwidth}

\usepackage{fancyhdr}
\fancyhfoffset{4 mm}
\fancypagestyle{ARTTITLE}{%
\fancyhf{} 
\lhead{\small{Proceedings of the 2018 CERN--Accelerator--School course on {\it Beam Instrumentation}, Tuusula, (Finland)}} 
\lfoot{Available online at \url{https://cas.web.cern.ch/previous-schools}}
\rfoot{\thepage\hspace*{3mm}}
 
}

\begin{document}
\title{Beam Diagnostic Requirements: an Overview}

\author {G. Kube}

\institute{Deutsches Elektronen Synchrotron (DESY), Hamburg, Germany}

\begin{abstract}
Beam diagnostics and instrumentation are an essential part of any kind of accelerator. There is a large variety of parameters to be measured for observation of particle beams with the precision required to tune, operate, and improve the machine. In the first part, the basic mechanisms of information transfer from the beam particles to the detector are described in order to derive suitable performance characteristics for the beam properties. However, depending on the type of accelerator, for the same parameter, the working principle of a monitor may strongly differ, and related to it also the requirements for accuracy. Therefore, in the second part, selected types of accelerators are described in order to illustrate specific diagnostics needs which must be taken into account before designing a related instrument.
\end{abstract}


\maketitle 

\thispagestyle{ARTTITLE}

\section{Introduction}

Nowadays particle accelerators play an important role in a wide number of fields, the number of accelerators worldwide is of the order of 30000 and constantly growing. While most of these devices are used for industrial and medical applications (ion implantation, electron beam material processing and irradiation, non-destructive inspection, radiotherapy, medical isotopes production, $\ldots$), the share of accelerators used for basic science is less than 1~\% \cite{Hamm12}.

In order to cover such a wide range of applications different accelerator types are required. As an example, in the arts, the Louvre museum utilizes a 2~MV tandem Pelletron accelerator for ion beam analysis studies \cite{AGLAE}. Cyclotrons are often used to produce medical isotopes for positron emission tomography and single photon emission computed tomography. For electron radiotherapy, mainly linear accelerators (linacs) are in operation, while cyclotrons or synchrotrons are additionally used for proton therapy \cite{Wieszczycka01}. Third generation synchrotron light sources and diffraction limited storage rings are based on electron synchrotrons, while free electron lasers operating at short wavelengths are electron linac based accelerators. Neutrino beams for elementary particle physics are produced with large proton synchrotrons, and in linear or circular colliders different species of particles (hadrons and/or leptons) are brought into collision. In recent years, the development of advanced accelerator concepts using electron and proton beams has been pushed which aim to increase the gradient of accelerators by orders of magnitude, using new power sources (e.g., lasers and relativistic beams) and new materials (e.g., dielectrics, metamaterials, and plasmas) \cite{Colby16}.

As can be seen from this short compilation there exist a large number of accelerator types with different properties, and as a consequence, the demands on beam diagnostics and instrumentation vary depending on machine type and application. To cover all these cases is out of the focus of this report, more details can be found in specific textbooks or lecture notes as in Refs.~\cite{Minty03,Strehl06,Smaluk09,Forck17}. Linear and circular accelerators for high energy physics and synchrotron radiation applications will be the primary concern, but nevertheless links to other types of accelerators are provided. However, before going into detail and describing specific accelerator examples, the first part of the report is devoted to the basic mechanisms of information transfer from the beam particles to the measurement device. A description is given how this information can be used to derive suitable performance characteristics for the beam properties.

\section{Electromagnetism and relativity}

\begin{figure}[!h]
\begin{center}
\includegraphics[width=7cm]{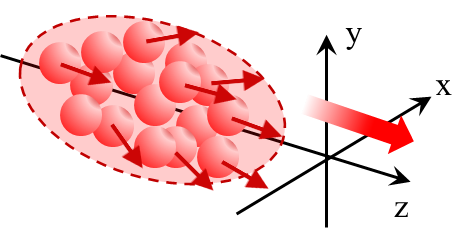}
\caption{A particle bunch consisting of $N$ identical particles travels along the $z$--axis}
\label{fig:S2intro}
\end{center}
\end{figure}

A typical particle bunch as schematically depicted in \Fref{fig:S2intro} consists of a large number of $N$ identical particles, e.g., $\Pp, \APproton, \Pepm, \Pn, \Pgamma, \Pmupm$, or heavy ions. The motion of the individual particles is described in the six-dimensional phase space which is bases on the canonical coordinates ($x, y, z; p_x, p_y, p_z$). Assuming decoupled motion, the projections onto the three orthogonal planes are usually used instead of the full six-dimensional phase space. If the momentum $p_0$ is preserved, it is common use in accelerator physics to represent the particle motion in the trace space rather than the canonical phase space, i.e., the momentum is replaced by the divergence (\eg $p_x \rightarrow x' = p_x/p_0$).

The particle beam represents a statistical ensemble which is characterized by the moments of the underlying distribution function. First and second statistical moments are of peculiar interest, they define mean and variance of the distribution. In the case of a particle beam, the corresponding moments can be translated into beam parameters \cite{Streun17}:
\newline

\begin{tabular}{ll}
	\textbf{1$^{st}$ order:  beam centroid} \hspace*{1.5cm} & \textbf{2$^{nd}$ order:  beam distribution}\\
	mean values $<r_i>$  & rms values $<r_i^2>$ and correlations $<r_i r_j>$	\\
	- beam momenta $p_x, p_y, p_z$ & - momentum spread $\sigma_{\Delta p/p}$\\
	\hspace*{0.2cm} movement along $z$--axis: & \\
	\hspace*{0.4cm} $\Rightarrow~~p_z \approx p_0 \gg p_x, p_y$ & \\
	- beam location $z(t)$ & - bunch length $\sigma_{\Delta z}$\\
	- beam positions $x, y$ & - beam sizes $\sigma_{x, y}$\\
	- beam angles $x' = p_x/p_0$, $y'$ & - beam divergences $\sigma_{x', y'}$\\
	& \hspace*{0.2cm} correlations$\ldots$
\end{tabular}

The task of beam instrumentation is to extract information from the beam in order to measure the statistical moments. For that purpose a connection between the measuring device (monitor) and the beam has to be established, and information must be transferred from the beam particles to the monitor. Each information transfer is usually expressed in terms of an interaction which for beam instrumentation applications should preferably
\begin{itemize}
	\item[(a)] be non-disturbing for the beam;
	\item[(b)] be strong in order to provide good signal quality;
	\item[(c)] act over a long range such that measuring devices are placed in larger distances from the beam.	
\end{itemize}

\begin{table}[!t]
	\begin{center}
		\caption{Comparison of fundamental interactions.}
		\label{tab:Interaction}
		\begin{tabular}{lllll}
			\hline\hline
			& \textbf{Gravitational}  & \textbf{Weak} & \textbf{Electromagnetic} & \textbf{Strong}\\
			\hline
			Acting on: & mass-energy & flavour & electric charge & colour charge\\
			Particles experiencing: & all particles  & quarks, leptons & electrically charged & quarks, gluons\\
			& with mass & & particles & \\
			Exchange particle: & Graviton & \PWpm, \PZzero & \Pgamma & \Pg\\
			Relative strength: & 6 $\times$ 10$^{-39}$ & 10$^{-5}$ & 1/137 & 1\\
			Range / m: & $\infty$ & 10$^{-18}$ & $\infty$ & 10$^{-15}$\\
			\hline\hline
		\end{tabular}
	\end{center}
\end{table}
\Tref{tab:Interaction} shows a compilation of the four fundamental interactions. Comparing their relative strengths and their interaction ranges, it is obvious that the electromagnetic force is well suited for beam instrumentation. Therefore almost all monitors which are in use for accelerator beam diagnostics rely on the exploitation of the particle electromagnetic field. However, the electromagnetic force acts on the electric charge of a particle. Thus the discussion about beam instrumentation will be restricted to charged particle beams in the remainder of this report, instrumentation for \Pn- and \Pgamma--beams will not be covered.

\subsection{Maxwell's equations and Lorentz transformation}

This section briefly summarizes the main aspects of electromagnetism and relativity in order to help in understanding the following explanation. For a deeper understanding the reader is referred to specialized textbooks.

Maxwell's equations are a set of four fundamental equations that form the theoretical basis for describing classical electromagnetism. They are expressed below both in differential and integral form in SI units:
\begin{align}	
	\vec{\nabla} \cdot \vec{E}(\vec{r}, t) & = \frac{\rho(\vec{r}, t)}{\varepsilon_0} &
	\oiint_{\partial\Omega} \vec{E} \cdot \mathrm{d}\vec{S} & = \frac{1}{\varepsilon_0} \iiint_{\Omega} \rho\,\mathrm{d}V
	\label{eq:S2gauss1}\\
	\vec{\nabla} \cdot \vec{B}(\vec{r}, t) & = 0 &
	\oiint_{\partial\Omega} \vec{B} \cdot \mathrm{d}\vec{S} & = 0
	\label{eq:S2gauss2}\\
	\vec{\nabla} \times \vec{E}(\vec{r}, t) & = -\frac{\partial\vec{B}}{\partial t} (\vec{r}, t) &
	\oint_{\partial\Sigma} \vec{E} \cdot {\rm d}\vec{l} & = - \frac{\mathrm{d}} {\mathrm{d}t} \iint_{\Sigma} \vec{B} \cdot {\rm d}\vec{S}
	\label{eq:S2faraday}\\
	\vec{\nabla} \times \vec{B}(\vec{r}, t) & = \mu_0 \vec{J}(\vec{r}, t) + \frac{1}{c^2} \frac{\partial\vec{E}}{\partial t}  &
	\oint_{\partial\Sigma} \vec{B} \cdot \mathrm{d}\vec{l} & = \mu_0 \iint_{\Sigma} \vec{J} \cdot {\rm d}\vec{S} + \frac{1}{c^2} \frac{\mathrm{d}} {\mathrm{d}t} \iint_{\Sigma} \vec{E} \cdot \mathrm{d}\vec{S}
	\label{eq:S2ampere}
\end{align}
Equations~(\ref{eq:S2gauss1},\ref{eq:S2gauss2}) describe Gauss' law in electrostatics and of magnetism, Eq. ~(\ref{eq:S2faraday}) Faraday's law of induction, and Eq.~(\ref{eq:S2ampere}) Amp\`ere's law with Maxwell's addition of the displacement current.

In order to apply this set of equations in view of beam instrumentation, a point-like particle in the accelerator with charge $Q$ moving at constant velocity $\vec{v}$ is considered. As input for Maxwell's equations the particle properties are used, i.e., charge and current density which are expressed in the following form:
\[
\rho(\vec{r}, t) = Q \, \delta[\vec{r}(t)] \qquad \qquad
\vec{J}(\vec{r}, t) = Q \, \vec{v} \, \delta[\vec{r}(t)] \; .
\]
Output of the equations is the particle electromagnetic field which acts as the \emph{information carrier}, linking together particle and measuring device, thus providing information about the beam properties. In a modern accelerator however, the particles move with a velocity $|\vec{v}| \approx c$ close to the speed of light, and consequently relativistic effects have to be taken into account. Due to the velocity dependence of the current density, relativistic effects will also affect the particle electromagnetic field. This influence is widely exploited for beam instrumentation purposes, especially in the case of high energy electron accelerators.

The following paragraph briefly summarizes basic ideas about special relativity and helps understanding the electromagnetic field properties of relativistic particles which will be derived afterwards.

\subsubsection{Special relativity at a glimpse}

\begin{figure}[!h]
	\begin{center}
		\includegraphics[width=8cm]{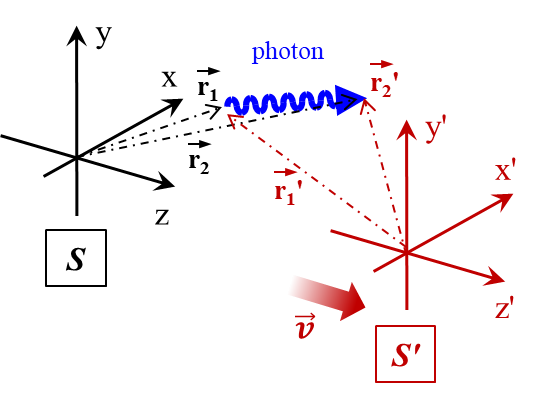}
		\caption{Observation of photon emission from two reference frames. The primed frame $S'$ moves with constant velocity $\vec{v}$ in the $z$-direction with respect to the fixed reference frame $S$.}
		\label{fig:S2relativity}
	\end{center}
\end{figure}

The theory of special relativity can be derived formally from a small number of postulates; Einstein formulated his theory on the basis of the following two:
\begin{description}
	\item[principle of relativity]: the laws of physics are invariant under a transformation between two coordinate frames moving at constant velocity with respect to each other (Lorentz invariance);	
	\item[invariance of speed of light]: the speed of light $c$ is the same for all observers, independent of the relative motion of the source.
\end{description}
By way of illustration two reference frames are considered, as shown in \Fref{fig:S2relativity}. The primed frame $S'$ moves with constant velocity $\vec{v}$ in the $z$-direction with respect to the fixed reference frame $S$. Two observers sitting in the origin of each frame follow the photon emission from a source point.
Due to the invariance of the speed of light $c$ the photon propagation velocity is the same in both frames and can be expressed as
\[
c = \frac{|\vec{r}_2 - \vec{r}_1|}{t_2 - t_1} = \frac{|\vec{r}^{~\prime}_2 - \vec{r}^{~\prime}_1|}{t^{~\prime}_2 - t^{~\prime}_1} = \left|\frac{\mathrm{d}\vec{r}}{\mathrm{d}t}\right| = \left|\frac{\mathrm{d}\vec{r}^{~\prime}}{\mathrm{d}t^{\prime}}\right| \stackrel{!}{=} \text{const.}
\]
From this expression the following invariant is derived
\[
\mathrm{d}(ct)^2 - \mathrm{d}x^2 - \mathrm{d}y^2 - \mathrm{d}z^2 = 0 \, ,
\]
which gives the same value ($= 0$) in each frame. Based on this invariant, the \emph{Lorentz transformation} of four-vectors can be deduced, resulting in the following set of transformation equations for the space--time coordinates:
\begin{align}
	x^{\prime} & = x \, , & z^{\prime} & = \gamma (z - \beta\;ct) \, , \label{eq:S2LorentzTrafo}\\
	y^{\prime} & = y \, , & (ct)^{\prime} & = \gamma (ct - \beta\;z) \, , \nonumber
\end{align}
with
\begin{align}
	\beta & = \frac{|\vec{v}|}{c} & \qquad\text{the reduced velocity} \label{eq:S2beta}\\
	\intertext{and}
	\gamma & =  \frac{1}{\sqrt{1-\beta^2}}  & \qquad\text{the Lorentz factor.} \label{eq:S2gamma}
\end{align}
For the derivation of the Lorentz transformation equations it was assumed that (1) the primed frame $S'$ moves with constant velocity $\vec{v}$ in the $z$-direction with respect to the fixed reference frame $S$ as before, (2) both reference frames coincide at the time $t = t' = 0$, and (3) any point $z'$ is moving with the primed frame. The reverse transformation from the moving frame $S'$ to the fixed reference frame $S$ can simply be performed by the exchange of $\beta \rightarrow -\beta$.

In accelerator physics, the Lorentz factor and reduced velocity are widely in use and have to be calculated rather often. Below a short list of kinematical parameters is given which is useful in order to calculate $\beta$ and $\gamma$:
\begin{align}
	&\text{particle momentum:} & \vec{p} & = m \vec{v} = \gamma m_0 \vec{v} =  \gamma m_0 \vec{\beta} c \, , & & \\
	&\text{total energy:} & E & = m c^2 = \gamma m_0 c^2 & \Rightarrow\hspace*{0.5cm}	& E^2  = (pc)^2 + (m_0c^2)^2 \, ,
	\label{eq:S2MassShell}\\
	&\text{kinetic energy:} & E & = T_{kin} + m_0c^2  & \Rightarrow\hspace*{0.5cm} & T_{kin}  = m_0c^2 (\gamma - 1) \, .
\end{align}
Here $m_0$ is the particle rest mass and $m_0c^2$ the rest mass energy. From this compilation it follows that the  Lorentz factor and reduced velocity can simply derived from
\begin{equation}
\beta = \frac{pc}{E} \, , \qquad\mathrm{and}\qquad \gamma = \frac{E}{m_0c^2} = 1 + \frac{T_{kin}}{m_0c^2} \, . \label{eq:S2bg}
\end{equation}
As an example, $\beta$ is calculated for a proton with total energy of $E$ = 1\UTeV{}. The proton rest mass energy amounts to $m_0c^2 =$ 938\UMeV{}, and according to Eq.~(\ref{eq:S2bg}) the Lorentz factor is $\gamma = \frac{E}{m_0c^2} = 1066.1$. With the help of Eq.~(\ref{eq:S2gamma}) $\beta$ is rewritten as $\beta = \sqrt{1-\gamma^{-2}} = 0.99999956 \approx 1$, i.e., the proton is fully relativistic.

So far the particle dynamics can be calculated, i.e., it is possible to transform the particle trajectory from a rest frame $S(x, y, z, ct)$  into a moving frame $S'(x', y', z', ct')$ and vice versa according to the Lorentz transformation and with knowledge of $\beta$ and $\gamma$. However, the transformation of the \emph{information carrier}, the electromagnetic field, is of special interest and will be summarized in the subsequent paragraph.

\subsubsection{Transformation of the electromagnetic field}

As before, the primed frame $S'$ moves with constant velocity $\vec{v}$ in the $z$-direction with respect to the fixed reference frame $S$. For the electromagnetic field, vector potential $\vec{A}(\vec{r}, t)$ and scalar potential $\phi(\vec{r}, t)$ are considered. Similar to the space--time coordinates $(ct, \vec{r})$ both potentials can be combined to form a four-vector which transforms in analogy to Eq.~(\ref{eq:S2LorentzTrafo}). Exploiting the relation
\begin{equation}
\vec{B} (\vec{r}, t) = \vec{\nabla} \times \vec{A}(\vec{r}, t)\, , \qquad\qquad \vec{E} (\vec{r}, t) = -\vec{\nabla} \phi(\vec{r}, t) - \frac{\partial}{\partial t} \vec{A}(\vec{r}, t)
\label{eq:S2potential}
\end{equation}
between the potentials and the electromagnetic field, the transformation for the latter can be expressed as
\begin{align}
	E^{\prime}_x & = \gamma \left[ E_x - v B_y \right] \, , & B^{\prime}_x & = \gamma \left[ B_x + \frac{v}{c^2} E_y \right] \, , \nonumber \\
	E^{\prime}_y & = \gamma \left[ E_y + v B_x \right] \, , & B^{\prime}_y & = \gamma \left[ B_y - \frac{v}{c^2} E_x \right] \, , \label{eq:S2trafoEB}\\
	E^{\prime}_z & = E_z\, , & B^{\prime}_z & = B_z \, . \nonumber
\end{align}
The comparison of Eq.~(\ref{eq:S2trafoEB}) and Eq.~(\ref{eq:S2LorentzTrafo}) indicates a fundamental difference in the structure of the transformation equations. This is due to the fact that the electromagnetic field vector cannot form a four-vector as it is the case for the space--time coordinates: the $\vec{E}$--vector has polar, the $\vec{B}$--vector axial symmetry.

Furthermore it is interesting to note that the Lorentz transformation Eq.~(\ref{eq:S2trafoEB}) combines electric and magnetic fields. This may cause the existence of a field type in one of the frames even if it is absent in the other one. This aspect will be discussed in the example described in the next section.

\subsection{Electromagnetic field of a moving charge}

\begin{figure}[!h]
	\begin{center}
		\includegraphics[width=12.cm]{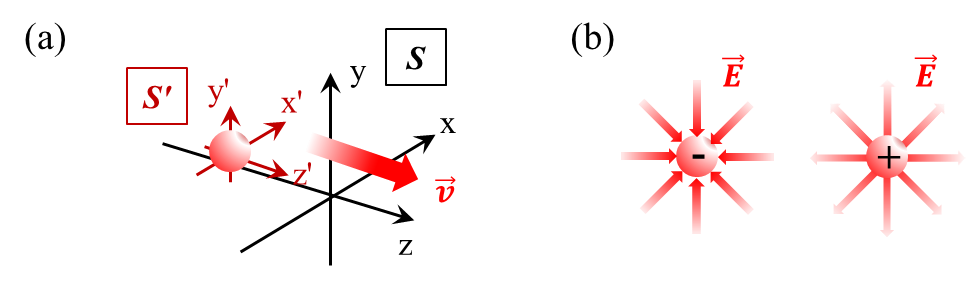}
		\caption{(a) Sketch of geometry: the moving frame $S'$ is the rest frame of the point charge, the fixed one $S$ the laboratory frame. (b) The point charge has a radial symmetrical field distribution in its rest frame $S'$.}
		\label{fig:S2pcharge}
	\end{center}
\end{figure}

In the following a point charge is considered which moves with constant velocity $\vec{v}$ in the $z$-direction in the laboratory frame (fixed reference frame $S$), see \Fref{fig:S2pcharge}(a). The moving frame $S'$ is the rest frame of the point charge, the field in this frame is the Coulomb field, i.e., in $S'$ there is a pure electric and no magnetic field.
The task is to calculate the field of the moving charge in the laboratory frame. This requires the transformation of the Coulomb field from the moving frame $S'$ to the fixed frame $S$, i.e., the transformation equations Eqn.~(\ref{eq:S2LorentzTrafo}) and (\ref{eq:S2trafoEB}) have to be applied using $\vec{v} \rightarrow -\vec{v}$.

\begin{figure}[!t]
	\begin{center}
		\includegraphics[width=13.cm]{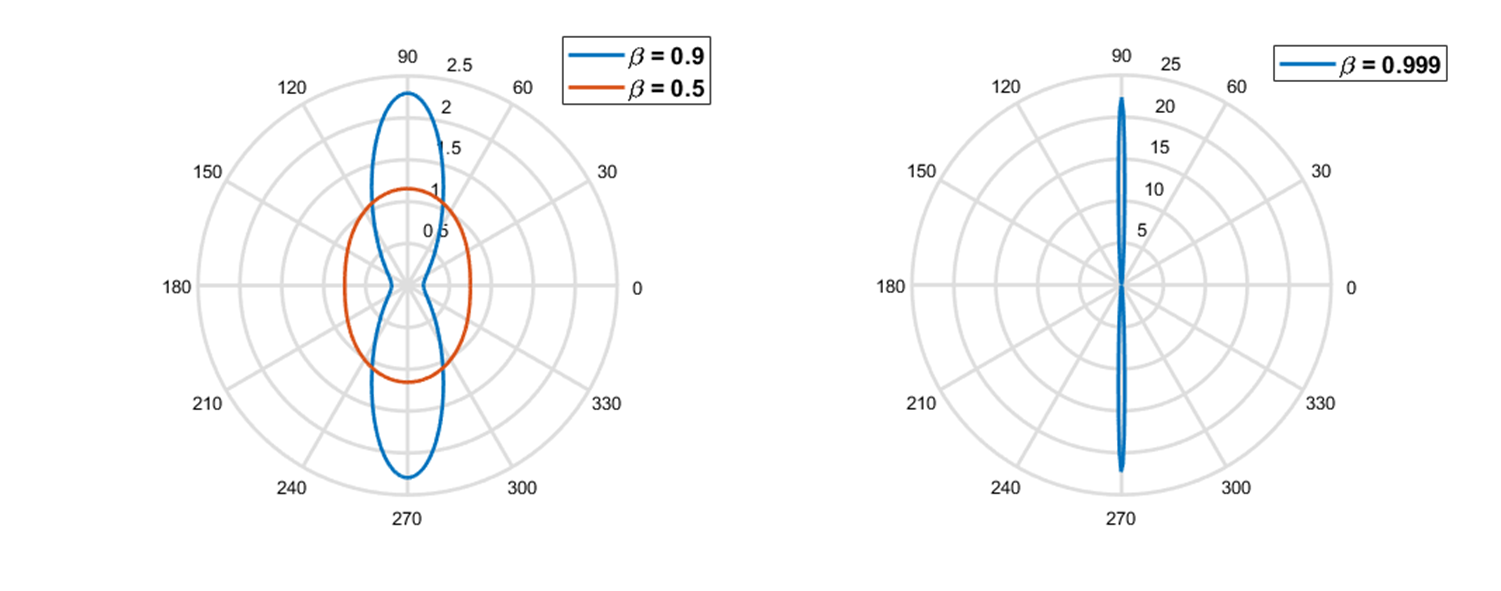}
		\caption{Relativistic modification factor according to Eq.~(\ref{eq:S2Coul}), plotted in polar coordinates for different particle velocities $\beta$. For small velocity the field has an almost rotational symmetry and becomes strongly squeezed in transverse direction with increasing $\beta$.}
		\label{fig:S2Rmodify}
	\end{center}
\end{figure}

The particle Coulomb field, expressed in its primed rest frame $S'$, is given by
\begin{equation}
\vec{E}^{~\prime}(\vec{r}^{~\prime}) = \frac{Q}{4\,\pi\,\varepsilon_0} \frac{\vec{r}^{~\prime}}{r^{\prime~3}}
= \frac{Q}{4\,\pi\,\varepsilon_0} \frac{1}{[x'^{~2} + y'^{~2} + z'^{~2}]^{3/2}}
\begin{pmatrix}
x' \\ y' \\ z'
\end{pmatrix}
\label{eq:S2Coulomb}
\end{equation}
and has radial symmetry as schematically depicted in \Fref{fig:S2pcharge}(b). In order to calculate the field in the laboratory frame, two consecutive Lorentz transformations have to be carried out. As the first step, the field is transformed according to Eq.~(\ref{eq:S2trafoEB}), but still expressed in the coordinates of the primed frame: $\vec{E}'(\vec{r}^{~\prime}) \rightarrow \vec{E}(\vec{r}^{~\prime})$. As the second step, the Lorentz transformation is applied for the space coordinates $\vec{E}(\vec{r}^{~\prime}) \rightarrow \vec{E}(\vec{r})$, resulting in
\begin{equation}
\vec{E}(\vec{r}, t) = \frac{Q}{4\,\pi\,\varepsilon_0} \frac{\gamma}{[x^2 + y^2 + \gamma^2 (z - vt)^2]^{3/2}}
\begin{pmatrix}
x \\ y \\ z - vt
\end{pmatrix}
~ .
\label{eq:S2Coulomb2}
\end{equation}
For the subsequent discussion the spatial field extension is more interesting rather than the spatio-temporal distribution. Therefore the static case for $t = 0$ is considered. Furthermore the angle $\vartheta$ is introduced as angle between direction of motion $\hat{z}$ and observation point $\vec{r}$ according to $\cos\vartheta = \frac{z}{r}$. Using this substitution the field in the laboratory frame is rewritten in the form
\begin{eqnarray}
\vec{E}(\vec{r}, t) & = & A_{rel} \cdot  \frac{Q}{4\,\pi\,\varepsilon_0} \frac{\vec{r}}{r^3} \label{eq:S2Coul}\\
\text{with} && A_{rel} = \frac{1 - \beta^2}{(1 - \beta^2 \sin^2\vartheta)^{3/2}} \;  .\nonumber
\end{eqnarray}
Equation~(\ref{eq:S2Coul}) represents the Coulomb field of a point charge with a relativistic modification factor $A_{rel}$ which depends on the beam energy resp. the reduced velocity $\beta$.
In \Fref{fig:S2Rmodify} $A_{rel}$ is plotted in polar coordinates with the particle velocity $\beta$ as parameter. As can be seen, with increasing particle speed the longitudinal field component is strongly suppressed while the amplitude of the transverse one is steadily increasing.
If the velocity $v$ approaches the speed of light $c$, the electromagnetic field has a pure transverse characteristics. This can easily be proved by Eq.~(\ref{eq:S2Coul}) if $\vartheta$ is set to $0^{\circ} (180^{\circ})$ in case of the longitudinal component, resp. to $90^{\circ} (270^{\circ})$ in case of the transverse one, resulting in
\begin{equation}
E_{\|}(r) = \frac{1}{\gamma^2} \cdot \frac{Q}{4\,\pi\,\varepsilon_0} \frac{1}{r^2} \; ,
\qquad\qquad
E_{\bot}(r) = \gamma \cdot \frac{Q}{4\,\pi\,\varepsilon_0} \frac{1}{r^2} \; .
\label{eq:S2CoulField}
\end{equation}

\begin{figure}[!b]
	\begin{center}
		\includegraphics[width=12.cm,clip]{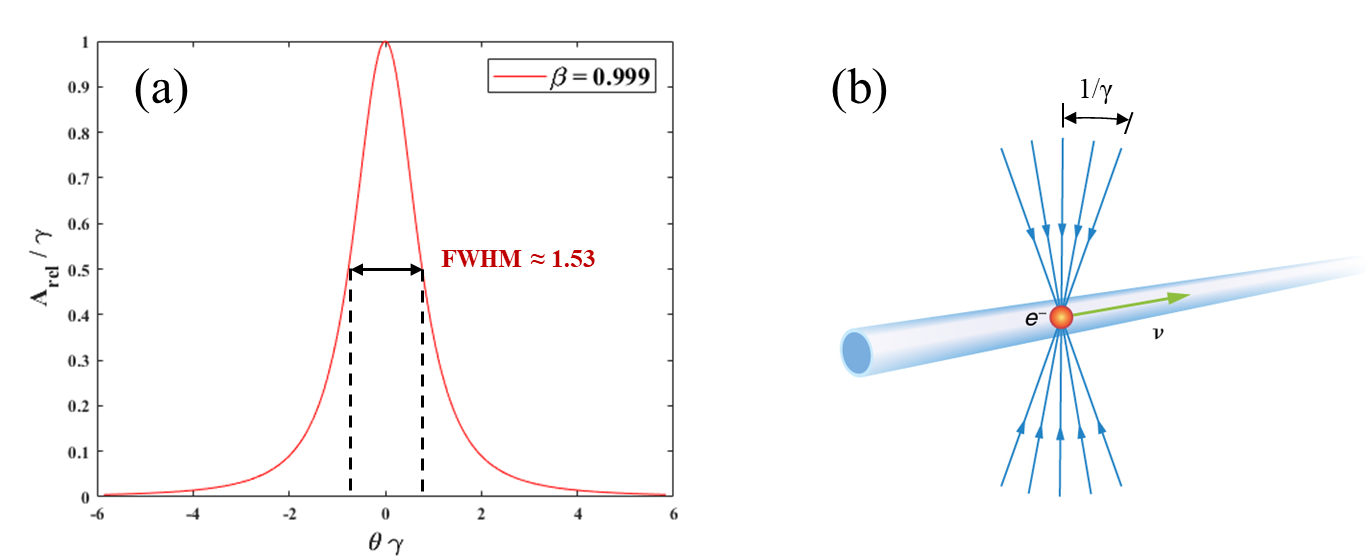}
		\caption{(a) Relativistic modification factor, normalized to its maximum value according to Eq.~(\ref{eq:S2Coul}). (b) Illustrative picture to describe the Lorentz boost: the particle electric field is represented by a bundle of field lines with half opening angle $1/\gamma$. Picture taken from \Bref{Relativity}.}
		\label{fig:S2Arel}
	\end{center}
\end{figure}
In order to characterize the electric field squeeze caused by the Lorentz boost, the relativistic modification factor $A_{rel}$ is investigated. While the squeezed field is strongly concentrated perpendicular to the direction of motion, it is suitable to define the angle in Eq.~(\ref{eq:S2Coul}) with respect to this direction rather than to the longitudinal ($z$) one. Therefore, the angle $\vartheta$ is replaced by $\theta = \frac{\pi}{2} - \vartheta$ in Eq.~(\ref{eq:S2Coul}) . Plotting $A_{rel}$ normalized to its maximum value $\gamma$ as a function of $\theta$ results in a narrow distribution as shown in \Fref{fig:S2Arel}(a). A measure for the field squeeze is the full width half maximum (FWHM) of this distribution which can be approximated by
\begin{equation}
\Delta\theta_{FWHM} = 1.53 \, \frac{1}{\gamma} \, ,
\label{eq:S2width}
\end{equation}
in the case of $\gamma \gg 1$, i.e., with increasing particle energy the electric field is increasingly `flattened' towards the plane perpendicular to the direction of motion. This issue is widely described in an illustrative picture as shown in  \Fref{fig:S2Arel}(b): the particle electric field is simply described as bundle of field lines with characteristic half opening angle $1/\gamma$.

Next, the magnetic field is considered. While there is no field in the particle rest frame $S'$, the Lorentz transformation gives rise to a $\vec{B}$ field in the laboratory frame according to Eq.~(\ref{eq:S2trafoEB}). With the electric field $\vec{E}^{~\prime}(\vec{r}^{~\prime})$ given by Eq.~(\ref{eq:S2Coulomb}) and the magnetic field $\vec{B}^{~\prime}(\vec{r}^{~\prime}) = 0$ in $S'$, the Lorentz transformation results in
\begin{equation}
\vec{B}(\vec{r}, t) = \frac{\mu_0\,Q}{4\,\pi} \frac{v\,\gamma}{[x^2 + y^2 + \gamma^2 (z - vt)^2]^{3/2}}
\begin{pmatrix}
-y \\ x \\ 0
\end{pmatrix}
\label{eq:S2Bfield}
\end{equation}
for the field in the laboratory frame $S$. Analogous to the discussion about the electric field, for the magnetic field only the spatial field distribution (static case for $t = 0$) is considered. Then it is interesting to investigate the non-relativistic limiting case $\gamma \rightarrow 1$ which is given by
\begin{equation}
\vec{B}(\vec{r}) = \frac{\mu_0}{4\,\pi} Q \frac{1}{r^3}
\begin{pmatrix}
-v y \\ v x \\ 0
\end{pmatrix}
= \frac{\mu_0}{4\,\pi} Q \frac{\vec{v} \times \vec{r}}{r^3} \;  .
\label{eq:S2BioSavart}
\end{equation}
The latter equation is nothing else than the Biot--Savard law in magnetostatics.

Furthermore, the comparison of Eq.~(\ref{eq:S2Coulomb2}) and Eq.~(\ref{eq:S2Bfield}) indicates the following important relation between the fields of a moving charged particle:
\begin{equation}
\vec{B}(\vec{r}, t) = \frac{1}{c^2} \; \vec{v} \times \vec{E}(\vec{r}, t) \;  ,
\label{eq:S2EBfield}
\end{equation}
i.e., both fields are normal to each other. The same issue can be seen, using the fact that the scalar product of both fields is invariant under Lorentz transformation:
\[
\vec{E}(\vec{r}, t) \cdot \vec{B}(\vec{r}, t) =
\vec{E}^{~\prime}(\vec{r}^{~\prime}, t^{\prime}) \cdot \vec{B}^{~\prime}(\vec{r}^{~\prime}, t^{\prime}) \;  .
\]
According to this equation, if $\vec{E}$ or $\vec{B}$ is zero in one reference frame (as it is the case for the magnetic field in $S'$), in other frames both fields are automatically normal to each other.

\subsection{Conclusion}

In order to summarize this chapter the following items are of importance for particle beam instrumentation and should be kept in mind:
\begin{itemize}
	\item Responsible for the information transfer between beam particles and measuring device (monitor) is the electromagnetic interaction which couples to the electric charge of the particles. Therefore beam instrumentation for neutral particles (\Pn, \Pgamma $\ldots$) will not be covered here.
	\item The electromagnetic field of beam particles acts as information carrier and is utilized for the measurement of beam properties.
	\item For the description of the particle electromagnetic field a basic knowledge of Maxwell's equations and special relativity is necessary.
	\item The electromagnetic field of a point charge is strongly affected by relativistic effects. Characteristic parameter therefore is the Lorentz factor $\gamma = E/m_0c^2$. $\gamma$ rises either with increasing beam energy or with decreasing rest mass (energy). Consequently relativistic effects are especially pronounced for high-energy electron or positron beams.
	\item The electric field of an ultra-relativistic particle is almost transversal and scales with the Lorentz factor according to
	\[
	\vec{E}_{\|} \propto \gamma^{-2}\;, \qquad\qquad \vec{E}_{\bot} \propto \gamma \;  .
	\]
	The particle magnetic field which can be measured in the laboratory frame is a consequence of the Lorentz transformation.
\end{itemize}

\section{Measurement principles}

In this chapter the underlying measurement principles are introduced. A monitor which is used for charged particle beam instrumentation has to extract information from the beam particles and pass it to the detector part on the basis of electromagnetic interaction. This interaction can be applied in different ways:
\begin{enumerate}
	\item Coupling to the particle electromagnetic field which is carried by the moving charge.
	\item Coupling to the particle electromagnetic field which is separated from the moving charge and freely propagating as radiation.
	\item Exploiting the energy deposition due to the interaction of the particle electromagnetic field with matter.
	\item Exploiting the interaction of an external electromagnetic field with the charged particle.
\end{enumerate}
In the following these different mechanisms will be discussed more detailed.

\subsection{Coupling to the particle electromagnetic field carried by the moving charge}

This kind of interaction is widely applied e.g., for beam charge and beam current measurements, for beam position monitoring, but also for bunch length measurements.

Following Refs.~\cite{Belohrad10,Belohrad11}, in order to understand how the beam signal is generated, the concept of a wall image current is introduced. It is assumed that the charged particles travel through metallic vacuum chambers of the accelerator. These chambers are evacuated tubes, bounded by electrically conducting material with 
zero longitudinal resistance in the ideal case. Any moving charged particle creates an electromagnetic field. As explained in the previous section, the electric field is caused by the charge and the magnetic one by the charge movement. Due to the relativistic particle motion, the Lorentz boost contracts the electric field in the direction of motion,
\begin{figure}[ht]
	\begin{center}
		\includegraphics[width=5cm]{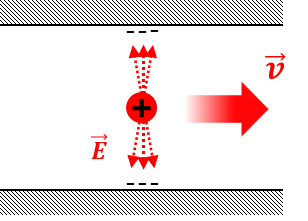}
		\caption{A moving charged particle induces image charges of opposite sign at the inner diameter of the vacuum chamber}
		\label{fig:S3WIC}
	\end{center}
\end{figure}
simply illustrated as a bundle of field lines with characteristic half opening angle $1/\gamma$. At the inner diameter of the vacuum chamber image charges of opposite sign are induced, cf. \Fref{fig:S3WIC}. As the beam particles travel, they are always accompanied by these mirror charges. The mirror charges are called the \emph{wall image current} (WIC) and form an inseparable counterpart to the beam current.

According to Gauss' flux theorem Eq.~(\ref{eq:S2gauss1}) 
\[
\oiint_{\partial\Omega} \vec{E}(\vec{r}, t) \cdot {\rm d}\vec{S} = \frac{1}{\varepsilon_0} \iiint_{\Omega} \rho(\vec{r}, t)\,\mathrm{d}V \;  ,
\]
charge and image charge neutralize each other outside the vacuum chamber, i.e., $\rho = 0$ if the integration volume $\Omega$ contains both charge contributions. As a consequence there is no electric field outside the beam pipe, and it is not possible to couple to the particle beam 
electric field.

An equivalent situation holds for the particle beam magnetic field which is described by Amp\`ere's law Eq.~(\ref{eq:S2ampere})
\[
\oint_{\partial\Sigma} \vec{B}(\vec{r}, t) \cdot \mathrm{d}\vec{l} = \mu_0 \iint_{\Sigma} \vec{J}(\vec{r}, t) \cdot {\rm d}\vec{S} \;  .
\]
Choosing an integration path $\partial\Sigma$ as a closed circle around the beam pipe outside the vacuum chamber and taking into account that the WIC has equal magnitude but opposite sign to the beam current (in 1$^{st}$ order), then the sum of beam and image current cancels out, i.e., $\vec{J} = 0$. As a consequence, the magnetic field outside the beam pipe is neutralized. Therefore it is also not possible to couple to the particle beam magnetic field outside the vacuum chamber.

More precisely, the electromagnetic fields will not vanish immediately outside the vacuum chamber. They are strongly attenuated with an attenuation factor which is specified in terms of the frequency dependent skin-depth length. However, for most practical applications in beam instrumentation, considering high resolution monitors mounted in an environment with non-magnetic vacuum chambers of high conductivity, the particle field outside the accelerator beam pipe is assumed to vanish.

In conclusion, it should be noted that there is no access to couple to the beam particle electromagnetic field outside a conventional metallic accelerator vacuum chamber. However, in order to generate useful information from the beam there are two possibilities for signal extraction:
\begin{itemize}
	\item coupling to the beam field inside the vacuum chamber;
	\item allowing the beam field to extend to the outside using materials with very low  conductivity.
\end{itemize}
Both methods are briefly described in the following.

\subsubsection{Beam field coupling inside vacuum chamber}

In order to circumvent the field cancellation outside the accelerator vacuum chamber, the antenna for signal extraction has to be placed between the primary charge/current (i.e., the particle beam) and the induced charge/current (i.e., in the metallic vacuum chamber) such that the integration paths in Eqs.~(\ref{eq:S2gauss1}) and (\ref{eq:S2ampere}) will not enclose both contributions. Thus, the coupling antenna has to be placed inside the vacuum chamber.

A moving charged particle possesses both an electric and a magnetic field, and it is possible to couple to both types. Depending on the way of coupling it is termed
\begin{description}
	\item[capacitive] if the particle electric field is used as signal source;
	\item[inductive] if the particle magnetic field is applied.
\end{description}
\begin{figure}[!t]
	\begin{center}
		\includegraphics[width=15.5cm]{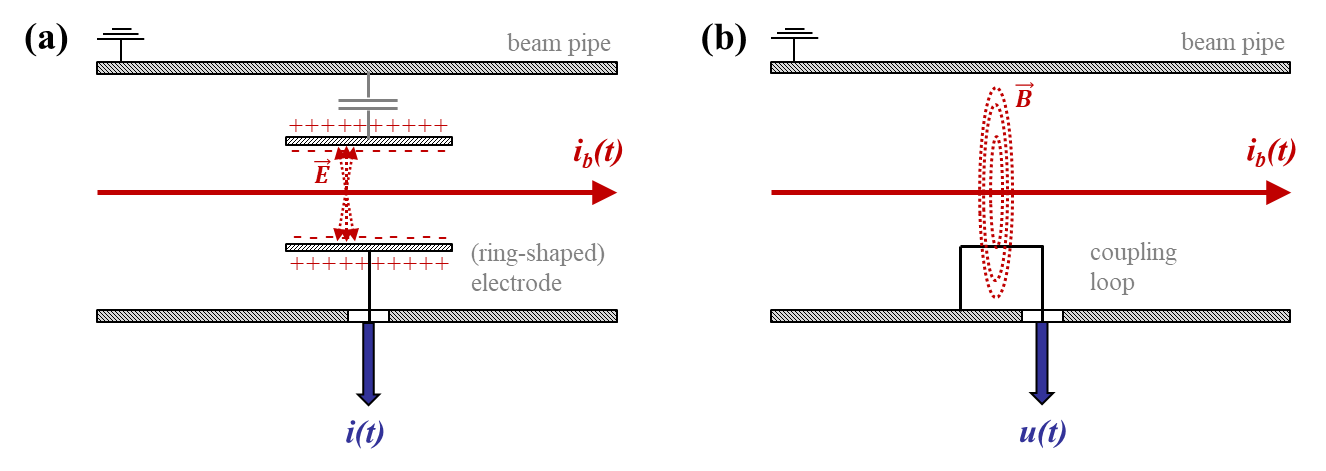}
		\caption{Operational principle of (a) capacitive and (b) inductive signal extraction. The left picture represents a cut through a ring-shaped electrode.}
		\label{fig:S3coupling}
	\end{center}
\end{figure}
Figure \ref{fig:S3coupling} shows the schematic view of both mechanisms. In case of capacitive coupling, the moving charged particle passes an electrode and induces time-varying mirror charges on the electrode's surface via its electric field, thus creating a displacement current $i = \oint_{\partial\Sigma} \vec{H} \cdot \mathrm{d}\vec{l}$ which is driven by the potential difference between electrode and vacuum chamber. The current is given by
\begin{equation}
i(t) = \varepsilon_0 \frac{\mathrm{d}} {\mathrm{d}t} \iint_{\Sigma_1} \vec{E} \cdot \mathrm{d}\vec{S} \;  ,
\label{eq:S3ampere}
\end{equation}
according to Amp\`ere's law, Eq.~(\ref{eq:S2ampere}). In case of inductive coupling, the moving charged particle passes the coupling loop and induces an induction voltage $u = \oint_{\partial\Sigma} \vec{E} \cdot {\rm d}\vec{l}$ via its time-varying magnetic field which is determined by
\begin{equation}
u(t) = -\frac{\mathrm{d}} {\mathrm{d}t} \iint_{\Sigma_2} \vec{B} \cdot {\rm d}\vec{S} \, ,
\label{eq:S3faraday}
\end{equation}
according to Faraday's law of induction Eq.~(\ref{eq:S2faraday}). 

Both monitor types are similar and can be used to extract the same information. Following the discussion in \Bref{Strehl06}, the signal ratios of both extraction methods are compared. For this, a cylindrical coordinate system is assumed with the particle motion $\vec{v} = v\,\hat{e}_z$ directed in $\hat{z}$-direction. Both fields obey a rotational symmetry with regard to the $z$-axis, and a further assumption is ultra-relativistic particle motion such that the electric field has only a component $E_r = |E\,\hat{e}_r|$. According to Eq.~(\ref{eq:S2EBfield}), the particle magnetic field in this case has only a component $B_{\vartheta} = |B\,\hat{e}_{\vartheta}|$ which is expressed as
\[
B_{\vartheta} = \frac{1}{c^2}\, v \, E_r = \frac{\beta}{c} \, E_r \;  .
\]
Inserting in Eqs.~(\ref{eq:S3ampere}) and (\ref{eq:S3faraday}), the signal ratio is
\[
\left|\frac{i(t)}{u(t)}\right| = \frac{c\,\varepsilon_0}{\beta}\,
\frac{\frac{\mathrm{d}}{\mathrm{d}t} \iint_{\Sigma_1} E_r \mathrm{d}S}
{\frac{\mathrm{d}}{\mathrm{d}t} \iint_{\Sigma_2} E_r \mathrm{d}S} \;  .
\]
$\Sigma_1$ is the area of the pickup electrode, $\Sigma_2$ that of the coupling loop. For most practical designs they are rather similar such that the ratio of both integral terms is of the order of one. In order to compare two voltages, broadband signal processing with an impedance of $R = 50~\Omega$ is assumed which results in
\begin{equation}
\left|\frac{R\,i(t)}{u(t)}\right| = \left|\frac{u_{cap}}{u_{ind}}\right| \approx \frac{R\,c\,\varepsilon_0}{\beta} = \frac{0.133}{\beta} \;   .
\label{eq:S3CapInd}
\end{equation}
From Eq.~(\ref{eq:S3CapInd}) it follows that inductive coupling has a higher sensitivity compared to the capacitive one. However, due to the high sensitivity of simple loop monitors to rapidly changing magnetic stray fields which are all the time present in an RF accelerator environment, usually capacitive pickups are used as beam monitors.

\subsubsection{Beam field extension to the outside}

\begin{figure}[!t]
	\begin{center}
		\includegraphics[width=15.5cm]{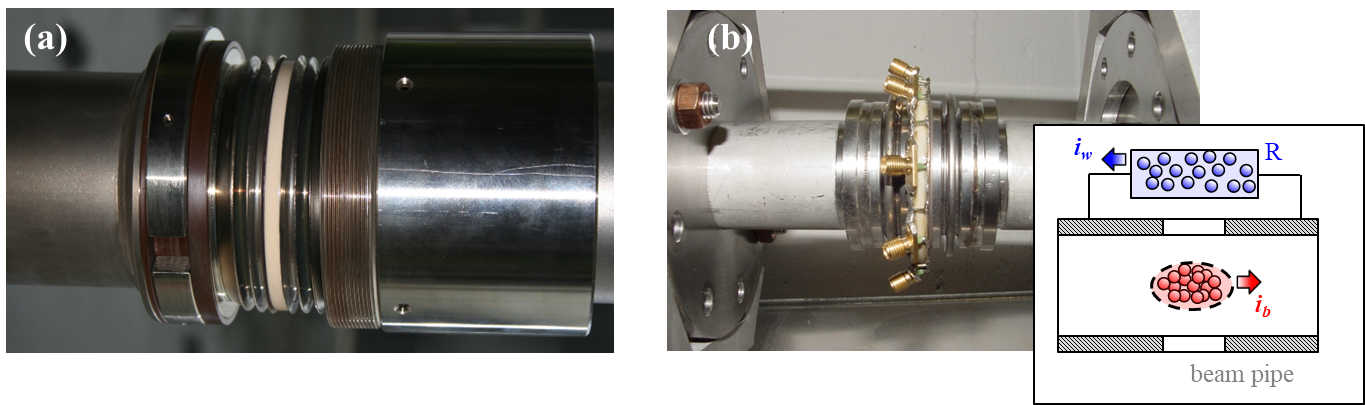}
		\caption{(a) Ceramic insert in the metallic vacuum chamber for a beam current monitor at HERAp (DESY Hamburg, Germany). (b) Principle and technical realization of a wall current monitor from FLASH (DESY Hamburg, Germany). Photos courtesy K. Knaack, R. Neumann, N. Wentowski (DESY).}
		\label{fig:S3outside}
	\end{center}
\end{figure}

Outside a metallic vacuum chamber the electromagnetic field of a moving charged particle is too weak to be exploited for beam instrumentation purposes. Nevertheless it is possible to couple from outside to the beam signal or to conduct it to the measurement device simply by breaking the conducting path in the chamber. Technically this is realized by inserting a non-conducting material, usually a ceramic, electrically in series with the metallic beam pipe. As an example \Fref{fig:S3outside}(a) shows a ceramic insert for a beam current monitor at the HERAp accelerator at DESY (Hamburg, Germany).

The missing metallic boundary allows the electromagnetic field to expand in the space outside of the vacuum chamber such that the filed coupling can be performed in air. Besides the longer field range it has the advantage of facilitating the monitor design because of several restrictions imposed by the vacuum environment (e.g., cleanliness and in-vacuum cooling) can be dropped. By installing a 
toroidal transformer close to the gap which consists of a material with high relative permeability, the AC component of the beam current can be measured. The monitor based on this concept is named AC or \emph{fast beam current transformer} (FBCT).

In the case of conducting the beam signal to the measurement device, the non-metallic interruption forces the WIC to find a new path. Favourable for beam instrumentation applications is that the alternative path for the WIC is under the control of the instrument designer. As shown in the example in \Fref{fig:S3outside}(b) the WIC (high-frequency part) can be guided to flow through a load resistance connected in series with the vacuum chamber such that a voltage drop proportional to the beam intensity can be measured across the gap. In addition, a photo of the technical realization of this monitor concept is shown, the so called \emph{wall current monitor} (WCM).

\subsubsection{Remarks}

Besides the signal extraction schemes discussed before, there exist alternative methods which are treated only briefly here. Further information can be found in Refs.~\cite{Wendt18,Gillespie18} in these proceedings.

Similar to the case of electromagnetic field expansion in the space outside of the vacuum chamber using a non-metallic insert, the field expansion can take place inside the vacuum chamber using an electromagnetic discontinuity in the metallic beam pipe. This principle is exploited in the case of cavity monitors to extract information about beam position and intensity. The field expansion results in an excitation of resonator modes within the cavity which is utilized as a passive, beam driven cavity monitor.

Another interesting application exploiting the particle electromagnetic field which is carried by the moving charge is to use it for environmental modifications. This principle is utilized for example in the case of bunch length measurements based on electro-optical principles. For these techniques it is essential that the charged particles which pass close to an electro-optical crystal (ZnTe or GaP) induce a change in the crystal refractive index (the so-called Pockels effect) via the interaction between the particle Coulomb field and the crystal. The information about the longitudinal profile is therefore encoded in a refractive index change which can be converted into an intensity variation by means of a combination of laser beam and polarizers.

\subsection{Coupling to the electromagnetic field emitted as radiation by the moving charge}

This kind of interaction is widely applied e.g., for beam size and profile measurements in the longitudinal and both transverse planes.

Radiation generated by high-energy particle beams is widely used for beam instrumentation. Depending on the mechanism of radiation generation, the emitted wavelength range extends from the THz up to the X-ray region, thus allowing us to measure beam profiles in the longitudinal and the transverse plane over a wide range. The information about the beam properties is generated from the electromagnetic fields which are separated from the charged particle itself. These freely propagating fields can be measured at large distances from the particle as radiation, even outside of the accelerator tunnel. Depending on the separation mechanism of the electromagnetic field, the process of radiation generation is named in a different way. Examples are synchrotron radiation, transition radiation, diffraction radiation, parametric X-ray radiation, Cherenkov radiation, and Smith--Purcell radiation. A comprehensive overview of the radiation generation from ultra-relativistic particles can be found for example in the textbooks of Refs. \cite{TerMikaelian72}--\cite{Potylitsyn11}. The subsequent discussion is mainly based on Refs.~\cite{Kube14,Kube07}, further information can be found there and in the references therein.

\subsubsection{Mechanism of radiation generation}

In the following the process of radiation generation is briefly explained in terms of a separation of the pseudo- or virtual photon field associated with the charged particle (the Weizs\"{a}cker--Williams approximation \cite{Weiz34,Will34}). In this picture, the various radiation processes appear as different ways to separate the virtual photons from the particle. 

Key point of the discussion is again the Lorentz contraction of the particle electromagnetic field which is characterized by the Lorentz factor $\gamma = E / m_0\,c^2$, see Eq.~(\ref{eq:S2bg}). In the case of ultra-relativistic particle energies the electric field is nearly transversal, the degree of contraction is described by the field opening angle 1/$\gamma$ and the field extension range scales proportional to $\gamma$, see \Fref{fig:S2Arel}(b) and Eq.~(\ref{eq:S2CoulField}).

In the limiting case $\gamma \rightarrow \infty$ the field would be completely transversal and correspond to a plane wave which is the classical description of a photon. This situation occurs either by considering a particle with zero rest mass (for example a real photon), or in the limiting case if the beam energy is increased into the ultra-relativistic regime. Due to the similarity between a real photon and the field of an ultra-relativistic particle, the action of this particle is described by so-called virtual or pseudo-photons. However, to measure radiation in the far field the virtual photon field bound to the beam particle has to be separated from the particle. In case of a circular accelerator this is achieved by a force acting on the charged particle which is caused by the magnetic field of accelerator (bending) magnets, and the resulting radiation is called {\it synchrotron radiation}. In the case of a linear accelerator, per definition there is no particle bending, but the separation can be achieved by acting on the virtual photons itself via structures that diffract the particle electromagnetic field away from the particle. The analogy between real and virtual photons can be exploited for better understanding: real photons can be refracted resp. reflected at a surface, the same holds for virtual photons. In this case the radiation is named {\it forward/backward transition radiation}. In classical optics the effect of edge diffraction is known, in the case of virtual photons the radiation effect is called {\it diffraction radiation}. Real photons can be diffracted at a grating, the same holds for virtual photons and the effect is called {\it Smith--Purcell radiation}. Finally, highly energetic real photons (X-rays) are diffracted at the 3D structure of a crystal, and if a charged particle beam traverses such crystal {\it parametric X radiation} is emitted.

At this point it has to be emphasized that hadrons have a comparatively large rest mass. As a consequence, $\gamma$ is much smaller than that for electrons. Therefore radiation based signal extraction is the exception rather than the rule at hadron accelerators.

It is very illustrative to consider the mechanism of radiation generation using the mass shell concept. This is a synonym for the mass hyperboloid (the hyperboloid in the energy momentum space) describing the solutions to Eq.~(\ref{eq:S2MassShell})
\[
E^2  = (pc)^2 + (m_0c^2)^2 \;  .
\]
\begin{figure}[!t]
	\begin{center}
		\includegraphics[width=10.cm]{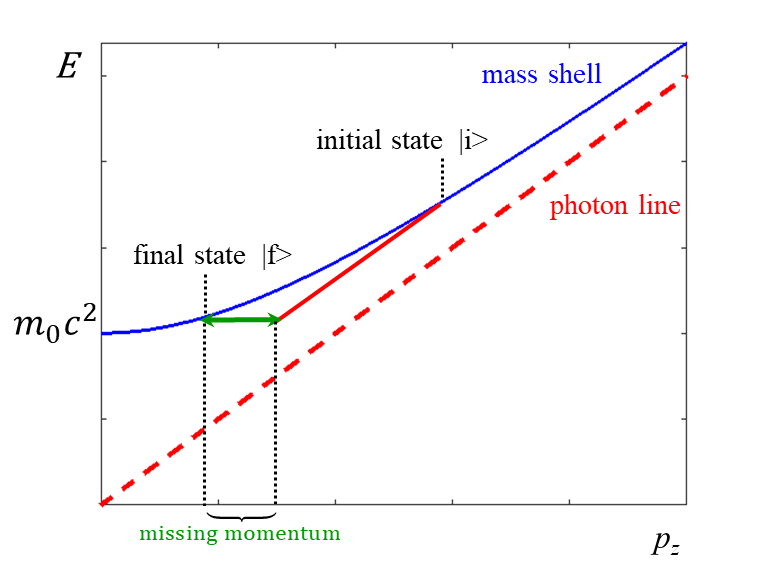}
		\caption{Energy momentum conservation and the concept of the mass shell. The behaviour of every real particle having mass is governed by Eq.~(\ref{eq:S2MassShell}), indicated by the solid mass shell line. Photons as massless particles follow the dashed photon line. A particle cannot undergo a direct transition from the initial state $\ket{i}$ to the final state $\ket{f}$ via radiation emission because energy and momentum conservation are not fulfilled simultaneously, a missing momentum has to be provided externally.}
		\label{fig:S3MassShell}
	\end{center}
\end{figure}
For simplification, in the ensuing discussion only the one-dimensional projection (onto the direction of particle motion in $\hat{z}$) will be considered, cf. \Fref{fig:S3MassShell}. Each real particle does satisfy this relation and is termed \emph{sitting on the mass shell}. If a particle looses energy it undergoes a transition from the initial state $\ket{i}$ to the final one $\ket{f}$. A photon as massless particle is also described by the energy--momentum relation. However, because of the missing mass term the expression can be simplified written as
\[
E  = pc = \hbar\omega \, ,
\]
which corresponds to the dashed line in \Fref{fig:S3MassShell}. If the particle energy loss is caused by radiation emission, then energy and momentum conservation must be ensured for the system consisting of particle and photon. As can be seen from \Fref{fig:S3MassShell}, a direct transition from  $\ket{i}$ to $\ket{f}$ is not possible under exclusive photon emission. A missing momentum remains which has to be provided externally, either as a radial force (synchrotron radiation in circular accelerators) or as diffractive material structure (radiation generation in linear accelerators).

A unique situation is the case of Cherenkov radiation. This radiation type is emitted in matter where the speed of light is determined by $c/n$ with $n(\omega)$ the refractive index of the material. If $n(\omega) > 1$ then the photon line slope in \Fref{fig:S3MassShell} is decreased such that it can intersect the mass shell in two points. In other words, in the case of Cherenkov radiation a direct transition from  $\ket{i}$ to $\ket{f}$ is possible without external momentum.

In the following sections, some of the radiation mechanisms which are frequently used for beam diagnostic applications in circular and linear accelerators are briefly described.

\subsubsection{Circular motion: synchrotron radiation}

Synchrotron radiation (SR) is a versatile tool for beam profile measurements due to its non-destructive nature. While in principle SR from insertion devices or bending magnets can be utilized for monitoring beam parameters, in reality most accelerators use bending magnet radiation based profile monitors because of space limitations. Due to the relativistic energy of the particles, the generated light has superior properties \cite{Hofmann04}: the process of radiation generation is non-invasive and the radiation spectrum is continuous from infrared up to X-rays. As consequence the photon energy can be freely chosen according to the monitoring problem. Typically the spectrum is characterized by the critical energy
\begin{equation}
\hbar\omega_c = \frac{3}{2}\hbar c \frac{\gamma^3}{\rho} \, ,
\label{eq:S3critEnergy}
\end{equation}
with $\gamma$ the Lorentz factor and $\rho$ the dipole bending radius. The natural divergence of the radiation which depends on the polarization state is very small with a vertical opening angle of about $1/\gamma$ in case of horizontal polarization.

\begin{figure}[!h]
	\begin{center}
		\includegraphics[width=9cm]{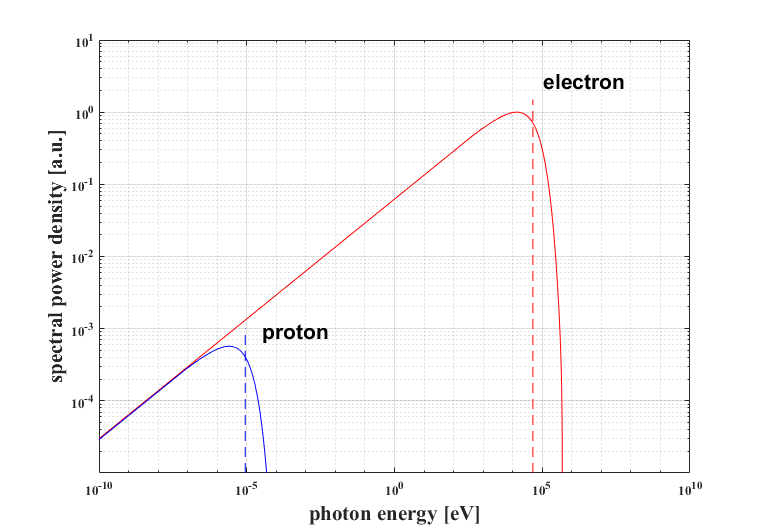}
		\caption{SR spectral power density for an electron and a proton having the same kinetic energy $T$ = 20\UGeV{}. The bending magnet radius is $\rho$ = \Unit{370}{m}. The dashed lines indicate the critical energy according to Eq.~(\ref{eq:S3critEnergy}).}
		\label{fig:S3SRspec}
	\end{center}
\end{figure}

Figure \ref{fig:S3SRspec} shows the calculated spectral power density for an electron and a proton having the same total energy. As can be seen from this comparison, the radiation emission from the proton is strongly suppressed because of the larger rest mass and the correspondingly smaller Lorentz factor according to Eq.~(\ref{eq:S2bg}).

\begin{figure}[!t]
	\begin{center}
		\includegraphics[width=12cm]{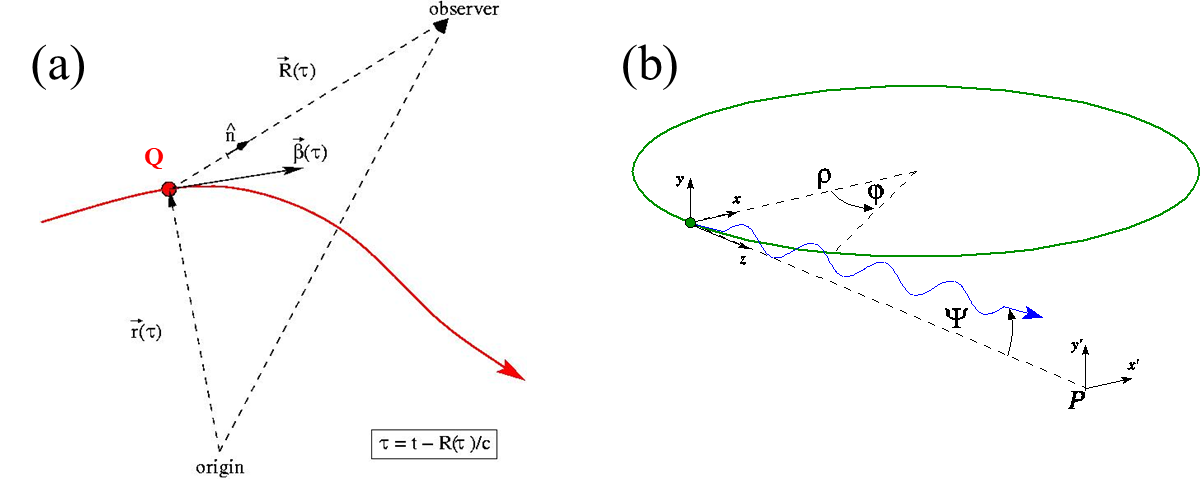}
		\caption{Geometry for the description of the Li\'{e}nard--Wiechert fields for a particle in arbitrary (left) and linear (right) motion}
		\label{fig:S3SyncGeo}
	\end{center}
\end{figure}

SR is typically used in order to image the beam and measure the transverse beam size for beam emittance determination. In this context the monitor resolution is of interest which requires  knowledge of the electromagnetic field of the source. In the following the field derivation will briefly be outlined, starting from the Li\'{e}nard--Wiechert potentials. They describe the classical electromagnetic effect of an electric point charge $Q$ in arbitrary motion in terms of a vector potential and a scalar potential in Lorentz gauge and are the basis of the complete, relativistically correct, time-varying electromagnetic fields. With the geometry depicted in \Fref{fig:S3SyncGeo}(a), the Li\'{e}nard--Wiechert potentials are expressed as
\begin{equation}\label{eq:S3LenWpot}
\phi(\vec{R}, t) = \frac{Q}{4\pi\,\varepsilon_0}\,\left(\frac{1}{R (1 - \hat{n}\cdot \vec{\beta})}\right)_{\tau}, \qquad \vec{A}(\vec{R}, t) = \frac{\mu_0\,Q}{4\pi}\, \left(\frac{\vec{\beta}c}{R (1 - \hat{n}\cdot \vec{\beta})}\right)_{\tau} \; .
\end{equation} 
The index $\tau$ indicates that the potentials have to be evaluated at the retarded time $\tau = t - R(\tau)/c$. With knowledge of these potentials, the field are derived according to Eq.~(\ref{eq:S2potential}). The common way found in most textbooks about electrodynamics is to deduce the fields in the time domain
\begin{eqnarray}\label{eq:S3LenWf}
&& \hspace*{-1.cm} \vec{E}(\vec{R}, t) = \frac{Q}{4\pi\,\varepsilon_0}\,
\left(\frac{(1 - \beta^2)(\hat{n}-\vec{\beta})}
{R^2(1-\hat{n}\cdot\vec{\beta})^3} +
\frac{\hat{n}\times[(\hat{n}-\vec{\beta}) \times \dot{\vec{\beta}}]}
{cR(1-\hat{n}\cdot\vec{\beta})^3}\right)_{\tau}\\
&& \hspace*{2cm} \vec{B}(\vec{R}, t) = \frac{1}{c}\,(\hat{n}\times\vec{E})_{\tau} \nonumber 
\; .
\end{eqnarray}
The first term in Eq.~(\ref{eq:S3LenWf})  which does not depend on the acceleration (the so-called velocity term) indicates constant particle motion. It can be shown that this term and Eqs.~(\ref{eq:S2Coulomb2},\ref{eq:S2Bfield}) derived on the basis of Lorentz transformed fields are equivalent. In far field approximation, the velocity term is usually omitted because it scales quadratically with the distance to the observer. In addition, to get rid of the retarded time, the fields are transformed in the Fourier domain, resulting in
\begin{eqnarray*}
\vec{E}_{\omega}(\omega) & \approx & \frac{i\, \omega Q}{4\pi\sqrt{2\pi}\varepsilon_0 c R} \int\limits_{-\infty}^{+\infty} {\mathrm{d}\tau} \; [\hat{n} \times [\hat{n} \times \vec{\beta}]] \; e^{-i\omega(\tau - R(\tau)/c)}\\
\vec{B}_{\omega}(\omega) & = & \frac{1}{c} \; [\hat{n} \times \vec{E}_{\omega}(\omega)] \; .
\end{eqnarray*}
If the special case of particle motion on a circular orbit is considered as depicted in \Fref{fig:S3SyncGeo}(b), the fields can be expressed in the following way
\begin{align}
E_{\omega, x}(\omega) & = & E_{\sigma} & = & -\frac{\sqrt{3} Q \gamma}{(2\pi)^{3/2}\varepsilon_0 c R} \; \frac{\hbar\omega}{2\hbar\omega_c}\; (1 + \gamma^2\Psi^2) \;
\mathrm{K_{2/3}}\left[\frac{\hbar\omega}{2\hbar\omega_c}\, (1 + \gamma^2\Psi^2)^{3/2}\right] \, , \label{eq:S3sync}\\
E_{\omega, y}(\omega) & = & E_{\pi} & = & i\frac{\sqrt{3} Q \gamma}{(2\pi)^{3/2}\varepsilon_0 c R} \frac{\hbar\omega}{2\hbar\omega_c} \; \gamma\Psi \sqrt{1 + \gamma^2\Psi^2} \;
\mathrm{K_{1/3}}\left[\frac{\hbar\omega}{2\hbar\omega_c}\, (1 + \gamma^2\Psi^2)^{3/2}\right] \, . \nonumber
\end{align}
K$_{1/3}$ and K$_{2/3}$ are modified Bessel functions of the second kind, $\hbar\omega$ the photon energy, and $\Psi$ the photon emission angle in the vertical plane as indicated in  \Fref{fig:S3SyncGeo}(b). Equation~(\ref{eq:S3sync}) is the standard representation for the fields which is usually used in textbooks about SR, see e.g., \Bref{Hofmann04}. The horizontal field component is denoted as $\sigma$-polarization, the vertical one as $\pi$-polarization. The advantage of this derivation method is that it gives an analytical formula for the SR fields. However, the radiation fields deduced in this way are only approximative because of the far field approximation which was used in the derivation. Furthermore, in this approach the emission is considered to originate from a single point, additional resolution broadening effects as depth-of-field and orbit curvature have to be introduced additionally, see e.g., Refs.~\cite{Andersson95,Kube04}.

Based on the work described in \Bref{Chubar95} there is an alternative approach for SR field calculations becoming increasingly widespread in beam diagnostics. The advantage is that the method is exact in the sense that it does not rely on the far field approximation and effects like depth-of-field and orbit curvature are directly included. Starting point are again the Li\'{e}nard--Wiechert potentials, but this time they are directly Fourier transformed and the fields are derived in the frequency domain, resulting in an integral equation
\begin{equation}
\vec{E}_{\omega}(\omega) = \frac{i \omega Q}{4\pi \varepsilon_0 c} \int\limits_{-\infty}^{+\infty} \mathrm{d}z \; \frac{1}{R(z)} \left[\vec{\beta}(z) - \left(1 + \frac{i c}{\omega R(z)} \right) \hat{n}(z) \right] e^{i \omega \tau(z)} \, ,
\label{eq:S3nearfield1}
\end{equation}
with the retarded time expressed in terms of the electron longitudinal position $z$ as
\begin{equation}
\tau = \int_0^z \frac{\mathrm{d}z}{c \beta_z} = \frac{1}{c} \int_0^z \mathrm{d}z \; \left[1 + \frac{1 + (\gamma \beta_{\bot})^2}{2 \gamma^2}\right] \; .
\label{eq:S3nearfield2}
\end{equation}
With knowledge of the particle orbit the fields are directly accessible according to Eqs.(\ref{eq:S3nearfield1}) and (\ref{eq:S3nearfield2}). The integration has to be performed numerically which can be done with high accuracy using e.g., numerical near field calculations \cite{Chubar95} in order to study resolution broadening effects. Codes like SRW \cite{Chubar98} or SPECTRA \cite{Tanaka01} are freely available allowing computations preserving all phase terms that are necessary for further propagation of the radiation through optical components. In SRW, even propagation is implemented in the frame of scalar diffraction theory applying the methods of Fourier optics.

Besides the numerical near field calculation, there exist also analytical approaches in which the disturbed wave front is characterized by an additional phase factor such that orbit curvature and depth-of-field influence can directly be included in the radiation field description \cite{Bosch99,Chubar99,Geloni05}.

\subsubsection{Constant linear motion}

\begin{figure}[!b]
	\begin{center}
		\includegraphics[width=13cm]{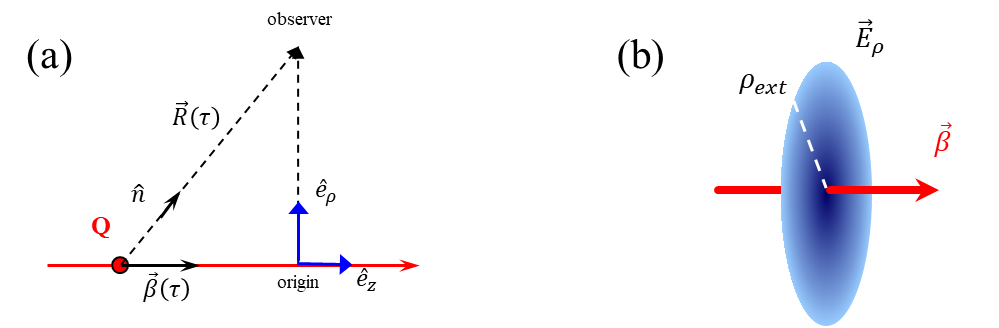}
		\caption{(a) Geometry for the description of the Li\'{e}nard--Wiechert fields for a particle in constant linear motion. (b) The field of the moving particle is considered as a virtual photon field disc having a radius $\rho_{ext} \approx \gamma\lambda$.}
		\label{fig:S3LinGeo}
	\end{center}
\end{figure}
For the discussion about radiation generation in linear accelerators, the electromagnetic field of a point charge $Q$ in constant linear motion is considered. Again the particle field is given by the Li\'{e}nard--Wiechert fields in Eq.~(\ref{eq:S3LenWf}), however in this situation it is the second term in the sum (acceleration term) which vanishes because there is no acceleration per definition. Because of the rotational symmetry it is convenient to describe the geometry in the cylindrical coordinate system as shown in \Fref{fig:S3LinGeo}(a). In this system, the electric field can be expressed as \cite{TerMikaelian72}
\begin{eqnarray}
\vec{E}_{\omega}(\rho, z, \omega) & = & \frac{1}{4\pi \varepsilon_0} \frac{Q \kappa}{\pi v} \; e^{i\frac{\omega}{v}z} \left( \mathrm{K_1} (\kappa\rho) \; \hat{e}_{\rho} - \frac{i}{\gamma} \; \mathrm{K_0}  (\kappa\rho) \; \hat{e}_z \right) \nonumber \\
&& \mathrm{with} \qquad \kappa = \frac{2 \pi}{\lambda \beta \gamma} \, ,
\label{eq:S3PseudoPhot}
\end{eqnarray}
and K$_0$, K$_1$ modified Bessel functions of second kind. The field representation is nothing other than the Fourier transform of the field of the moving point charge from Eq.~(\ref{eq:S2Coulomb2}). While K$_0$ is already smaller than K$_1$, according to Eq.~(\ref{eq:S3PseudoPhot}) in the ultra-relativistic limit $\gamma \rightarrow \infty$ the contribution from the longitudinal component can be completely neglected and the particle field exhibits a pancake-like structure. It is this field which is associated with the pseudo-photons in order to describe the different radiation generation mechanisms.

With increasing distance $\rho$ from the beam orbit, the field shrinks following the K$_1$ dependency. It is convenient to assign a value to the radial field extension $\rho_{ext}$ by setting the argument of the Bessel function equal to one, i.e.,
\begin{equation}\label{eq:S3FieldExt}
\rho_{ext} = \frac{\lambda \beta \gamma}{2 \pi} \approx \gamma \lambda \; .
\end{equation}
In a descriptive way the virtual photon field is interpreted as a radial field disc with the radius $\rho_{ext}$, see \Fref{fig:S3LinGeo}(b). The angular distribution of the virtual photon field is given by
\begin{equation}\label{eq:S3WVang}
\frac{\mathrm{d}^2 W}{\mathrm{d}(\hbar\omega) \mathrm{d}\Omega} = \frac{\alpha}{\pi^2} \frac{\theta^2}{(\gamma^{-2} + \theta^2)^2} \; ,
\end{equation}
with $\alpha$ the fine structure constant and $\theta$ the angle between the virtual photon wave vector and the charged particle direction of motion.
\begin{figure}[!h]
	\begin{center}
		\includegraphics[width=8cm]{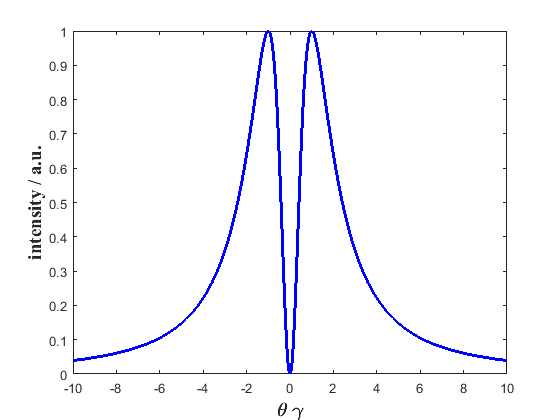}
		\caption{Calculated angular distribution of the virtual photon field according to Eq.~(\ref{eq:S3WVang})}
		\label{fig:S3virtPhot}
	\end{center}
\end{figure}
Figure \ref{fig:S3virtPhot} shows the calculated virtual photon angular distribution associated to the charged particle according to Eq.~(\ref{eq:S3WVang}). As can be seen, it possesses a characteristic double lobe structure with a central minimum, the intensity maxima are close to the angle $\pm \frac{1}{\gamma}$. It is interesting to note that it is this structure which is imprinted to the real photons (radiation) when the field is separated from the beam, i.e., all radiation phenomena which originate from the virtual photon field have an intensity minimum in the direction which is connected to the motion of the charged particle beam.

In the following some of the radiation phenomena will briefly be described in view of their field separation mechanism and applications for particle beam diagnostics.

\subsubsubsection{Transition radiation}

If a charged particle passes the boundary between two media with different dielectric constants, a broad band electromagnetic radiation is produced which is named transition radiation. For beam diagnostic purposes the visible part of the radiation (optical transition radiation, OTR) is predominantly used and an observation geometry in backward direction is mainly chosen such that the screen has an inclination angle of 45$^{\circ}$ with respect to the beam axis, and observation is performed under 90$^{\circ}$. In a typical monitor set-up the beam is imaged via OTR using standard lens optics, and the recorded intensity profile is a measure of the particle beam spot. OTR has the advantage that it allows fast single shot beam profile measurements, and the radiation output scales linearly with the bunch intensity (neglecting coherent effects).

\begin{figure}[!t]
	\begin{center}
		\includegraphics[width=10cm]{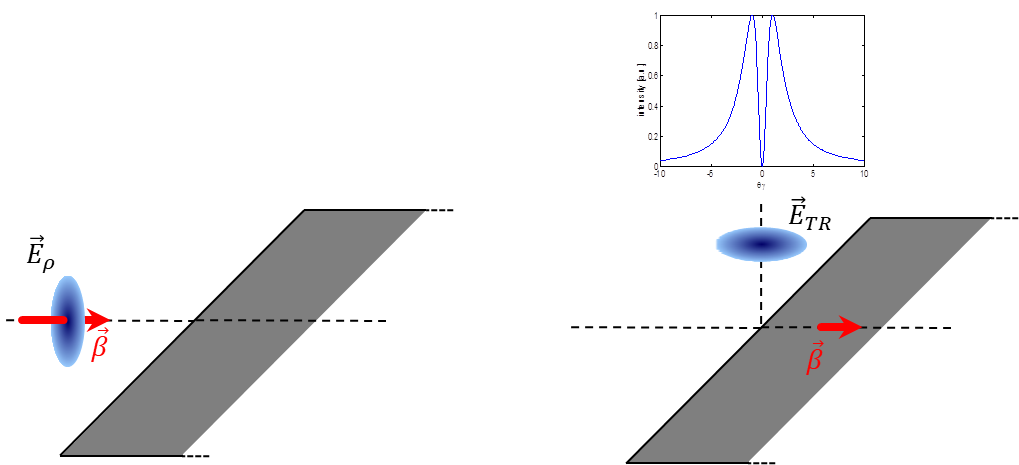}
		\caption{Illustration of the field separation mechanism for backward OTR. An ultra-relativistic particle with typical pancake field distribution traverses a metallic screen. The field is reflected from the screen surface and can be measured as radiation.}
		\label{fig:S3TR}
	\end{center}
\end{figure}
The separation mechanism for backward emitted OTR corresponds to the direct reflection of pseudo-photons at the screen surface which acts as a mirror and which is assumed to be a perfect conductor for simplicity (otherwise the Fresnel coefficients have to be taken into account). In this reflection process, the virtual photons absorb momentum from the screen and are released from the charged particle, transformed into real photons (radiation) which can be measured at large distances as OTR. The reflection does not modify the field properties, therefore the incoming virtual and outgoing real photons are described by the fields Eq.~(\ref{eq:S3PseudoPhot}) and consequently have the same angular distribution Eq.~(\ref{eq:S3WVang}).

Transverse beam profile imaging in electron linacs is widely based on OTR as standard technique \cite{Fiorito94}. Imaging resolution studies based on the 
propagation of the pseudo-photon field through the optical system can be found in Refs.\cite{Castellano98}--\cite{Stupakov11} for different cases and examples.

\subsubsubsection{Diffraction radiation}

OTR beam size diagnostics has the disadvantage that it requires the beam interaction with the screen. Owing to the high power density of modern high brightness beams, the energy deposition in the screen may lead to a damage of the device. Therefore the development of non-intercepting methods is essential. In this context optical diffraction radiation (ODR) is an interesting candidate. This kind of radiation is generated if a charged particle beam passes close to a diffracting structure like an edge or a slit, and the physics of DR is well known in the literature, see e.g., Refs.~\cite{TerMikaelian72,Potylitsyn10} and the references therein. Similar to OTR, only backward emitted ODR will be considered because it is more convenient for beam diagnostic applications.

\begin{figure}[!t]
	\begin{center}
		\includegraphics[width=14cm]{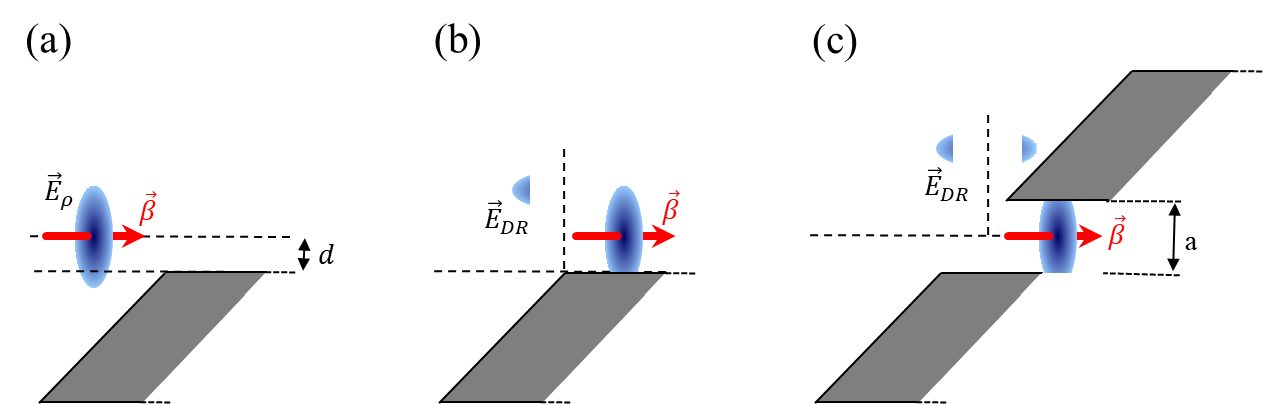}
		\caption{Illustration of the field separation mechanism for backward ODR. (a) An ultra-relativistic particle with typical pancake field distribution passes close to (b) a metallic edge or (c) a slit in a metallic screen. Parts of the field are diffracted away and can be measured as radiation.}
		\label{fig:S3DR}
	\end{center}
\end{figure}

The mechanism of radiation generation is similar to the one of OTR and sketched in \Fref{fig:S3DR}. But in the case of ODR it is not the complete pseudo-photon field which is diffracted away, only a part of it is released from the electron. Therefore the ODR intensity will be lower than the one of OTR. Keeping in mind the radial field extension Eq.~(\ref{eq:S3FieldExt}), it is obvious that the distance from the electron to the edge resp. the slit size $a$ in \Fref{fig:S3DR} should be within the range of $\rho_{ext}$ in order to efficiently generate ODR. Furthermore, in the limit $a \ll \rho_{ext}$ there is no difference between ODR and OTR.

In principle ODR can be generated at any kind of aperture. Nevertheless the use of rectangular slit shapes is advantageous because the mathematical description is simplified due to the translational invariance with respect to one coordinate, and the slit size itself can be considered as infinitely long with respect to $\rho_{ext}$. As consequence, the beam size in only one dimension can be deduced from an ODR measurement.

In order to deduce beam size information, instead of using the ODR image the angular distribution from a slit can also be exploited. Information about the beam size can be extracted from a measurement of the visibility, i.e., the ratio between the maximum intensity and the intensity in the central minimum which is smeared out due to the non-zero beam size. However, the ODR angular distribution is not only influenced by the beam size, but also by the beam offset from the slit centre and by the beam divergence. Different schemes are proposed as discussed e.g., in Refs.~\cite{Fiorito01,Tasha03}. In order to overcome this ambiguity, in the pioneering experiment of Ref. \cite{Karataev04} a beam with very low divergence was used such that only the additional position dependence had to be taken into account which could be controlled by independent beam position measurements. With this method the authors measured beam sizes down to about 10 $\mu$m \cite{Karataev05}. Optical diffraction radiation interferometry (ODRI) is another promising method for high resolution beam profile measurements using ODR \cite{Cianchi12,Cianchi14}.

\subsubsubsection{Parametric X-ray and Smith--Purcell radiation}

\begin{figure}[!t]
	\begin{center}
		\includegraphics[width=15.5cm]{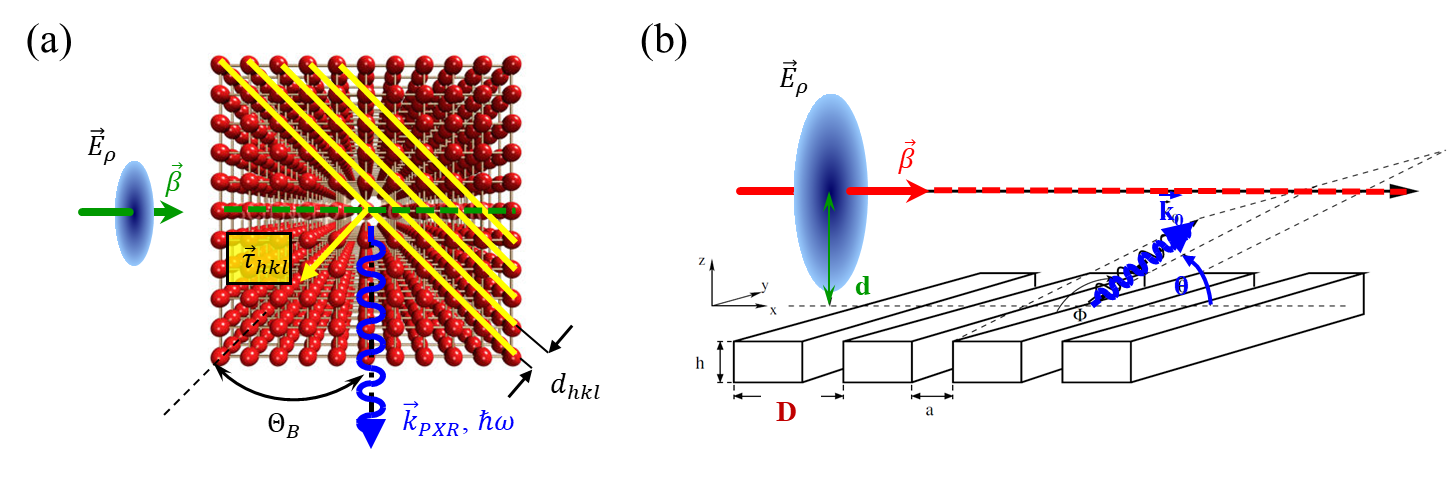}
		\caption{Illustration of the field separation mechanism for (a) parametric X-ray radiation (PXR) and (b) Smith--Purcell radiation (SPR). PXR can be understood as virtual photon diffraction at crystallographic planes, SPR as the corresponding diffraction at a grating structure.}
		\label{fig:S3PXRSP}
	\end{center}
\end{figure}

Parametric X-Ray radiation (PXR) is emitted when a relativistic charged particle beam crosses a crystal. The radiation process can be understood as diffraction of the virtual photon field associated with the particles at the crystallographic planes, see \Fref{fig:S3PXRSP}(a). As result, radiation is emitted in the vicinity of directions satisfying the Bragg condition. Because of the discrete momentum transfer from the crystal planes PXR exhibits a line spectrum.

PXR for beam diagnostics was independently proposed in Refs.~\cite{Gogolev12,Takabayashi12}. Besides the smaller radiation wavelength and the better resolution, the usage of PXR is advantageous because it is emitted from crystallographic planes inside the radiator which usually have a certain inclination angle with respect to the crystal surface, thus allowing a spatial separation from a possible coherent OTR background which is directly generated at the surface. Disadvantage is the PXR radiation yield which is typically 1--2 orders of magnitude smaller than the one from transition radiation. Experimental tests in view of particle beam instrumentation indicate that indeed the low radiation intensity might hamper a meaningful application of PXR for this purpose \cite{Takabayashi12}--\cite{Kube13}.

Smith--Purcell radiation (SPR) is emitted when an electron beam passes a diffraction grating at a fixed distance close to its surface. The radiation mechanism can be understood as diffraction of the incoming pseudo-photon field at the grating structure. The grating with spacing $D$ represents a one-dimensional Bravais structure, thus offering a discrete momentum which results in the dispersion relation
\[
n \lambda =  D (\beta^{-1} - \cos\theta)
\] 
with $n$ the diffraction order and $\theta$ the observation angle as measured between grating surface and outgoing photon, cf. \Fref{fig:S3PXRSP}(b).

A general overview about SPR in view of particle beam diagnostics is given in \Bref{Kube03}. Furthermore, in \Bref{Doucas01} the use of SPR as high-resolution position sensor for ultra-relativistic electron beams was proposed, but the most promising application seems to be for longitudinal profile diagnostics in frame of coherent radiation diagnostics, cf. \Bref{Gillespie18}.

\subsection{Particle electromagnetic field interaction with matter}

This kind of interaction is widely applied e.g., for beam loss monitoring, for intercepting beam current measurements (Faraday cup), but also for beam profile measurements e.g., with wire scanners, scintillators, secondary emission monitors, or ionization profile monitors.

In the main it is the charged particles energy deposition in a part of the monitor which is used in order to derive information about the beam properties. Beam particles transmit some of their energy to the particles in the medium, resulting in excitations of medium particles either by ionization or by excitation of optical states. At the level of particle--particle interaction there are a number of important modes of interaction which can be subdivided as follows:
\begin{itemize}
	\item elastic scattering, i.e., an incident particle scatters off a target particle and the total kinetic energy of the system remains constant;
	\item inelastic scattering, i.e., an incident particle excites a target atom to a higher electronic or nuclear state;
	\item annihilation, i.e., an incident particle collides with its respective antiparticle to produce a new kind of particle (as e.g., \Pep \Pem $\rightarrow$ \Pgamma);
	\item Bremsstrahlung emission, i.e., an incident particle is accelerated in the Coulomb field of a target atom nucleus and emits a photon;
	\item Cherenkov or transition radiation emission as explained in the previous section.
\end{itemize}
In the following electromagnetic reaction channels will briefly be reviewed for different particle species in view of beam instrumentation applications. Again it is the particle rest mass which causes the main difference in the interaction channels for different particle species. Therefore one has to distinguish between heavy particles (i.e., particles with atomic number $A \ge 1$ as e.g., \Pp, $\alpha$, ions$\ldots$) and light particles (\Pep and \Pem).

In the following both particle species will be discussed with respect to their energy loss and their range in matter.

\subsubsection{Interaction of heavy charged particles}

Heavy charged particles have two electromagnetic channels by which they can interact with surrounding matter. The first one is the classical \emph{Rutherford or Coulomb scattering} as an elastic scattering process which describes the interaction between an incident particle and a target nucleus via Coulomb force. However, this type of interaction is of less relevance for particle beam instrumentation and will not be considered in the following.

The second one is summarized as \emph{passage of particles through matter} and describes a number of electronic and nuclear mechanisms through which a charged particle can interact with the medium atoms. Net result of all individual interactions however is a reduction of the primary particle energy. While the underlying individual interaction mechanisms are rather complicated, it was proven to predict the rate of energy loss fairly accurate by semi-empirical relations. These relations are of relevance for beam instrumentation and will be discussed in the following.

As starting point the energy transfer from a projectile particle to a target is considered. This energy transfer is dominated by elastic collisions with the shell electrons.
\begin{figure}[!h]
	\begin{center}
		\includegraphics[width=11cm]{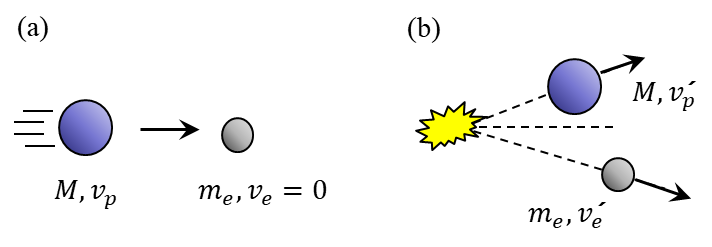}
		\caption{Illustration of the kinematics for head-on collision of a beam particle and a shell electron at rest (a) before and (b) after the collision}
		\label{fig:S3coll}
	\end{center}
\end{figure}
In the present case the projectile is a beam particle having mass $M$ and moving with velocity $v_p$, the target is an atomic shell electron with mass $m_e$ initially being at rest, see \Fref{fig:S3coll}. After the collision both particles will have velocities of $v_p'$ and $v_e'$. The maximum energy transfer occurs for head-on collisions. Considering momentum and energy conservation simply for non-relativistic particle motion, the maximum relative energy transfer $\Delta E_{max}$ can be written as
\begin{equation}\label{eq:S3coll}
\frac{\Delta E_{max}}{T_{kin}} = 4 \, \frac{m_e \, M}{(m_e + M)^2}
\quad \xrightarrow[M \gg m_e]{} \quad 4 \; \frac{m_e}{M} \, ,
\end{equation}
with $T_{kin} = \frac{M}{2}v_p^2$ the non-relativistic kinetic energy of the projectile. For a proton beam with $M = 1836\,m_e$ for example, the maximum relative energy transfer would amount to $\frac{\Delta E_{max}}{T_{kin}} \approx \frac{1}{500}$ which is a very small value. In a single collision, the beam particle will transfer only a small amount of energy to the target, and as consequence the particle trajectory will nearly be unaffected. Therefore the beam particle will move along an almost straight line through matter.

\subsubsubsection{Energy loss of heavy charged particles}

\begin{figure}[!t]
	\begin{center}
		\includegraphics[width=13.5cm]{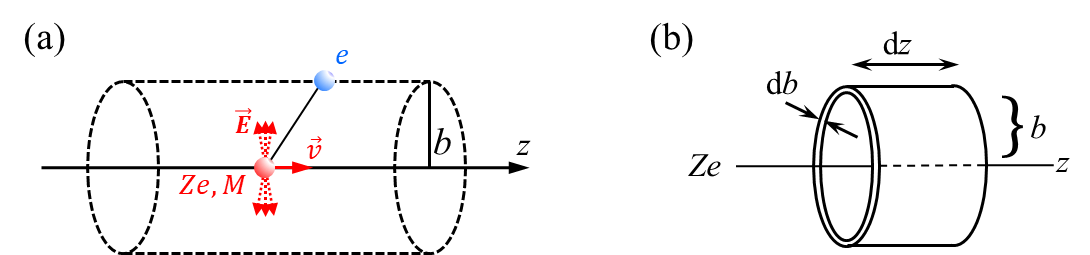}
		\caption{Illustration of the geometry for the derivation of the classical energy loss of heavy charged particles according to Niels Bohr \cite{Bohr13}}
		\label{fig:S3Eloss}
	\end{center}
\end{figure}

The classical non-relativistic formula for the energy loss was deduced by Niels Bohr in 1913 \cite{Bohr13}. His derivation shows very descriptively how the particle electromagnetic field is involved and will briefly be outlined:

A beam particle with charge $Ze$ and mass $M$ is considered to move with velocity $\vec{v} = v\,\hat{e}_z$ through a medium with electron density $n$, cf. \Fref{fig:S3Eloss}(a). It passes a shell electron at a distance $b$ which is assumed to be unbound and initially at rest.
Due to the Lorentz contracted transverse particle electric field there is a transverse momentum transfer to the shell electron in the form
\[
\Delta \vec{p}_{\bot} = \int \mathrm{d}t \, \vec{F}_{\bot} = \int \mathrm{d}z \, \vec{F}_{\bot} \, \frac{\mathrm{d}t}{\mathrm{d}z} = \int \vec{F}_{\bot} \frac{\mathrm{d}z}{v} = e \int \vec{E}_{\bot} \frac{\mathrm{d}z}{v}  \; ,
\]
the longitudinal momentum transfer $\Delta\vec{p}_{\|}$ averages to zero because of symmetry reasons. Applying Gauss's flux theorem Eq.~(\ref{eq:S2gauss1}) results in an expression for the field
\[
\int_S \vec{E} \cdot \mathrm{d}\vec{S} = \int \vec{E}_{\bot} \cdot 2 \pi b \, \mathrm{d}z = \frac{Ze}{\varepsilon_0} \qquad\qquad \Rightarrow \qquad \int \vec{E}_{\bot} \mathrm{d}z = \frac{Ze}{2 \pi b \varepsilon_0}
\]
such that the momentum transfer is
\[
\Delta \vec{p}_{\bot} = \frac{Ze^2}{2 \pi b v \varepsilon_0} \; .
\]
For non-relativistic motion the energy transfer to a single shell electron, located at a distance $b$ away from the particle orbit, is written as
\[
\Delta E(b) = \frac{\Delta \vec{p}^{~2}_{\bot}}{2\,m_e} \quad = \quad \frac{Z^2 e^4}{8 \pi^2 \varepsilon_0^2 b^2 m_e v^2} \; .
\]
In order to get an expression for the overall energy loss the integration over all electrons in the medium has to be carried out. As shown in \Fref{fig:S3Eloss}(b) a cylindrical barrel containing $N_e$ electrons in considered with $N_e = n \, 2 \pi b \, \mathrm{d}b \, \mathrm{d}z$. The energy loss $\Delta E$ per path length $\mathrm{d}z$ for a distance between $b$ and $b + \mathrm{d}b$ in the medium with electron density $n$ is given by
\[
-\mathrm{d}E(b) = \frac{\Delta p^2_{\bot}}{2\,m_e} \, N_e \quad = \quad \frac{Z^2 e^4}{4 \pi \varepsilon_0^2 m_e v^2} \, n \, \frac{\mathrm{d}b}{b} \, \mathrm{d}z
\]
which results in
\begin{equation}\label{eq:S3Bohr}
-\frac{\mathrm{d}E}{\mathrm{d}z} = \frac{Z^2 e^4}{4 \pi \varepsilon_0^2 m_e v^2} \, n \int\limits_{b_{min}}^{b_{max}} \, \frac{\mathrm{d}b}{b} \quad = 
\quad  4 \pi N_A r_e^2 m_ec^2 \frac{Z_t}{A_t} \, \rho_t \frac{Z^2}{\beta^2} \ln \left(\frac{b_{max}}{b_{min}}\right) \; .
\end{equation}
In the equation above the electron density $n$ was replaced by the properties of the target material, i.e., $n = N_A \rho_t \frac{Z_t}{A_t}$ with $N_A$ Avogadro's number. Furthermore the classical electron radius $r_e$ was introduced.

In the literature there exist different assumptions for the impact parameter limits $b_{min}$ and $b_{max}$. $b_{min}$ can be estimated for example from the uncertainty principle such that impact parameters below the electron de Broglie wavelength are not relevant ($b_{min} = \lambda_e$), and $b_{max}$ from the principle of adiabatic invariance, i.e., the assumption that the interaction time must be shorter than the electron revolution time to guarantee relevant energy transfer. However, the discussion will not be continued. Instead of the mean energy loss per distance according to Bethe \cite{Bethe30,Bethe32}, based on a quantum mechanical derivation using  first-order Born approximation is quoted below in the form as published by the \emph{Particle Data Group} \cite{PDG18}
\begin{equation}\label{eq:S3BBloch}
    \left\langle -\frac{\mathrm{d}E}{\mathrm{d}z}\right\rangle  = 4 \pi N_A \, r_e^2 \, m_e c^2 \, \frac{Z_t}{A_t} \, \rho_t \, \frac{Z^2}{\beta^2} \,
    \left[ \frac{1}{2}
    \ln\frac{2 m_e c^2 \beta^2 \gamma^2 W_{max}}{I^2} - \beta^2 - \frac{\delta(\beta\gamma)}{2}
    \right] \, ,
\end{equation}
with $I$ the ionization potential and the maximum energy transfer in a single collision
\begin{equation}\label{eq:S3Wmax}
W_{max} = \frac{2 m_e c^2 \beta^2 \gamma^2}{1 + 2\gamma\frac{m_e}{M} + (\frac{m_e}{M})^2} \; .
\end{equation}
It should be noted that in literature often the low-energy approximation for the maximum energy transfer $W'_{max} = 2 m_e c^2 \beta^2 \gamma^2$ is quoted instead of Eq.~(\ref{eq:S3Wmax}) which is valid in the case $2 \gamma m_e \ll M$. Using this expression instead of $W_{max}$, Eq.~(\ref{eq:S3BBloch}) is expressed in a form which is commonly found in textbooks, see e.g., Refs.~\cite{Jackson99,Ahmed15}. Furthermore the function $\delta(\beta\gamma)$ describes the \emph{density effect} correction to the ionization energy loss which is important for high beam energies and caused by saturation polarization of the target atoms resulting in screening of the electric particle field, see e.g., \Bref{Sternheimer84}. Additional correction terms (Barkas-Anderson–Bloch corrections) are recommended for the low energy region in \Bref{ICRU49} which are not quoted here.

Comparing Eq.~(\ref{eq:S3Bohr}) and Eq.(\ref{eq:S3BBloch}) one can see that the general form of both equations is rather similar. They depend on natural constants ($N_A$: Avogadro number, $m_e c^2$: electron rest mass, $r_e$: classical electron radius), on target material properties ($\rho_t$: material density, $A_t, Z_t$: atomic mass and nuclear charge, $I$: mean excitation energy, $\delta$: density effect correction), and on beam particle properties ($Z$: projectile charge, $\beta = v/c$: reduced particle speed), but not explicitly on the projectile mass $M$. 

Equation~(\ref{eq:S3BBloch}) is termed the \emph{Bethe} or \emph{Bethe--Bloch} equation, the negative sign indicates that the particles loose energy. Instead of energy loss (which will be termed $-\frac{\mathrm{d}E}{\mathrm{d}x}$ in the following instead of $-\frac{\mathrm{d}E}{\mathrm{d}z}$ in accordance with the common literature), the term \emph{stopping power} $S = -\frac{\mathrm{d}E}{\mathrm{d}x}$  is extensively used. This notion not only includes the collision or electronic part, but also energy losses due to radiation and nuclear reactions. However, if not stated otherwise in the following stopping power stands for the collision losses described by Eq.~(\ref{eq:S3BBloch}).

\begin{figure}[!t]
	\begin{center}
		\includegraphics[width=14.5cm]{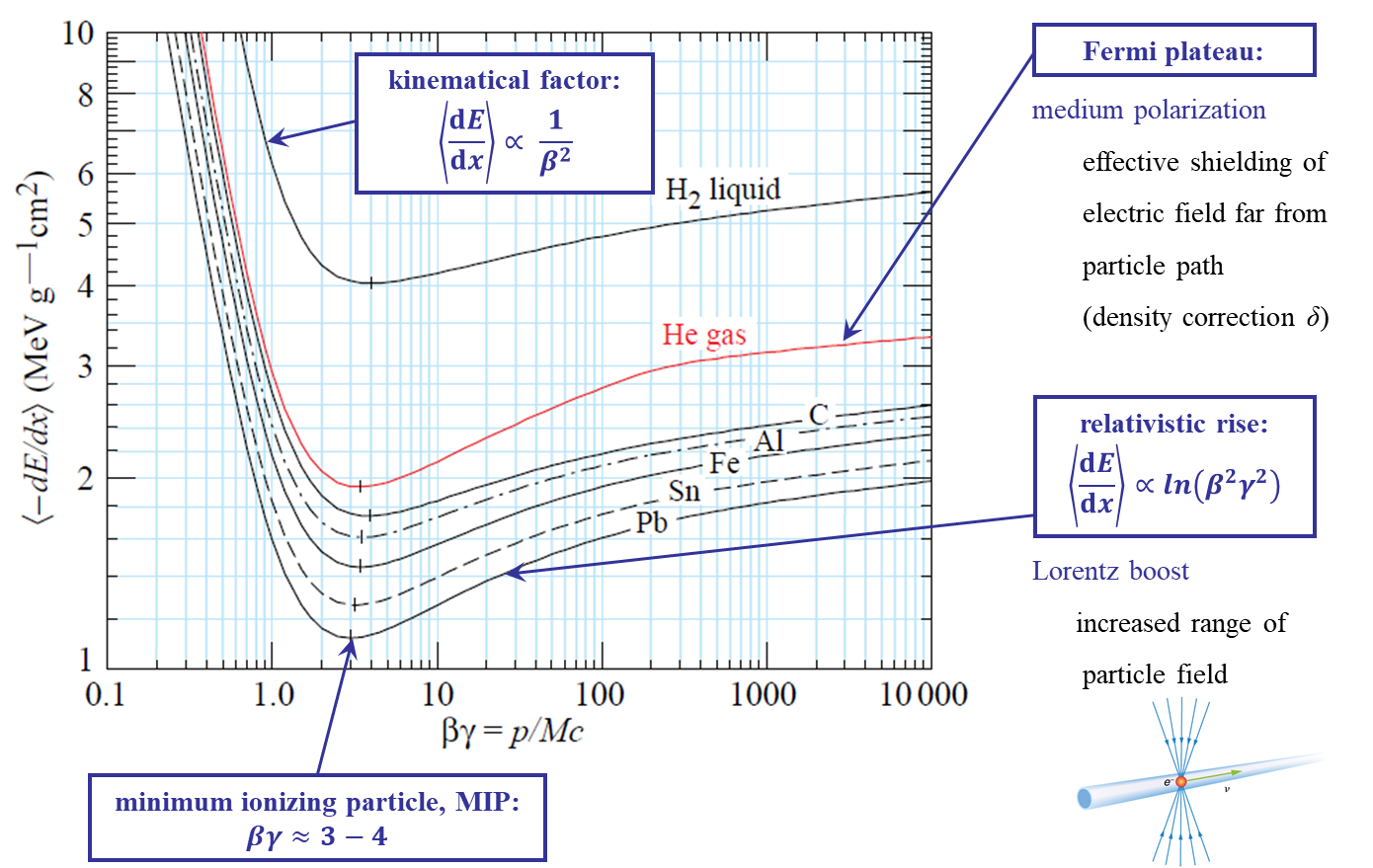}
		\caption{Calculated mass collision stopping power of heavy charged particles for different materials. The four kinematic regions of $S_{col}$ are highlighted. Underlying picture taken from \Bref{PDG18}.}
		\label{fig:S3BB}
	\end{center}
\end{figure}

It is common practice to divide Eq.~(\ref{eq:S3BBloch}) by the target density $\rho_t$ and to name this quantity the \emph{mass collision stopping power}, measured in the units $[\mathrm{MeV}\,\mathrm{g}^{-1}\,\mathrm{cm}^2]$. Figure \ref{fig:S3BB} shows mass collision stopping powers for various materials. Neglecting the density correction contribution, the general shape of these curves is simply characterized by the dependency
\[
S \propto \frac{Z_t}{A_t} \, \frac{Z^2}{\beta^2} \ln(a \beta^2 \gamma^2) \, ,
\]
with $a$ a proportional factor. While $\frac{Z_t}{A_t} \approx 0.5$ for most common materials, the dependency of the mass stopping power on the target material properties is rather weak. As can be seen from \Fref{fig:S3BB}, the mass collision stopping power can be subdivided into four regions:
\begin{itemize}
	\item At small $\beta\gamma$ the stopping power is quickly decreasing because it is dominated by the kinematical factor $\beta^{-2}$ (precisely $\beta^{-5/3}$) as described by the Bohr model, see Eq.~(\ref{eq:S3Bohr}). The reason is that slower particles experience the electric field for a longer time and consequently have larger energy losses.
	\item At $\beta\gamma \approx 3-4$ the stopping power has a minimum in ionization. Particles there are named \emph{minimum ionizing particles} (MIP). Because the stopping power weakly depends on the absorber material properties, the energy loss for a MIP is usually estimated to amount to $\sim 2~\mathrm{MeV}\,\mathrm{g}^{-1}\,\mathrm{cm}^2$.
	\item For larger $\beta\gamma$ the curve is dominated by the relativistic rise $\propto \ln(\beta^2\gamma^2)$. The cause is the increase in the transverse electric field due to the Lorentz boost Eq.~(\ref{eq:S2CoulField}), and therefore increased loss  contributions from larger impact parameters $b$.
	\item The so called \emph{Fermi plateau} at larger $\beta\gamma$ is connected to the target density effect. Real media are polarized from the projectile particle electric field, resulting in an effective shielding of the field far from the particle path. As consequence the shielding effectively reduces the long range contributions to the relativistic rise.
\end{itemize}

\subsubsubsection{Particle range}

Besides the energy loss, the range of particles in matter is of interest when designing an interceptive instrument for beam diagnostic measurements. As discussed in the previous section the mean energy loss due to ionization and excitation is well described by Eq.~(\ref{eq:S3BBloch}) for all charged particles, the only exception is the interaction of \Pepm beams with matter as will be pointed out in the subsequent section.

Speaking about the particle range, it is defined as the average distance a heavy charged particle will travel in matter. The particle energy loss is a statistical process, and as demonstrated in 
Eq.~(\ref{eq:S3coll}) heavy particles loose only a small fraction of their energy in collisions with the shell electrons of the target material. As a consequence they experience only a slight deflection in the scattering with atomic electrons and travel in nearly straight lines through the target. Due to the small gradual amount of energy transferred from a beam particle to the target the particle passage through matter can be treated as a continuous slowing down process.

In the \emph{continuous slowing down approximation} (CSDA), the range a particle beam will travel in a medium is calculated by integrating the stopping power over the kinetic energy $T$
\begin{equation}\label{eq:S3csda}
r_{CSDA} = \int\limits_0^T \mathrm{d}T \left[-\frac{\mathrm{d}E}{\mathrm{d}x}(T)\right]^{-1} \; .
\end{equation}
This CSDA range is a very close approximation to the average path length travelled by a charged particle as it slows down to rest.
\begin{figure}[!t]
	\begin{center}
		\includegraphics[width=14.5cm]{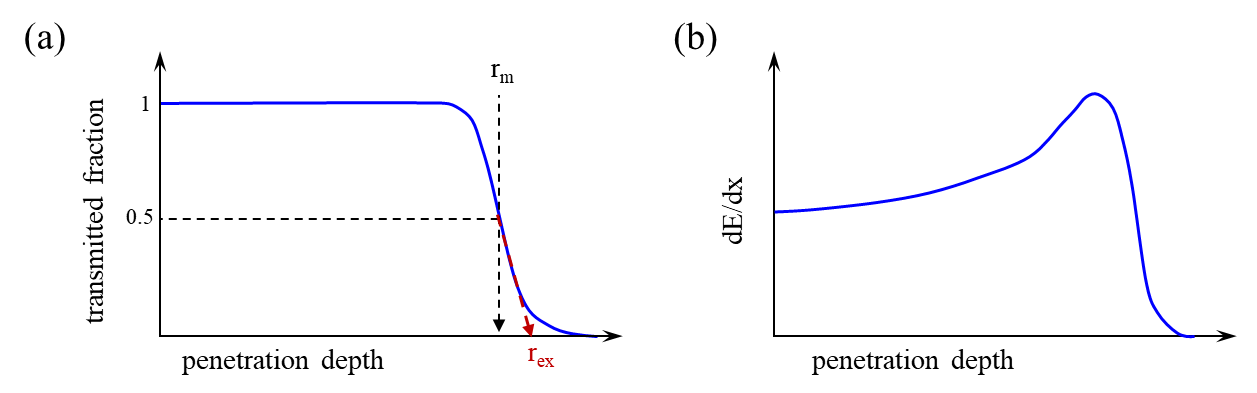}
		\caption{(a) Sketch of the transmission probability of normally incident heavy charged particles impinging a target. Range definitions for the mean range $r_m$ and the extrapolated range $r_{ex}$ are shown in addition. (b) Illustration of the energy dissipation as function of the particle penetration depth. The strong increase of the local energy deposition short before heavy charged particles are stopped is known as the Bragg peak.}
		\label{fig:S3range}
	\end{center}
\end{figure}
In Eq.~(\ref{eq:S3csda}) it is assumed that the rate of energy loss at every point along the track is equal to the total stopping power, energy-loss fluctuations are neglected.

Besides $r_{CSDA}$ other measures are in use for the range. Since the nature of energy loss events is statistical, the number of collisions required to bring the particle down to rest within the medium varies slightly with each particle. Hence, there will be a small variation in the range, also known as straggling which is usually a small effect. For 100\UMeV{} protons in biological material for example it is $\sim 1\%$. According to \Fref{fig:S3range}(a) the mean range $r_m$ is defined as the penetration depth at which half of the projectile particles are stopped.

The extrapolated range $r_{ex}$ also shown in this figure is commonly defined as the penetration depth at which the extrapolation of the almost straight descending portion of a transmission curve intersects the $x$-axis. In some cases however, this straight portion cannot be well defined. Therefore, a generalized definition of $r_{ex}$ is given as the point where the tangent at the steepest point on the transmission curve intersects the $x$-axis, see \Bref{Tabata02}.

Because of this ambiguity in range a number of experimentalists have turned to experimental means of measuring this quantity and modelling the range on the basis of their results. The Bragg--Kleeman rule for example allows us to compute the range of a particle in a medium if its range in another medium is known \cite{BK05},
\[
\frac{r_1}{r_2} = \frac{\rho_2}{\rho_1} \sqrt{\frac{A_1}{A_2}}
\]
with $\rho_i$ the density and $A_i$ the atomic mass number in both media.

For comparison \Fref{fig:S3range}(b) shows the energy loss (stopping power) as function of the penetration depth. As can be seen, a peak occurs short before the projectile particles are stopped, i.e., in a situation where they have lost already a major part of their energy due to ionization losses. This so-called \emph{Bragg peak} originates from the kinematical factor $\beta^{-2}$ in the Bohr resp. Bethe--Bloch equation: the slower the particle speed (energy), the longer they experience the electric field and consequently have larger energy losses. The possibility of creating such a localized high energy loss is exploited for example in particle therapy for cancer to concentrate the effect of light ion beams on the region of the tumour being treated while minimizing the effect on the surrounding healthy tissue \cite{Durante16}.

\subsubsection{Interaction of electrons and positrons}

In case of \Pepm beams the situation is quite different. Due to the relatively small rest mass energy of $m_0c^2 = 0.511\UMeV$ relativistic effects have to be taken into account in order to deduce meaningful results. In addition, large energy transfers to the shell electrons of the target are possible. For better understanding the simple example of energy transfer in a head-on collision for non-relativistic particle motion from Eq.~(\ref{eq:S3coll}) is considered, but this time projectile and target shell electron have the same mass. This leads to
\[
\frac{\Delta E_{max}}{T_{kin}} = 4 \, \frac{m_e \, M}{(m_e + M)^2}
\quad \xrightarrow[M = m_e]{} \quad 1 \; ,
\]
i.e., in a single collision electrons and positrons can transfer all of their kinetic energy to the target.

Some interesting consequences follow especially for electron beams. Due to the fact that incident beam and target electron are indistinguishable particles, it is convention to assume that the electron with higher energy after the collision was formerly the beam electron. As consequence, the maximum energy transfer a beam electron will experience in a single collision corresponds to half of its initial kinetic energy $T_{kin}/2$. In contrast to electrons, for a positron beam target electron and beam particle are clearly distinguishable after the collision, and the maximum possible energy transfer is the initial kinetic energy $T_{kin}$. As a result, the energy loss for electrons and positrons is different.

Furthermore, due to the large energy transfer in a single collision large angular deviations from the initial trajectory are possible. As consequence, \Pepm trajectories have a rather curled shape compared to the ones from heavy charged particles.

Finally, due to the small rest mass energy radiative losses caused by the emission of Bremsstrahlung have to be taken into account.
\begin{figure}[!h]
	\begin{center}
		\includegraphics[width=13.cm]{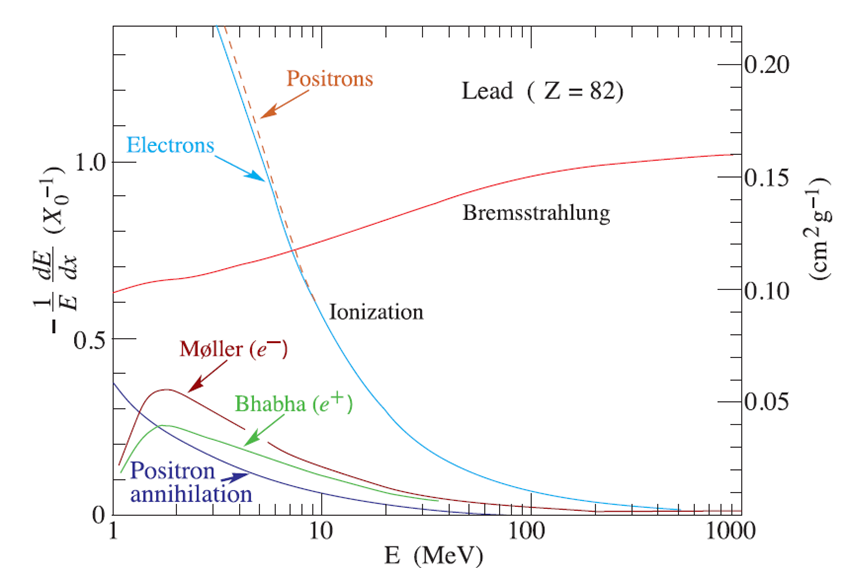}
		\caption{Fractional energy loss per radiation length in lead as a function of electron or positron energy. Figure taken from \Bref{PDG18}.}
		\label{fig:S3emode}
	\end{center}
\end{figure}

Below the different interaction modes of \Pepm are briefly summarized on the level of particle--particle interaction. In addition they are plotted in \Fref{fig:S3emode} as a function of the particle energy.
\begin{itemize}
	\item Ionization losses which include distant collisions having a small energy transfer. This interaction channel is comparable to the losses described by Eq.~(\ref{eq:S3BBloch}).
	\item M{\o}ller scattering: this process includes scattering events with close collisions, i.e., with a large energy transfer taking into account relativistic, spin, and exchange effects. The beam particle is scattered at the same particle species, i.e., \Pepm $\rightarrow$ \Pepm. Due to the fact that the atomic shells of the target atoms do not contain positrons, M{\o}ller scattering will occur only with electron beams.
	\item Bhabha scattering: this process is similar to M{\o}ller scattering, but this time the scattering event takes place between a particle and antiparticle \Pepm $\rightarrow$ \Pemp. With the same argument as before, Bhabha scattering will occur only with positron beams.
	\item Electron--positron annihilation, i.e., the annihilation between particles and antiparticles with the generation of a new particle. In the case of \Pepm annihilation the new particle is usually a \Pphoton, at higher energies even \PZ production and more could appear. Again, annihilation as primary process is only possible with positron beams.
	\item Emission of Bremsstrahlung. This refers to the process in which beam particles, decelerated in the Coulomb field of a target nucleus, emit electromagnetic radiation.
\end{itemize}
From this short compilation it is possible to conclude that there exist different interaction modes for electron and positron beams, therefore their energy loss in matter is different, as can be seen also in \Fref{fig:S3emode}. However, the stopping power for both particle species is not dramatically different. In addition \Fref{fig:S3emode} shows that ionization losses (collision stopping power $S_{col}$) are dominant for small particle energies while radiative losses due to Bremsstrahlung emission (radiative stopping power $S_{rad}$) are the dominating process at higher beam energies.

Following \Bref{ICRU37} the mass collision stopping power for electrons and positrons is expressed as
\begin{equation}\label{eq:S3Scol}
S_{col}^{\mp} = \left\langle -\frac{1}{\rho_t}\,\frac{\mathrm{d}E}{\mathrm{d}x}\right\rangle  = 4 \pi N_A \, r_e^2 \, m_e c^2 \, \frac{Z_t}{A_t} \, \frac{1}{\beta^2} \,
\left[\ln\left(\frac{T}{I}\right) + \frac{1}{2} \ln\left(1 + \frac{\tau}{2}\right) + F^{\mp}(\tau) - \frac{\delta(\beta\gamma)}{2}
\right] \, ,
\end{equation}
with $T$ the kinetic particle energy, $\tau = \frac{T}{m_ec^2}$ the normalized kinetic energy, and
\begin{align*}
F^-(\tau) & = \frac{1-\beta^2}{2} \, \left[1 + \frac{\tau^2}{8} - (2\tau + 1) \, \ln 2\right] & \text{for electrons,}\\
F^+(\tau) & = \ln 2 - \frac{\beta^2}{24} \, \left[23 + \frac{14}{\tau + 2} + \frac{10}{(\tau + 2)^2} + \frac{4}{(\tau + 2)^3} \right] & \text{for positrons.}
\end{align*}
The remaining parameters are the same as in Eq.~(\ref{eq:S3BBloch}). Equation~(\ref{eq:S3Scol}) not only includes the inelastic impact ionization process but also other scattering mechanisms, such as M{\o}ller and Bhabha scattering.

High-energy electrons and positrons predominantly lose energy in matter by Bremsstrahlung emission which is usually described in the frame of quantum electrodynamics. The first relativistic quantum mechanical theory was formulated by Bethe and Heitler in the Born approximation, considering free-particle wave functions perturbed to the first order in Z (so-called Bethe--Heitler theory) \cite{BH34}. Subsequent developments included various corrections to the Born approximation that account for atomic screening and Coulomb effects not included in the original theory \cite{Koch59}, see also the review \Bref{Tsai74}. In high-energy approximation the radiation losses due to Bremsstrahlung emission in the screened field of a nucleus are  expressed as \cite{Potylitsyn11}
\begin{eqnarray}\label{eq:S3Srad}
S_{rad} & = & \left\langle -\frac{1}{\rho_t}\,\frac{\mathrm{d}E}{\mathrm{d}x}\right\rangle =  \frac{1}{X_0} E \\
\mathrm{with} && X_0^{-1} = 4 \alpha N_A  \Big( \underbrace{\frac{e^2}{4 \pi \varepsilon_0 m c^2}}_{r_e} \Big)^2 \, \frac{Z_t (Z_t + 1)}{A_t} \, \ln\left(\frac{183}{Z_t^{1/3}}\right) \, \left[1 + 0.12 \left(\frac{Z}{82}\right)^2\right] \; . \nonumber
\end{eqnarray}
The parameter $X_0$ is named \emph{radiation length} and plays an important role in the description of \Pepm and photon losses in matter due to Bremsstrahlung and pair creation. It is usually normalized to the target density $\rho_t$  and measured in [g/cm$^2$]. Besides the equation given above for $X_0$, different expressions can be found in the literature, see e.g., Refs.~\cite{PDG18,Tsai74} or the engineering formula
\[
X_0 = 716.4 \; \mathrm{g \, cm^{-2}} \; \frac{A}{Z_t (Z_t + 1) \ln\frac{287}{\sqrt{Z_t}}} \; 
\]
according to \Bref{Eidelman04}. Inspecting Eq.~(\ref{eq:S3Srad}) it can be considered as a simple differential equation with the solution
\[
E(x) = E(0) e^{-x/X_0}
\]
for the energy of electrons and positrons which have emitted Bremsstrahlung while crossing a thin target of thickness $x$. From the general shape of the radiative stopping power
\[
S_{rad} \propto Z_t^2 \frac{E}{m^2}
\]
according to Eq.~(\ref{eq:S3Srad}), it is obvious that radiative losses are especially important for light masses $m$ and for high energies $E$ of the projectile particles. Therefore, Bremsstrahlung emission is strongly suppressed apart from \Pepm beams.

Furthermore, the comparison between $S_{col}$ according to Eq.~(\ref{eq:S3Scol}) and $S_{rad}$ indicates that the collisional losses scale $\propto \ln E$ while the radiative ones $\propto E$, i.e., at high energies $S_{rad}$  dominates as already concluded from \Fref{fig:S3emode}.
The energy at which both types of losses become equal is called the \emph{critical energy} $E_c$, it is defined according to  $S_{col}(E_c) \stackrel{!}{=} S_{rad}(E_c)$. $E_c$ is sometimes used to compare material properties in view of particle energy losses. Different approximations can be found in literature, see e.g., the definition according to \Bref{Berger64},
\begin{align*}
E_c & = \frac{610\UMeV}{Z_t + 1.24} & \text{for solids and liquids,}\\
 & = \frac{710\UMeV}{Z_t + 0.92} & \text{for gases.}
\end{align*}

\begin{figure}[!t]
	\begin{center}
		\includegraphics[width=16.cm]{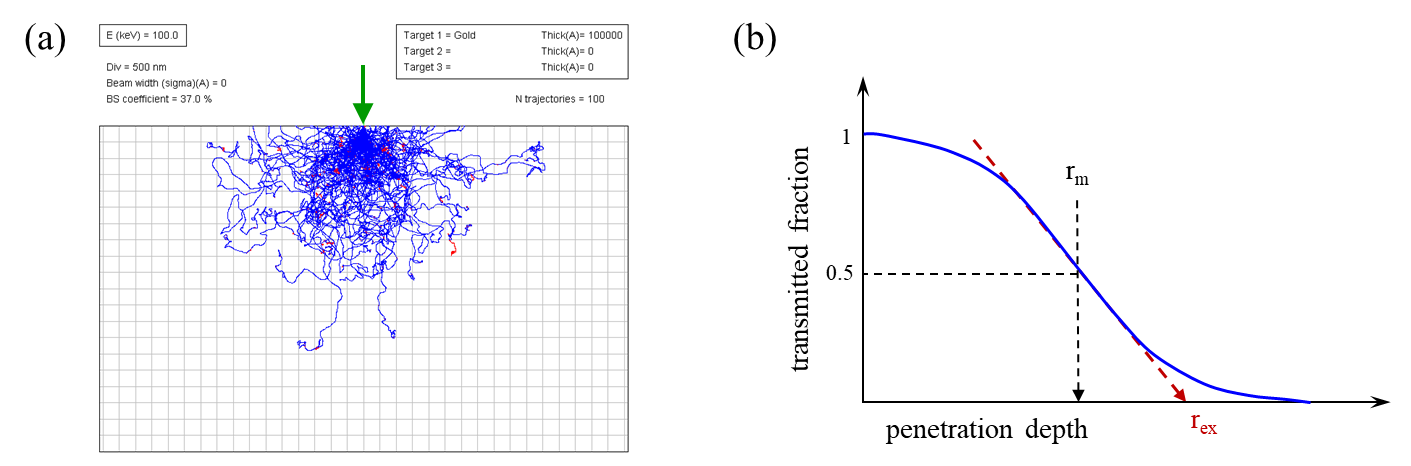}
		\caption{(a) Simulated electron trajectories for a 100~\UkeV{} pencil-like \Pem beam hitting a \Unit{10}{$\mu$m} thick gold target. The incoming beam is shown in green, the primary electron trajectories in blue, the secondary ones in red. The simulation was performed with the \emph{EISS} Monte Carlo simulator taken from \Bref{eiss}. (b) Sketch of the transmission probability of normally incident \Pepm impinging a target. Range definitions are the same as in \Fref{fig:S3range}(a).}
		\label{fig:S3eRange}
	\end{center}
\end{figure}
\begin{figure}[!b]
	\begin{center}
		\includegraphics[width=10.cm]{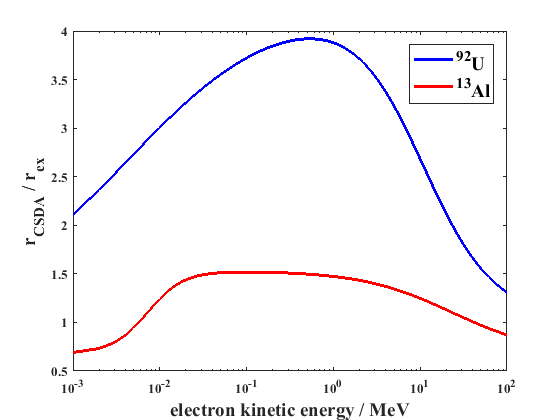}
		\caption{Comparison of extrapolated range $r_{ex}$ and CSDA range $r_{CSDA}$ for $^{92}$U and $^{13}$Al as target material, plotted as a function of the electron kinetic energy. As can be seen the CSDA range usually results in larger values for the penetration depth. The calculation was performed according to \Bref{Tabata96}.}
		\label{fig:S3eRange2}
	\end{center}
\end{figure}
In contrast to heavy charged particles, the range of \Pepm is difficult to treat mathematically. The primary reason for this difficulty is that the particle trajectory cannot be considered as straight line anymore. Due to the low particle mass, large angular deviations caused by scattering are possible, and a non-negligible fraction of the particle kinetic energy may be lost in single collisions. Penetration depth and trajectory length are random, therefore \Pepm beams show a pronounced range straggling. The example shown in \Fref{fig:S3eRange}(a) illustrates this aspect. Different trajectories were simulated for a 100\UkeV{} pencil-like electron beam impinging on a gold target. Figure \ref{fig:S3eRange}(b) shows a sketch of the related range distribution. In contrast to heavy charged particles as plotted in \Fref{fig:S3range}(a), after entering the target surface some of the electrons are immediately stopped and their number is continuously decreasing. In addition, the same range definitions as for heavy charged particles are shown in \Fref{fig:S3eRange}(b). There is a large discrepancy between the mean range $r_m$ and the extrapolated one $r_{ex}$, and it is questionable which range definition is most suitable. Besides the definitions mentioned before, various alternatives can be found in literature such as maximum range, median range, transmission range, or CSDA range, see e.g., \Bref{Iskef83}. The CSDA range, Eq.~(\ref{eq:S3csda}), which is commonly used for heavy charged particles has to be applied with the total stopping power $S_{tot} = S_{col} + S_{rad}$ in case of \Pepm beams. However, it usually overestimates the penetration depth as can be seen from the comparison in \Fref{fig:S3eRange2} which was calculated according to \Bref{Tabata96}. One of the range definitions widely in use is the extrapolated range $r_{ex}$, parametrizations can be found for example in Refs.~\cite{Inguimbert16,Tabata96}. 

Due to the ambiguity in particle range definitions and similar to the situation with heavy charged particle beams, a number of authors proposed empirical range--energy expressions describing the \Pepm extrapolated ranges in various materials predominantly for low-energetic beams. Examples of this type of parametrization can be found in Refs.~\cite{Katz52,Kobetich69}.

As can be seen from this compilation, the way in which heavy charged particles and light leptons interact with matter is quite different, and the semi-empirical equations which are necessary for the calculations of energy losses and ranges are rather cumbersome. Fortunately there exist freely accessible stopping power and range tables for electrons, protons, and He ions \cite{NIST} which are based on Refs.~\cite{ICRU49,ICRU37} and which enable a fast calculation of losses and CSDA ranges. However, the discussion above showed that some definitions are not consistent, especially for the \Pepm range due to the large range straggling. It should be pointed out that the semi-empirical expressions and parametrizations discussed so far are good for a first insight. If deeper understanding is required, Monte Carlo based simulation tools are recommended as for example Geant4 \cite{GEANT4a,GEANT4b,GEANT4c}, FLUKA \cite{FLUKAa,FLUKAb}, or EGS5 \cite{EGS}. Meanwhile the field of particle matter interaction is a domain of simulation toolkits, and depending on the task to solve and the strategy of the laboratory, one of the simulation codes should be used, all of which have their advantages and disadvantages.

\subsection{Interaction of external electromagnetic fields with charged particles}

The last measurement principle introduced in the beginning of this chapter is the interaction of external electromagnetic fields with charged particles. This kind of interaction is being deployed less frequently. Applications can be found in bunch length and transverse bunch profile measurements. On the one hand, external electromagnetic fields can act as signal source in the sense that photons from a laser are scattered at beam particles and the beam shape is scanned. On the other hand, external fields may be used in order to manipulate the beam and prepare it for a subsequent measurement. This beam manipulation can be an atomic excitation of an ion beam or an external force induced by the electromagnetic field which acts on the charged particle beam. Both variants are briefly described in the following.

\subsubsection{External electromagnetic fields as signal source}

\begin{figure}[!b]
	\begin{center}
		\includegraphics[width=15.cm]{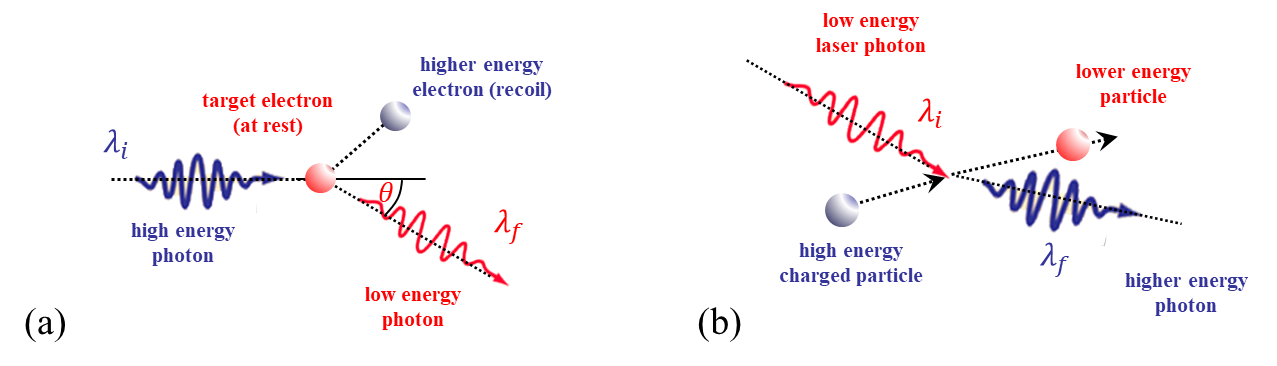}
		\caption{Schematic diagram of (a) classical Compton effect and (b) inverse Compton effect}
		\label{fig:S3Compton}
	\end{center}
\end{figure}

The inelastic scattering of photons on charged particles is named the \emph{Compton effect}. In classical Compton scattering it is assumed that  photons are scattered on a \emph{quasi-free} atomic l electron
\[
 \Pgamma + \mathrm{atom} \quad \longrightarrow \quad \Pgamma + \Pem + \mathrm{ion^-} \; ,
\]
and the photon energy is large compared to the electron's binding energy. In the inelastic scattering process, which is schematically depicted in \Fref{fig:S3Compton}(a), the photon is deflected and its wavelength changes due to energy transfer, i.e., the photon loses energy and the electron gains recoil energy.

The Compton differential cross-section is described by the Klein--Nishina formula \cite{KleinNishina29}
\begin{equation}\label{eq:S3KNishina}
    \frac{\mathrm{d}\sigma_c}{\mathrm{d}\Omega} = \frac{1}{2} \,
    \Big( \underbrace{\frac{e^2}{4 \pi \varepsilon_0 m c^2}}_{r_e} \Big)^2 \,
    \left\lbrace\frac{1}{1 + \varepsilon (1 - \cos\theta)} \right\rbrace^2 \,
    \left[ 1 + \cos^2\theta + \frac{\varepsilon^2 (1 - \cos\theta)^2}{1 + \varepsilon (1 - \cos\theta)} \right] \, ,
\end{equation}
with $\varepsilon = \frac{\hbar\omega}{mc^2}$ the incident photon energy normalized by the particle rest mass and $\theta$ the polar angle as defined in \Fref{fig:S3Compton}(a). As can be seen from Eq.~(\ref{eq:S3KNishina}) the cross-section scales like
\[
\frac{\mathrm{d}\sigma_c}{\mathrm{d}\Omega} \propto \left(\frac{1}{m c^2}\right)^2 \, ,
\]
i.e., Compton scattering is strongly suppressed for heavy charged particles and plays only a role in case of \Pepm beams.

However, in case of light scattering at electrons or positrons in an accelerator the situation is quite different, the scattering particles are not at rest and the ultra-relativistic \Pepm energy is large compared to the low photon energy from an optical laser. In contrast to the conventional Compton effect, the low photon energy is boosted at the expense of beam particle energy, cf. \Fref{fig:S3Compton}(b). Due to this inverse situation the process is named the \emph{inverse Compton effect}.

The differential cross-section for inverse Compton scattering has to be calculated in three steps. In the first step the incoming photon is Lorentz transformed from the laboratory system to a reference frame in which the electron is stationary. In the second step formulae for the classical Compton effect are applied which require that the electron be stationary. Finally, the third step is to switch back to the laboratory frame. The differential cross-section for inverse Compton scattering defining the energy spectrum of the emerging gamma rays is expressed as \cite{Arutyumian63,Milburn63,Shintake92}
\begin{figure}[!b]
	\begin{center}
		\includegraphics[width=10.cm]{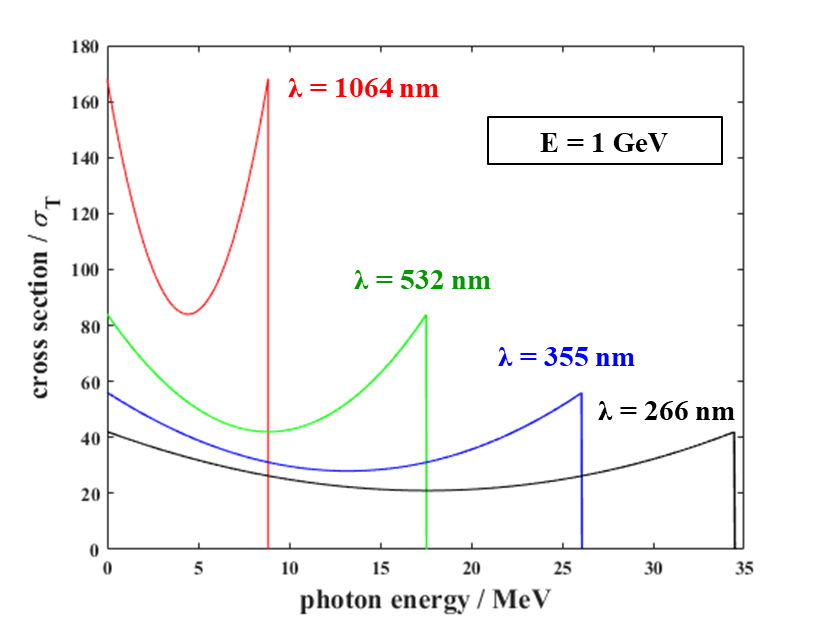}
		\caption{Calculated cross-sections for inverse Compton scattering according to Eq.~(\ref{eq:S3invCompton}), using the first four harmonics of a Nd:YAG laser with a 1\UGeV electron beam}
		\label{fig:S3invCompton}
	\end{center}
\end{figure}
\begin{equation}\label{eq:S3invCompton}
\frac{\mathrm{d}\sigma_{ic}}{\mathrm{d}\varpi} = \frac{3}{8} \, \frac{\sigma_T}{\epsilon_1} \,
\left[\frac{1}{1 - \varpi} + 1 - \varpi + \left\lbrace \frac{\varpi}{\epsilon_1 (1 - \varpi)} \right\rbrace^2
- \frac{2 \varpi}{\epsilon_1 (1 - \varpi)}
\right] \, ,
\end{equation}
with
\begin{alignat*}{2}
\sigma_T & = \frac{8 \pi}{3} r_e^2 & \hspace*{1cm} &\text{Thomson cross section,}\\
\epsilon_1 & = \frac{\gamma \hbar \omega_0}{m_e c^2} & \hspace*{1cm}  &\text{normalized energy of laser photons,}\\
\varpi & = \frac{\hbar \omega_{\gamma}}{E} & \hspace*{1cm}  &\text{normalized energy of emitted photons.}
\end{alignat*}
Figure~\ref{fig:S3invCompton} shows calculated cross-sections for inverse Compton scattering according to Eq.~(\ref{eq:S3invCompton}). As can be seen, due to the scattering with the ultra-relativistic electron beam the laser photons are boosted in energy up to several MeV.

With such high photon energies the signals from the scattered photons can well be discriminated against the background radiation always present in an accelerator environment. Therefore inverse Compton scattering is a very suitable process for signal generation from light lepton beams and utilized e.g., for a laser wire scanner in high resolution transverse beam profile diagnostics.

\subsubsection{External electromagnetic fields for beam manipulation}

External fields may be used in order to manipulate the beam and prepare it for a subsequent measurement. Two examples for this interaction are given in the following.

While Compton signal generation is strongly suppressed for heavy charged particles, nevertheless a laser beam may interact via different reaction channels. Practical applications can be found in the case of \PHm beams from ion sources which are favoured for many proton accelerator applications, in particular because of the possibility of applying charge exchange injection. Usually thin carbon foils are used as strippers for this kind of injection into high intensity proton rings. However, they become radioactive and produce uncontrolled beam losses, which is one of the main factors limiting beam power in high intensity proton rings. Instead, laser stripping of \PHm ion beams is successfully applied as alternative method \cite{Danilov07}. 
\begin{figure}[!h]
	\begin{center}
		\includegraphics[width=15.cm]{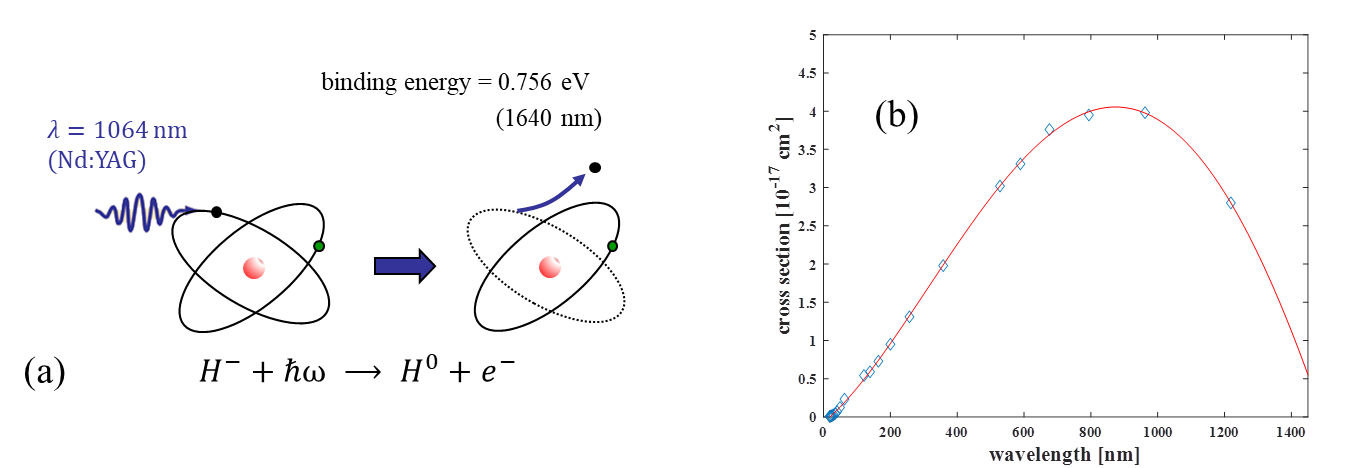}
		\caption{(a) Illustration of the process of photo neutralization of an \PHm ion using a Nd:YAG laser. (b) Calculated cross-section for \PHm photo neutralization as function of photon wavelength $\lambda$. Data are taken from a table in \Bref{Broad76}.}
		\label{fig:S3Hstrip}
	\end{center}
\end{figure}

In case of beam diagnostics and instrumentation, photo neutralization of \PHm ions is successfully used as basis for a laser wire scanner, see e.g., \Bref{Connolly12} and \Fref{fig:S3Hstrip}(a). The binding energy of the additional electron is 0.756\UeV{} which allows a neutralization by a photon with wavelength $\lambda$ < \Unit{1.64}{$\mu$m} (for example with a Nd:YAG laser). Since the detached electron is boosted into an energy continuum the cross-section versus photon wavelength is a broad curve with maximum $\lambda \approx$ 900 nm, cf. \Fref{fig:S3Hstrip}(b). Using this method, ion-beam profiles are measured by scanning the laser across the \PHm beam and measuring the laser-stripped electron charge versus the laser position.

The second example is the transverse deflecting structure (TDS) which has become a widely utilized diagnostics component for investigations of the longitudinal phase space in high brightness electron beam applications \cite{Emma00,Roehrs09}. It is an iris loaded RF waveguide structure which is designed to provide hybrid deflecting modes (HEM$_{1,1}$), i.e., a linear combination of TM$_{1,1}$ and TE$_{1,1}$ dipole modes. As a result it provides a transverse force that acts on the synchronously moving relativistic particle beam. For beam diagnostic applications it is operated as a RF deflector, the operation principle can be understood as an intra-beam streak camera. Further information about the TDS working principle can be found e.g., in \Bref{Gillespie18}.

\subsection{Conclusion}

From the discussion in this chapter it is obvious that there is a fundamental difference in the way  signals are generated from heavy charged particle and from \Pepm beams. The cause of this difference is the particle rest mass. In the case of electromagnetic fields carried by the beam the mass difference translates into a difference in the Lorentz factor $\gamma = E / m_0c^2$ which determines the field line compression and their transverse range. In the case of scattering or particle matter interaction it is the particle rest mass which significantly determines the kinematics.

As a consequence of the difference in beam signal generation, the underlying physical processes applied for beam instrumentation in an \Pepm and a hadron accelerator are not the same. This may require completely different monitor concepts even for the measurement of the same beam parameter---and if the concept is the same the technical realization may look quite different. However, it is not only the beam instrumentation which is affected, but also the beam dynamics is quite different for different particle species. At the one hand, due to the large $\gamma$ the electromagnetic fields of an \Pepm beam have a large transverse range and may easily interact with the surrounding vacuum chamber causing instabilities. At the other hand, in \Pepm accelerators synchrotron radiation is emitted which causes radiation damping. Both mechanisms are suppressed for hadron machines. The different conditions in light lepton and heavy charged particle accelerators may impose additional restrictions on design and performance of the beam instrumentation.

Therefore the focus of the second part of this report which follows in the subsequent chapters lies on the specific diagnostics needs of hadron and \Pepm accelerators. It is based on the lecture \Bref{Kube08b} which was given at the CERN accelerator school in 2008 in Dourdan (France), the subsequent chapters will briefly summarize and update the former report.

\section{General aspects of beam diagnostics and instrumentation}

In the following an overview of primary beam parameters together with examples for values which can be deduced is given. This list is far from complete, more details can be found in specific textbooks or lecture notes as in Refs.~\cite{Minty03,Strehl06,Smaluk09,Forck17}.

The \emph{beam intensity} is one of the most interesting accelerator parameters. A measurement of the beam intensity could mean the determination of the bunch current (charge), of the dc current, or even of both. With knowledge of the intensity it is possible to determine lifetime and coasting beam in circular machines, or transfer efficiencies in linacs and transfer lines.

The precise determination of the beam centroid transverse \emph{position} with respect to the target position (so-called \emph{golden orbit}) is of utmost importance because position measurements give access to a wide number of very important accelerator parameters. The most fundamental one is the beam orbit from which the lattice parameters can be deduced. Position measurements are also required for tune measurements and the determination of the chromaticity; they are a fundamental part of feedback systems and more.

\emph{Beam profiles} in both transversal and longitudinal dimensions are of special interest. Beam size measurements are fundamental for the determination of the beam emittance; time resolved beam size studies give information about injection mismatch (betatron and dispersion matching) via the observation of turn-by-turn shape oscillations, or about dynamical processes as for example the study of beam blow up of individual bunches under collision in a particle collider. Furthermore beam halo diagnostics rely mainly on measurements of the transverse beam size.

Another parameter of interest is the \emph{beam energy}, but mainly for users. In a lepton collider for example it defines the reaction energy which is available in order to produce new particles, while in synchrotron light sources it defines the spectral characteristics of the emitted radiation.

In a collider the \emph{luminosity} is the key parameter because it defines the count rate of the reaction channel under investigation. While an absolute online-luminosity determination is sometimes difficult to provide, the determination of a relative luminosity or simply a count rate which is proportional to it is a very important tool for the collision optimization (angle and position) of both beams via beam steering with local bumps.

In order to identify positions of beam losses, to prevent damage of accelerator as well as of facility components, and to optimize the daily accelerator operation, \emph{beam loss} monitors represent a very important diagnostic system.

In the next sections examples of these systems for different accelerator types will be given. Nevertheless, depending on the type of accelerator there might be even more beam parameters of interest. In heavy ion machines for example the determination of particle charge states and mass numbers are essential for the accelerator operation. However, these parameters are out of the focus of this report, more details can be found e.g., in Refs.~\cite{Strehl06,Forck17}.

\subsection{Operational modes}

Depending on the operational mode of an accelerator there exist different requirements for beam instrumentation. Sometimes they cannot be fulfilled with only one device. As a consequence two or more instruments are needed in order to measure the same beam parameter under different operational conditions because the dynamical range of a single device may not be sufficient.

One can roughly distinguish between two different modes of operation and summarize their impact on beam instrumentation:
\begin{enumerate}
\item Monitors for accelerator (section) commissioning
\begin{itemize}
\item applied in order to adjust the beam transport through different accelerator sections;
\item required for the characterization of the beam behind each accelerator section;
\item simple and robust devices with high sensitivity, allowing  operation with single or a few bunches of low intensity;
\item low or modest demands on accuracy;
\item applications of beam disturbing methods are possible.
\end{itemize}
\item Monitors for standard operation
\begin{itemize}
\item applied for precise beam characterization in order to control and improve the accelerator operation;
\item required for the diagnosis of unwanted errors and to trigger interlocks;
\item devices are typically based on more or less sophisticated schemes;
\item high demands on accuracy;
\item application of minimum beam disturbing schemes.
\end{itemize}
\end{enumerate}
As can be seen from this comparison, even one accelerator has already specific diagnostics needs for different operational modes. However, the aim of the following sections is to compare the diagnostics requirements for different accelerators. Besides the classification according to the particle mass in light lepton and heavy charged particle beams, an intuitive way to subdivide different accelerator types is to distinguish between linear and circular machines as discussed in the following section.

\subsection{Linear versus circular accelerators}

In order to judge this means of classification the basic differences of these accelerator concepts will be recalled below.

A linear accelerator has many accelerating cavities through which the beam passes once. One can consider a linac as an `open loop' system in the sense that there exists no possibility for an orbit feedback, and everything depends on the start parameters. Furthermore, a linac is a pulsed system. This means that the signals generated by the beam are a sequence of single events which may vary from shot to shot, the formation of an equilibrium state is not possible. In a linac emittance and energy are  functions of the location in the accelerator. Because the charge can be lost everywhere in the machine, many devices for transfer measurements are required. In case of beam loss in the machine the gun will supply charge until it is stopped e.g., by an interlock.

In contrast to that a circular accelerator or storage ring has only a small number of accelerating cavities. It can be understood as a `closed loop' system in the sense that there exists a periodical solution for the particle orbit, and  orbit feedback is possible. Due to the closed loop behaviour resonances can occur in the system. A storage ring is a continuous wave (cw) system, the signals from the beam are repetitive and stable in general for many turns. It is possible that the beam reaches a kind of equilibrium state, and with it also the beam generated signals. Therefore high precision can be achieved by averaging, and the signals are typically treated in the frequency domain. Emittance and beam current are non- or slowly varying parameters. In case of a beam loss there are no particles in the storage ring until the beam is transferred again through the injector chain.

From the preceding comparison and the discussion in the first chapters it is clear that there are fundamental differences in the way beam generated signals are treated and processed. This is underlined by the comparative analysis of several accelerator types in the following. 
First a hadron collider together with its injector chain will be considered, afterwards diagnostics and instrumentation for electron accelerators will be described. This includes diagnostics for lepton storage rings and for synchrotron light sources of 3$^{rd}$ and 4$^{th}$ generation.


There exist a number of particle colliders for high energy physics in the world, see e.g., the review article of Ref. \cite{Edwards08}. Examples for hadron colliders in Europe are the Large Hadron Collider (LHC) at CERN \cite{LHC04,Myers12} which is presently undergoing a luminosity upgrade (HL-LHC) \cite{LHC17}, the heavy ion facility FAIR at GSI (Germany) \cite{FAIR06,Spiller18} which is under construction, and the Hadron Electron Ring Anlage (HERA) at DESY \cite{Voss94}, an $\Pe\Pp$ collider which was shut down in 2007. References~\cite{Scandale14,Fischer14,BenZvi14} give a comprehensive review for $\Pp\Pp$ ($\Pp\APproton$), ion, and electron--hadron collider projects worldwide. Beyond that, new projects at the energy frontier are in discussion \cite{Benedikt18a}, such as for example the Future Circular Collider (FCC-hh) at CERN \cite{Benedikt18b} and the Super $\Pp\Pp$ Collider (SppC) in China \cite{SppC}.

Circular hadron colliders are known as discovery machines. Their discovery reach is determined by the beam energy which depends on the dipole magnetic field and the size of the collider. Besides the collision energy, a key parameter for collider exploitation is the luminosity $\mathfrak{L}$. The nominal values of collision energy and luminosity are both deeply related to a series of accelerator physics and technology issues. Typically the luminosity is used to quantify the collider performance. It is a relativistic invariant proportionality factor between the reaction cross-section $\sigma$ (a property of the interaction itself) and the event rate $\dot{N}$ which is the primary concern for the experiments
\begin{equation}\label{eq:S5Lumi}
\dot{N} = \mathfrak{L} \, \sigma \; .
\end{equation}
Apart from the necessity of measuring the luminosity, which requires an additional beam monitor, from a diagnostics point of view there is no difference between a collider and a storage ring. Therefore in the following no distinction will be made between these types of machines.

\subsection{General considerations}

Below some general aspects of beam diagnostics will be pointed out which are common for all high-energy hadron machines and their injector chains.

Firstly, a high beam energy requires superconducting magnets to achieve the required dipole magnetic field for particle bending. In order to obtain superconductivity, the magnets have to be cooled down with the consequence that parts of the beam instrumentation are located in the cold vacuum system. Furthermore, a superconducting environment requires reliable beam-loss monitors which are an integral part of the fast quench protection system needed to avoid an uncontrolled transition from the super-conducting to the normal-conducting state which may damage the magnets or, at least, stop the machine for some time to recover from the quench.

\begin{figure}[!h]
\begin{center}
\includegraphics[scale=.65,clip]{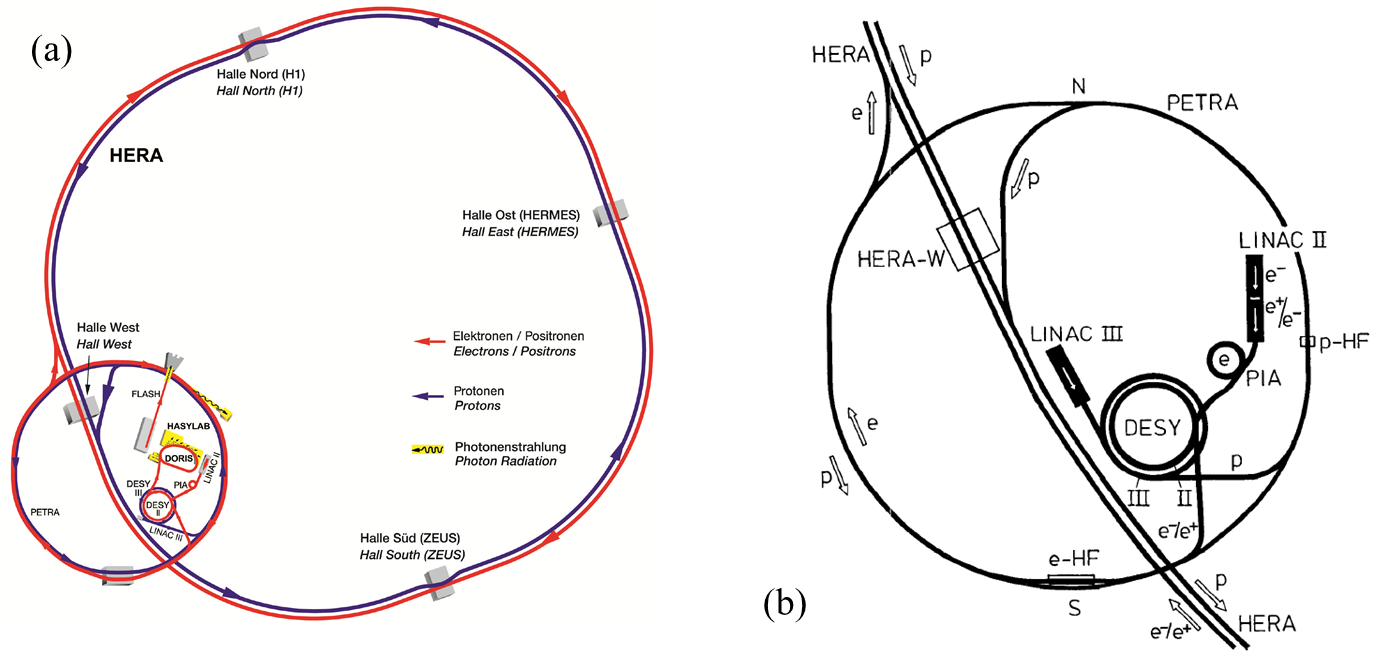}
\caption{(a) Overview of the former HERA accelerator complex at DESY with the $\Pe\Pp$ collider HERA which was shut down in 2007. (b) HERA injector chain. The proton injection system consisted of Linac III, the DESY III injector synchrotron and PETRA; the electron injection system consists of Linac II, the positron intensity accumulator PIA, the DESY II injector synchrotron, and the PETRA II ring which is used for both electrons and protons \cite{Voss94}.}
\label{fig:S5desy}
\end{center}
\end{figure}

On the other hand high particle energies cannot be reached in a single machine, i.e., typically a long injector chain is required in order to achieve the final beam energy. Figure \ref{fig:S5desy} gives an overview of the former $\Pe\Pp$ collider HERA at DESY \cite{Voss94} with 920 GeV proton energy and 27.5 GeV electron energy together with the injector complex. However, a long injector chain means that there exist several pre-accelerators with beam properties which involve typically different requirements for beam diagnostics and instrumentation.

From the demands on the high-energy storage ring it is possible to deduce directly several consequences for the injector chain.

In order to have a sufficient counting rate in  colliding beam experiments it is necessary to achieve optimum collider performance, i.e., the luminosity must be as high as possible. Assuming identical round counter-propagating beams of equal bunch populations $N$, the luminosity can be expressed as \cite{Herr06}
\begin{equation}\label{eq:S5Lumi2}
    \mathfrak{L} = \frac{\gamma N^2 n_b f}{4 \pi \varepsilon_n \beta^*} \, S \; ,
\end{equation}
where $n_b$ is the number of bunches per beam, $f$ the revolution frequency, $\gamma$ the Lorentz factor, $\beta^*$ the betatron function at the collision point, $\varepsilon_n$ the normalized beam emittance, and $S$ a reduction factor accounting for several effects (crossing angle, hourglass effect, $\ldots$). As can be concluded from Eq.~(\ref{eq:S5Lumi2}), the bunch populations $N$ should be as high as possible and the emittance  $\varepsilon_n$ should be as small as possible. However, the beam emittance in a circular machine is strongly connected to the emission of synchrotron radiation. In a lepton accelerator the radiation emission leads to radiation damping on the one hand and quantum excitation (random excitation of oscillations) on the other hand \cite{Walker94,Rivkin98}. The consequence is a formation of an equilibrium emittance, i.e., the emittance is determined by the storage ring itself. In a hadron machine the situation is completely different. Due to the large particle rest mass the synchrotron radiation emission is strongly suppressed. As a result the emittance is essentially determined already in the injector chain. This implies for the beam diagnostics that an accurate characterization of the beam is already required for the lowest-energy machines. Furthermore, only minimum-disturbing instrumentation can be used: a beam distortion in the injector chain is irreversible and may result in an emittance blow-up in the final machine, involving a reduction of the collider luminosity which has to be avoided.

According to Liouville's theorem the normalized emittance $\varepsilon_n$ is conserved during the acceleration process. However, it is the absolute emittance $\varepsilon$ which defines the beam size and which is therefore the parameter of interest. Both parameters are related according to
\begin{equation}\label{Eq:Emit}
    \varepsilon = \frac{\varepsilon_n}{\beta \gamma}
    \qquad
    \mathrm{with}
    \qquad
    \beta \gamma = \frac{p c}{m_0 c^2} \; ,
    \quad
    \mathrm{see~Eq.~(\ref{eq:S2bg}).}
\end{equation}
Therefore the absolute emittance (and with it the achievable beam size) shrinks adiabatically during the acceleration process. As an example, the LHC injector chain is considered: from the 50 MeV injector Linac II to the 450 GeV extraction from the SPS the absolute emittance shrinks by a factor of 1450, and the factor between Linac II and the LHC at full energy of 7 TeV amounts to 22000. The consequence for beam instrumentation is that beam sizes and divergences are rather large at the beginning of the injector chain. As a result, a tight mesh of focusing magnets is required in order to transport the beam, and usually there is only restricted space for instrumentation.

Furthermore, in the low energetic part of the injector chain the particle speed still changes with acceleration. This offers the possibility of using time of flight (TOF) measurements for some applications. Low-energy particles have also a small magnetic rigidity, i.e., they are easy to bend. Therefore magnetic spectrometers can be used for momentum determination. However, space charge effects are pronounced at low energies and may influence particle dynamics and/or beam diagnostic measurements (especially for heavy ion beams). Finally, according to the collision stopping power Eq.~(\ref{eq:S3BBloch}) the energy deposition in matter is high for low energetic particles. This may lead to a degradation or even the destruction of interceptive beam monitors like screens.

In the subsequent section the instrumentation of a hadron collider will be described together with its injector chain. For illustration the HERA injector accelerators are described which represent a standard injector chain for hadron accelerators. Starting from the linac a `walk' along the injector complex will be done. Specific properties of the individual accelerators together with the consequences for beam diagnostics are discussed. Further information about specific instrumentation and diagnostics for high-intensity hadron beams can be found in Ref.~\cite{Wittenburg13}.

\subsection{Source and injector linac}

\begin{figure}[!t]
	\begin{center}
		\includegraphics[scale=.7,clip]{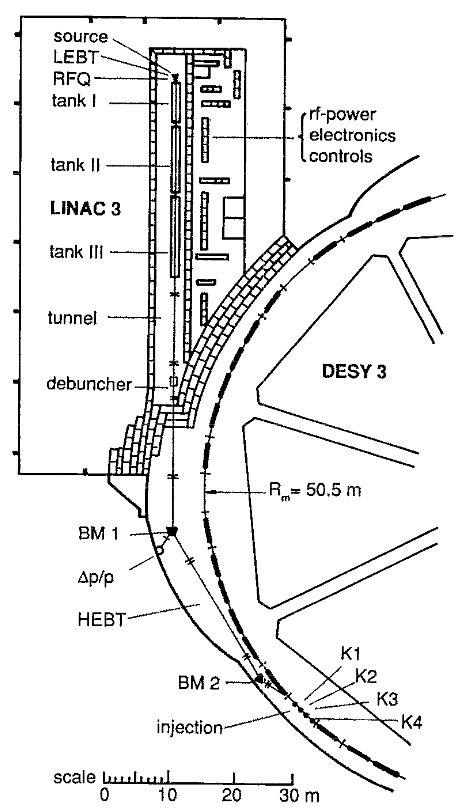}
		\caption{Overview of the 50\UMeV{} \PHm injector linac for HERA}
		\label{fig:S5linac3}
	\end{center}
\end{figure}

\begin{figure}[!h]
\begin{center}
\includegraphics[scale=.5,clip]{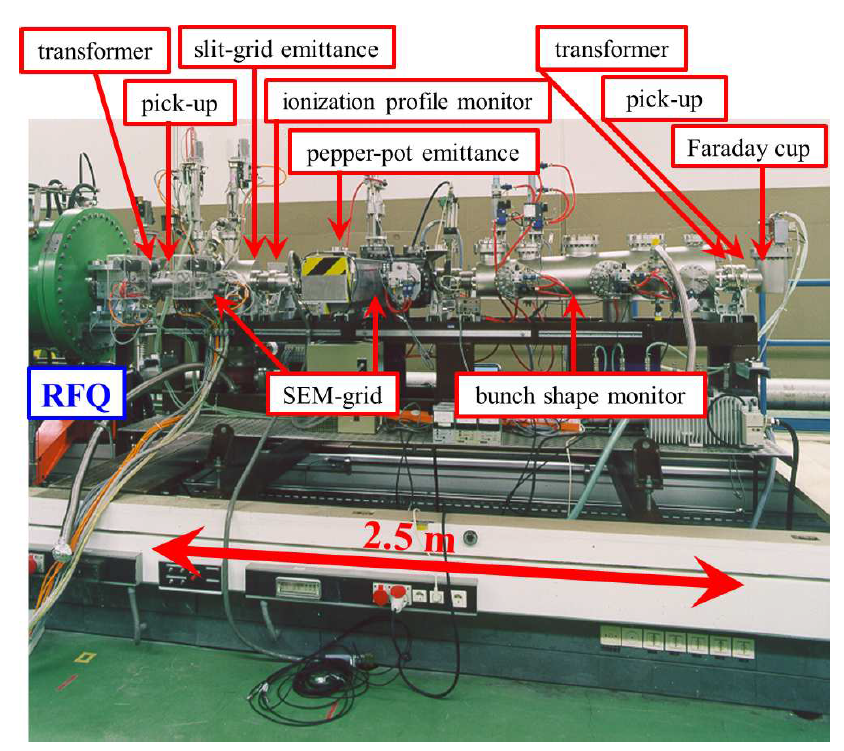}
\caption{Moveable test bench which was used during the commissioning of the high current RFQ at the GSI \cite{Forck99}. Picture taken from \Bref{Forck17}.}
\label{fig:S5GSIbench}
\end{center}
\end{figure}

A widely used scheme for a proton injector chain is to produce and accelerate negative hydrogen ions (\PHm) and inject them into a circular accelerator whereby charge exchange injection using a stripper foil the injection process is not limited by Liouville's theorem. Figure \ref{fig:S5linac3} gives an overview of the 50\UMeV{} \PHm injector Linac III for HERA. In the following a short description of the subsystems is given. More details can be found e.g., in \Bref{Linac3}.
\begin{itemize}
  \item \PHm sources:\\
  Electron capture of vibrationally excited hydrogen molecules and subsequent dissociation into a neutral and negative ion results in the formation of an \PHm beam. Two sources can be operated: either a 18 keV magnetron source or an RF-driven volume source.
  \item Low-energy beam transport (LEBT):\\
  the LEBT serves as matching section for the beam to the acceptance of the subsequent radio frequency quadrupole (RFQ).
  \item Radio frequency quadrupole:\\
  the RFQ accelerates the beam from 18\UkeV{} up to 750\UkeV{}.
  \item Medium energy beam transport (MEBT):\\
  the MEBT serves as matching section for the beam to the acceptance of the subsequent linac structures.
  \item \PHm linac (tank I-III):\\
  the linac is a conventional Alvarez type linac with end energy $T_{kin} = 50\UMeV$.
  \item High-energy beam transport (HEBT):\\
  the HEBT serves as matching section for the beam to the acceptance of the injector synchrotron. Furthermore an integrated diagnostic beamline (indicated by $\Delta p /p$ in \Fref{fig:S5linac3}) is used to measure the beam properties during linac tuning.
  \item Injection in synchrotron:\\
  an \PHm multi-turn injection scheme is applied, i.e., the hydrogen ions are converted to protons using a stripper foil.
\end{itemize}
From an operational point of view the injector linac needs key devices for the adjustment of the beam transport through the individual linac sections, for the setting of the RF system (i.e., the phases and amplitudes), and to indicate the operating status during standard operation. Apart from the beam monitors installed in the linac itself, usually several diagnostic instruments are grouped together in a permanently installed diagnostic beamline behind the linac sections. In addition, during the commissioning phase sometimes a moveable diagnostic test bench is used which allows a full six-dimensional phase space characterization behind each linac section. As an example, \Fref{fig:S5GSIbench} shows the test bench which was used during commissioning of the high current RFQ at the GSI facility in Darmstadt (Germany).

In the following key instruments for linac diagnostics will shortly be described. Extended descriptions about each monitor can be found in dedicated articles in this CERN report or in the former CERN Accelerator School about \emph{Beam Diagnostics} \cite{Dourdan08}. Further information and different examples of beam instrumentation and commissioning strategies in hadron linacs can be found e.g., in Refs.~\cite{Ostroumov98,Forck98,Raich05,Sieber14,Wang16,Lombardi17}.

\subparagraph{Current and transmission}

Two different types of monitors are widely  used, Faraday cups and beam current transformers. Faraday cups are destructive monitors, i.e., they are beam stoppers which are isolated from the beam pipe ground potential and connected to a current meter. A Faraday cup allows to measure very low intensities down to the pA region and provides absolute charge or current values.

Beam current transformers (or AC current transformer, ACCT) determine the current in a non-destructive way. In this type of monitor the beam acts as single turn primary winding of a transformer, and the AC component of the current is measured. In contrast to Faraday cups a  transformer requires a calibration.

Besides the report in these proceedings \cite{Peters18a} more information about beam current measurements can be found in \Bref{Denard08} and in the tutorials  Refs.~\cite{Webber94,Gelato94}.

\subparagraph{Beam position monitors (BPMs)}

Position information can be gained via the electric, the magnetic, or the electromagnetic fields. Specific for a hadron linac injector chain is that bunch lengths are rather large and acceleration frequencies are rather low. Therefore the corresponding beam spectrum contains only low frequencies (typically in the kHz up to the 100 MHz region). To provide a precise determination of the beam position a high pick-up sensitivity is required at these frequencies.

Furthermore, the induced signals from the non-propagating fields are rather small because of the small Lorentz factor $\gamma$ for particles at low energies with large rest masses, cf. Eq.~(\ref{eq:S2bg}). Therefore, in order to capture as many field lines as possible the pick-ups have typically large electrodes.

\begin{figure}[!h]
\begin{center}
\includegraphics[scale=.8,clip]{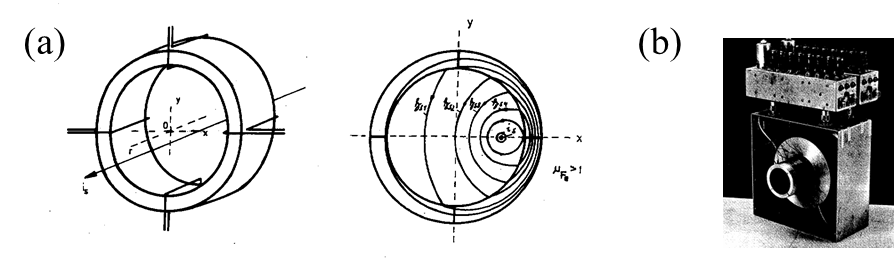}
\caption{(a) Working principle of an inductive pick-up together with the induced magnetic field distribution in the high permeability core. (b) Technical realization  according to \Bref{Kriens68}.}
\label{fig:S5LinacBPM}
\end{center}
\end{figure}

An example for a capacitive beam position monitor having large electrodes is the `shoe-box' type pick-up which is widely in use at heavy ion accelerators. Figure \ref{fig:S5LinacBPM} shows the realization of an inductive monitor \cite{Kriens68}. This pick-up type is seldom in use due to the high sensitivity to magnetic stray fields always present in an accelerator environment.

Comprehensive review articles about BPMs can be found in Refs.~\cite{Lambertson88,Shafer89,Schulte94,Smith97,Forck08} together with the report in these proceedings \cite{Wendt18}.

\subparagraph{Transverse beam profiles}

\begin{figure}[!t]
\begin{center}
\includegraphics[scale=.6,clip]{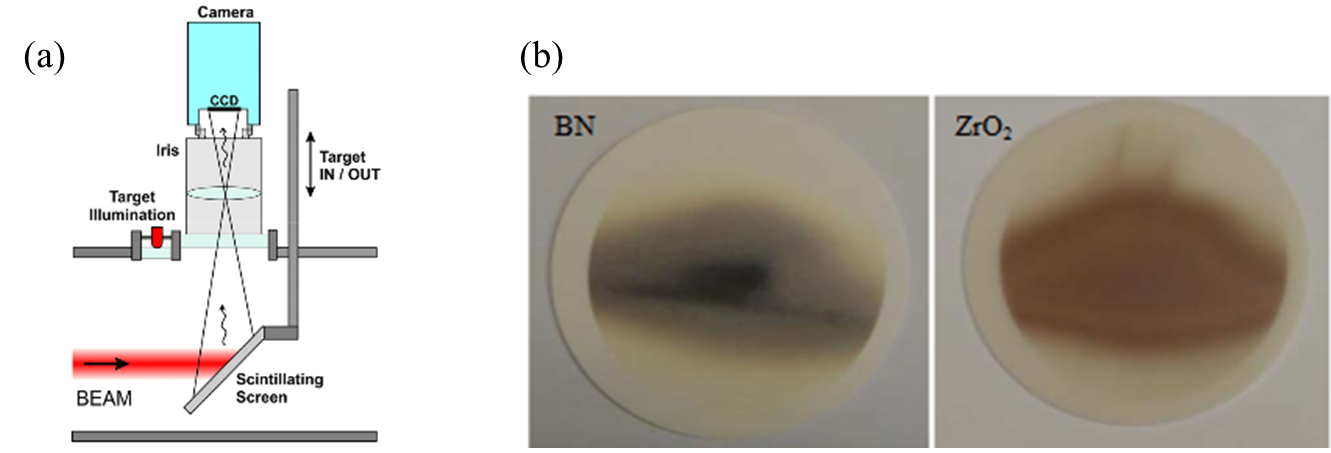}
\caption{(a) Scheme of intercepting scintillator screen set-up. Picture taken from \Bref{Beata11}. (b) Radiation damaged screens from a scintillating screen study for the LEIR/LHC heavy ion beams \cite{Bal05}.}
\label{fig:S5Screen}
\end{center}
\end{figure}

The determination of the transverse beam profile relies on the interaction of beam particles with matter. The method mostly applied is a destructive one where the particle beam hits a luminescent screen, cf. Fig.\ref{fig:S5Screen}(a). A part of the deposited energy results in excited electronic states which de-excite partially via light emission. Therefore the beam profile can be observed via a CCD camera.
Sometimes screens are even used instead of beam position monitors by analysing the centre of gravity of the measured light distribution. Care has to be taken because of the high energy deposition in material according to the Bethe--Bloch equation which is especially critical for heavy ion machines. This may lead to a degradation of the screen material as shown in Fig.\ref{fig:S5Screen}(b).

Less destructive methods are in use where the luminescent screen is replaced by a configuration of stretched wires or strips. If the particles hit the surface, secondary electrons are liberated. The secondary current from each individual wire/strip is converted to a voltage via a current-to-voltage amplifier, and the voltage distribution from all wires is a measure of the transverse beam profile. Configurations in use are the wire grid (i.e., stretched wires in both transverse planes), the harp (i.e., stretched wires in one transverse plane), or secondary emission monitors (SEMs) which consist of strips with a larger surface and therefore higher sensitivity.

A nearly non-destructive profile measurement relies on the creation of gas ions and free electrons in the beam interaction with residual gas in the beam pipe. The ionization products are accelerated via electrostatic guiding fields towards a microchannel plate for signal enhancement (secondary electron generation with a multiplication factor of up to 10$^7$), and signal readout is performed either optically (phosphor screen together with CCD camera) or electronically (wire array with guiding field). A variation of this type of monitor is the residual gas fluorescence monitor.

Supplemental to the report in these proceedings \cite{Bravin18}, comprehensive review articles about the monitor types described in this section can be found in Refs.~\cite{Jung88,Jung94,Plum04,Bravin08,Beata11,Beata12,Schwickert13}.

\subparagraph{Transverse emittance}

A method often applied in proton/heavy ion linacs is the slit-grid measurement. A slit formed by two metal blades produces a vertical slice in the transverse phase space. In the free-field drift space behind the slit (typical length 10 cm to 1 m), the angular distribution of the slice is transformed into a spatial one which can be scanned with a moveable intensity detector. Moving the slit across the phase space ellipse and repeating the procedure described above it is possible to scan the whole phase space. In order to reduce the number of measurements it is convenient to replace the moveable detector by a spatial resolving one like a SEM or profile grid. A further reduction of the measurements can be done with a \emph{pepper pot} which offers even single shot capability. It uses a viewing screen to observe the trajectory-angle distribution of the individual beamlets sampled with a pepper pot plate (matrix of small holes).

Besides the report about emittance diagnostics and instrumentation in these proceedings \cite{Bravin18b}, further information can be found e.g., in Refs.~\cite{McDonald88,Stockli06}.

\subparagraph{Longitudinal plane}

For the full six-dimensional characterization of the phase space diagnostics and instrumentation for the longitudinal plane is required. References~\cite{Webber90,Shishlo16} give an introduction to the concept and a survey of measurement techniques in this plane. The parameters of interest are particle energy resp. momentum and bunch length. Because of the small magnetic rigidity the determination of the particle momentum is usually done with a magnet spectrometer, where the dipole magnet of a permanently installed diagnostics beamline acts as spectrometer magnet. The spectrometer transforms momentum (spread) into position (spread) which is measured with a spatial resolving detector (screen, SEM $\ldots$) according to
\[
 \frac{\Delta x}{x_0} = \frac{\Delta p}{p_0} \; .
\]
In the case of non-relativistic energies, even time-of-flight measurements are sometimes applied, see Refs.~\cite{Forck17,Webber90} and the references therein.

For the determination of the bunch length a so-called bunch shape monitor (BSM) is applied in various hadron accelerators, see e.g., Refs.~\cite{Ostroumov98,Feschenko96,Feschenko97}. The principle of operation is based on the analysis of secondary electrons produced by the primary beam hitting a 0.1 mm diameter tungsten wire, to which a potential of typically \Unit{-10}{kV} is applied. The longitudinal charge distribution of the analysed beam is transformed into a spatial one of low-energy secondary electrons by synchronized transverse RF modulation. Readout can be performed electrically or optically via a CCD in combination with an MCP and a phosphor screen.

\subsection{Injector synchrotron}

\begin{figure}[!b]
	\begin{center}
		\includegraphics[scale=0.8,clip]{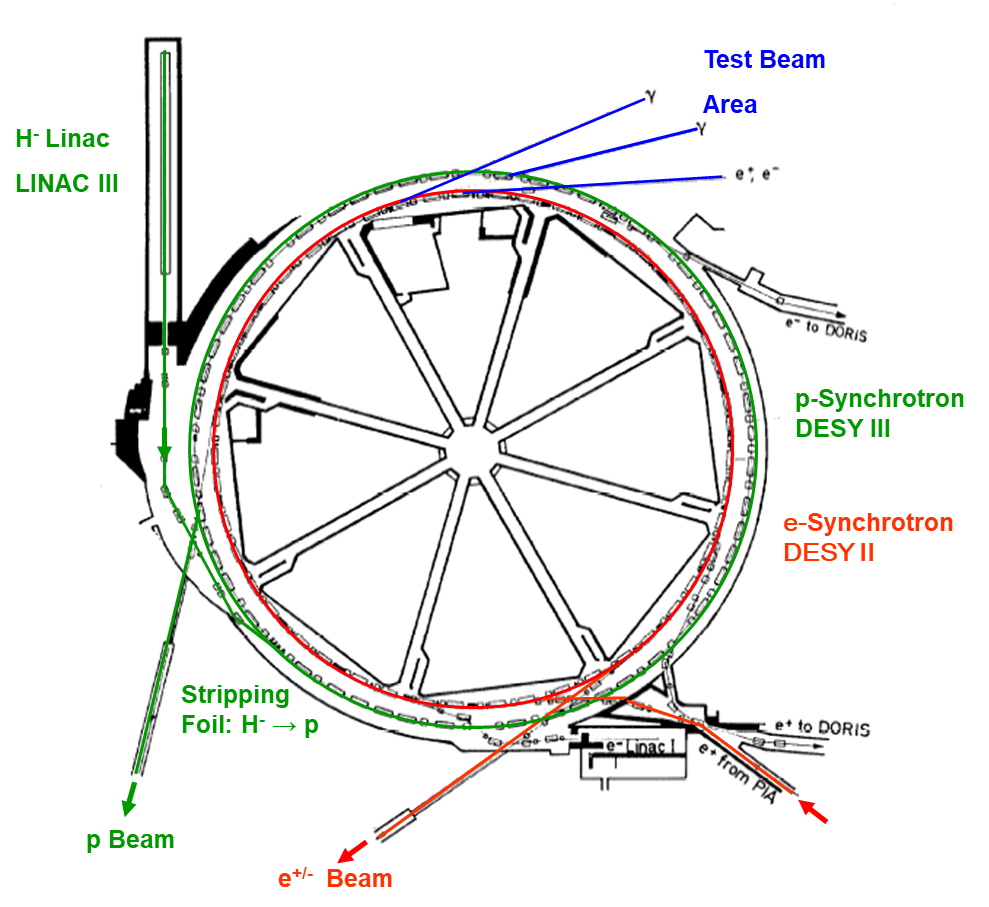}
		\caption{Overview of the proton injector synchrotron DESY III (mean radius 50.42 m). The electron injector synchrotron DESY II is situated in the same tunnel. Further information about both machines can be found in Refs.~\cite{Hemmie87a,Hemmie87b}.}
		\label{fig:S5DESY23}
	\end{center}
\end{figure}

After the description of the specific diagnostics needs of a hadron linac together with an overview of the typical instrumentation, this section briefly describes the peculiarities of the hadron injector synchrotron instrumentation. The discussion is based on the experience with the first proton synchrotron DESY III in the HERA injector chain, see \Fref{fig:S5DESY23}.

DESY III used a \PHm multi-turn injection with a stripper foil for conversion to protons at an injection \emph{energy} of \Unit{0.31}{GeV/c}. During acceleration up to the extraction \emph{energy} of  \Unit{7.5}{GeV/c}, the RF frequency increased from \Unit{3.27}{MHz} to \Unit{10.33}{MHz}. Further information can be found in Refs~.\cite{Hemmie87a,Hemmie87b}.

From the operational viewpoint the requirements are threefold. Firstly beam instrumentation is required for parameter control during the beam acceleration. Then, in the case of faulty operation, monitors are required for fault finding. Furthermore, signals for beam optimization must be provided in specific critical places in the machine, especially for injection and extraction.

Similar to the previous section, in the following, the main beam instrumentation used in the injector synchrotron is briefly described.

\subparagraph{Beam current}

Beam current measurements are required for optimization of the injection efficiency, for the measurement of single bunch charge and average current, and for determination of the coasting beam.

A monitor typically used for bunch current measurements is the beam transformer as in the case of linac instrumentation. For the determination of the average current, a parametric or DC current transformer (DCCT) is widely used. It relies on the extension of the transformer's bandwidth down to DC, typically realized in configuration of a zero flux magnetometer. Illustrative explanations about the monitor's working principle can be found in Refs. \cite{Schimckler06,Raich06} additionally to the references about beam current measurements listed before.

\subparagraph{Beam position}

For the measurement of beam parameters (closed orbit, position, oscillations, $\beta$-function$\ldots$) beam position monitors are required. In order to sample the closed orbit with sufficient accuracy, typically four monitors are used per betatron oscillation, i.e., they are located in a distance of about 90$^{\circ}$ phase advance. Due to the large bunch lengths and the low acceleration frequencies the beam spectrum contains only rather low frequencies, therefore a high pick-up sensitivity is required at these frequencies.

For DESY III inductive pick-ups were used. Other schemes are capacitive ones as for example shoe-box types. At higher acceleration frequencies and beam energies, even stripline BPMs are sometimes in use.

\subparagraph{Tune measurements}

\begin{figure}[!t]
\begin{center}
\includegraphics[scale=0.08,clip]{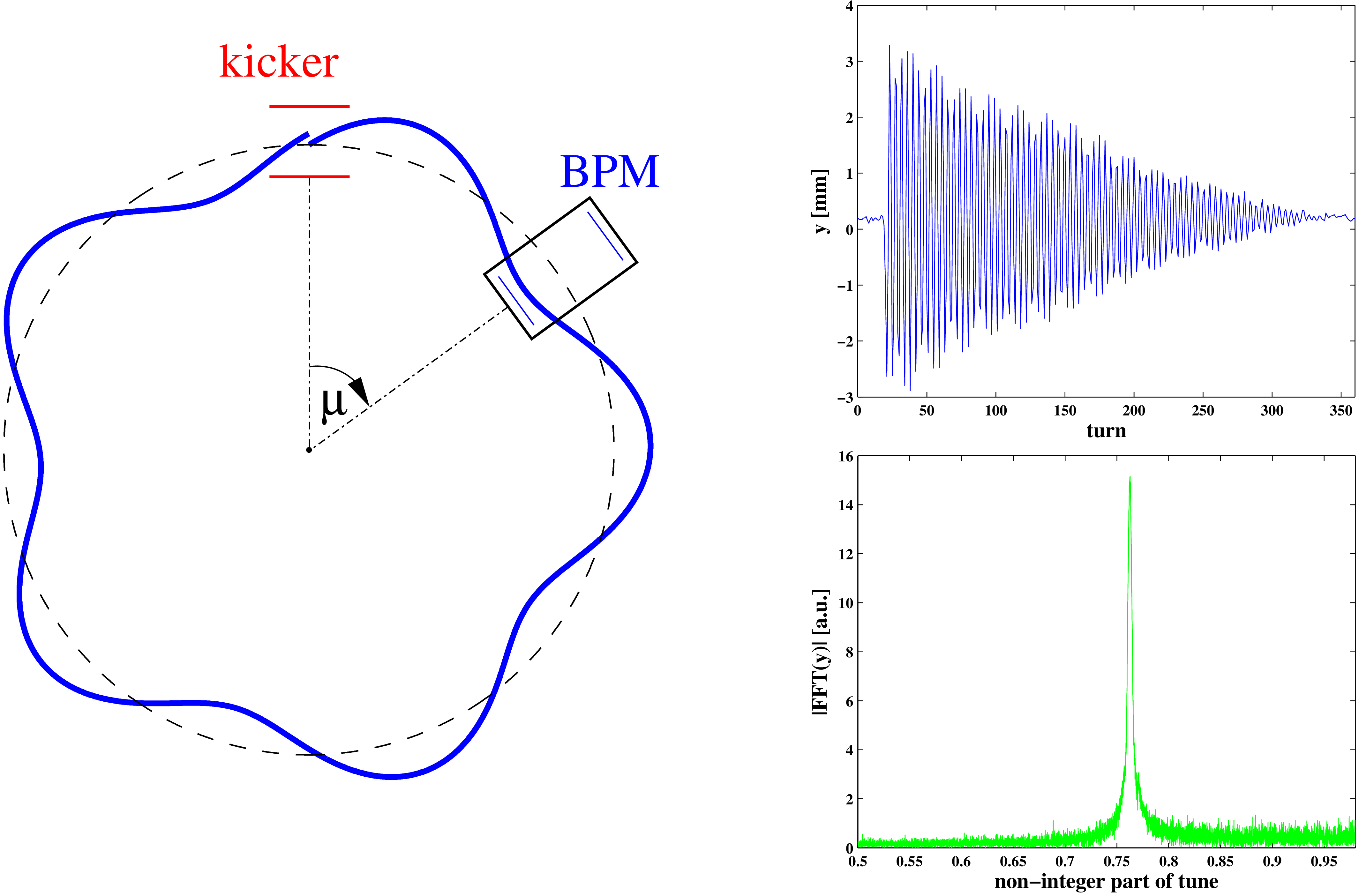}
\caption{Principle of a tune measurement. A kicker magnet excites a coherent betatron oscillation which is observed with a pick-up at fixed position. The BPM measures a sequence of turn-by-turn beam positions, a fast Fourier transform (FFT) of the position sequence gives the non-integer part of the tune. The turn-by-turn data were taken from a measurement at the positron storage ring DORIS at DESY. The fast oscillation damping is a result of the orbit feedback and radiation damping, for hadron machines the damping in the turn-by-turn displacement is much slower.}
\label{fig:S5syncTune}
\end{center}
\end{figure}
The tune is the eigenfrequency of the betatron oscillations in a circular machine. It is a characteristic frequency of the magnet lattice, produced by the strength of quadrupole magnets. Besides the report in these proceedings \cite{Jones18b}, Refs~.\cite{Serio91,Steinhagen08} give an introduction to the principles of tune diagnostics.

Figure~\ref{fig:S5syncTune} illustrates a simple scheme for a tune measurement. A coherent betatron oscillation is excited with a kicker, and the dipole moment due to the (coherent) transverse beam oscillation is observed with a pick-up. In order to have maximum sensitivity (betatron amplitude) the BPM is placed with a phase advance $\mu \approx$ 270$^{\circ}$ from the kicker. The primary observable in this case is a time sequence of turn-by-turn beam positions, from which the (fractional) tune can be deduced via a fast Fourier transform (FFT).

Due to the strongly suppressed radiation damping for hadron beams a permanent excitation with a kicker magnet may lead to an emittance blow-up. Therefore only very small excitations can be applied and the pick-up needs a very high sensitivity in order to detect small coherent beam oscillations. Furthermore, at injection energy a hadron synchrotron is often space-charge dominated so that the acceptance is fully occupied. In this case a beam excitation will lead immediately to particle losses and has to be avoided.

DESY III for example was space charge dominated at injection energy. For this machine it was decided to have no continuous tune measurement in standard operation. Tune measurements were performed only in dedicated machine studies in order to find a suitable working point in the tune diagram and a reproducible machine set-up.

\subparagraph{Transverse profile and emittance}

In a circular machine there exists a unique solution for the particle orbit and Twiss parameters. Therefore, the (absolute 1$\,\sigma$) emittance $\varepsilon$ can immediately be deduced with knowledge of transverse beam profile $\sigma$ and betatron function $\beta$ in one location of the machine according to
\begin{equation}\label{Eq:SyncEmitt}
    \varepsilon = \frac{\sigma^2}{\beta} \; .
\end{equation}
Equation~(\ref{Eq:SyncEmitt}) is simplified in the sense that the dispersion contribution to the beam size is neglected. Furthermore one should keep in mind that unfortunately several emittance definitions are in use, see e.g., Refs.~\cite{Bravin18b,Sander91}. However, in order to determine the emittance it is sufficient to perform a transverse profile measurement at one location. Therefore transverse emittance diagnostics can be reduced to the case of transverse profile diagnostics.

The simplest way to determine the transverse profile in an accelerator is to use a luminescent screen. In a circular machine however, due to the multiple passages through the screen, this method is completely destructive in the sense that the particle beam will be lost after several turns. Therefore this method is usually applied only during the machine commissioning or in case of fault finding, if there are doubts about signals from other monitors. A less destructive method for transverse profile measurements is to use wire scanners. Here a thin wire is quickly moved through the beam with a speed of about \Unit{1}{m/s}. A simultaneous detection of the intensity of the particle shower outside the vacuum chamber with a scintillator/photomultiplier assembly gives an image of the beam profile.

\begin{figure}[!t]
	\begin{tabular}{ll}
		(a)
		\includegraphics[scale=.8,clip]{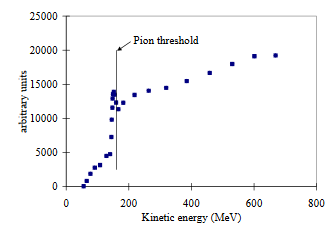}&
		(b)
		\includegraphics[scale=.75,clip]{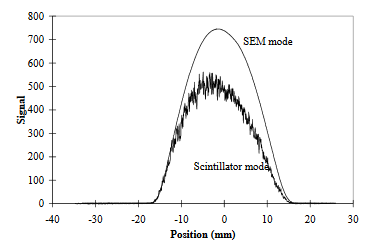}
	\end{tabular}
	\caption{(a) Secondary particle shower intensity as function of the primary beam energy \cite{Raich05}. (b) Transverse beam profile, recorded with a scintillator/photomultiplier assembly and via the secondary emission current from an electrically isolated wire \cite{Raich05}.}
	\label{fig:S5syncWS}
\end{figure}

However, in low-energy machines there is a limitation in the application of the wire scanner principle as described before due to the fact that the intensity of the secondary particle shower strongly depends on the primary beam energy \cite{Raich05}. As can be seen in \Fref{fig:S5syncWS}(a) at about 150\UMeV{} there is a steep increase in the shower intensity. This energy corresponds to the \Ppi threshold, i.e., the threshold to produce the lightest shower particle. Below this energy it is more suitable to measure the secondary electron emission current of an electrically isolated wire, see the signal comparison \Fref{fig:S5syncWS}(b).

Finally residual gas monitors are often used as transverse profile monitors in hadron machines. However, in a circular machine the vacuum pressure has to be much better (about \Unit{10^{-10}}{mbar}) than in a linac or a transfer line (\Unit{10^{-6}}{mbar} to \Unit{10^{-8}}{mbar}). Therefore the signal is much lower which can be compensated by a local pressure bump.

\subparagraph{Bunch length and time structure}

This type of instrumentation is required to investigate longitudinal beam parameters and bunch oscillations. A monitor widely used is the wall current monitor (WCM) as shown in \Fref{fig:S3outside}(b). Working principle and design of such a monitor are described in detail in \Bref{Webber90}. A WCM offers typically a bandwidth up to a few GHz. However, \Bref{DElia08} reports the design of a monitor with up to 20 GHz bandwidth.

\subparagraph{Loss detection}

Indications of beam losses are required in specific critical places, such as for example at injection/extraction for optimization purposes. Therefore, beam loss monitors are mandatory in a synchrotron. Detailed information about beam loss detection can be found in Refs.\cite{Wittenburg93,Shafer02,Wittenburg08} together with the report in these proceedings \cite{Wittenburg18}.

\subparagraph{Comment on \APproton and heavy ion machines}

In an antiproton or heavy ion machine the source emittance is worse compared to that in a proton accelerator, and the adiabatic shrinking of the emittance during acceleration is not sufficient to achieve the required final beam quality. In order to improve the emittance, electron cooling is often applied for bunched beams. However, one prerequisite for an efficient cooling process is a small cooling time which in the case of electron cooling is achieved at smaller beam energies. Therefore electron cooling is applied usually in a low energy synchrotron in the injector chain. More information about beam cooling and related topics can be found in Refs.\cite{Cooling93,Bosser95}.

\begin{figure}[!t]
	\begin{center}
		\includegraphics[scale=0.6,clip]{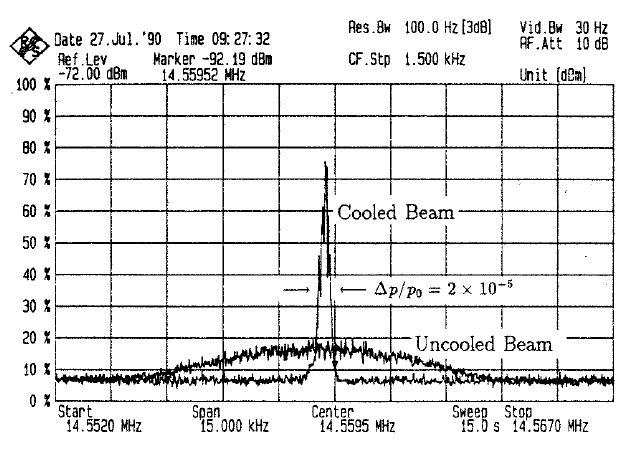}
		\caption{Longitudinal Schottky scan for an Ar$^{18+}$ ion beam at the GSI \cite{Forck17}. The broad curve is the frequency spectrum at injection with $\Delta p/p = 10^{-3}$, the narrow one after the application of electron cooling.}
		\label{fig:S5syncSchottky}
	\end{center}
\end{figure}
Schottky diagnostics is an important tool to control the cooling process. It relies on the exploitation of the individual particle behaviour (Schottky noise) in the beam spectrum of a bunched or un-bunched beam. The detection of these fluctuations with a very sensitive spectrum analyser allows for a non-destructive measurement of a variety of beam parameters like momentum distribution, tune, transverse emittance, and chromaticity. As an example, \Fref{fig:S5syncSchottky} from \Bref{Forck17} shows measurements of the momentum distribution where the momentum width of an ion beam has been reduced by two orders of magnitude via electron cooling.

Supplemental information to the Schottky diagnostics report in these proceedings \cite{Kowina18} can be found in Refs.~\cite{vdMeer88,Linnecar94,Caspers08}.

\subsection{Transfer line}

After the beam is ejected from the first synchrotron it has to be transported to the subsequent accelerator. A transfer line links both circular machines together while matching the optical beam parameters: due to the imposed periodicity in a circular machine the Twiss parameters are determined uniquely. This holds for the injection/extraction points of both accelerators, and the transfer line has to map the extraction Twiss parameters from the preceding synchrotron correctly onto the injection parameters of the subsequent one. Therefore the line has usually a regular cell structure (FODO) over the majority of the length with matching sections at either end. From the diagnostics point of view, instrumentation is required for
\begin{itemize}
  \item Adjustment of the beam transport:\\
  this includes (i) a control of the transfer efficiency via AC current transformers (at least at the entrance and the exit of the line), and (ii) the control of the beam position for orbit correction and steering via BPMs and/or luminescent screens.
  \item Beam quality determination:\\
  the beam quality parameter of interest in a transfer line is mainly the transverse beam emittance. It may be spoiled owing to a mismatch of the beam optics between the transfer line and the preceding accelerator. Similar to emittance measurements in a synchrotron it is measured by determining the transverse beam profiles, and monitors typically in use are luminescent screens.
  \item Machine protection:\\
  beam loss monitors are required to control the beam losses and keep them as low as possible in standard operation. The loss monitor system can be connected to the machine interlock in order to prevent injection resp. extraction under faulty or irregular conditions.
\end{itemize}
General transfer line instrumentation is described in \Bref{Raich98}, a specific example is given in Ref.\cite{Jensen05}. In the following, beam steering and emittance diagnostics will shortly be discussed with respect to their impact on the layout of a transfer line. More details and the specific features of a beam transfer line together with diagnostics requirements are described e.g., in \Bref{Bryant94}.

\subsubsection{Beam steering}

\begin{figure}[!t]
\begin{center}
\includegraphics[scale=0.33]{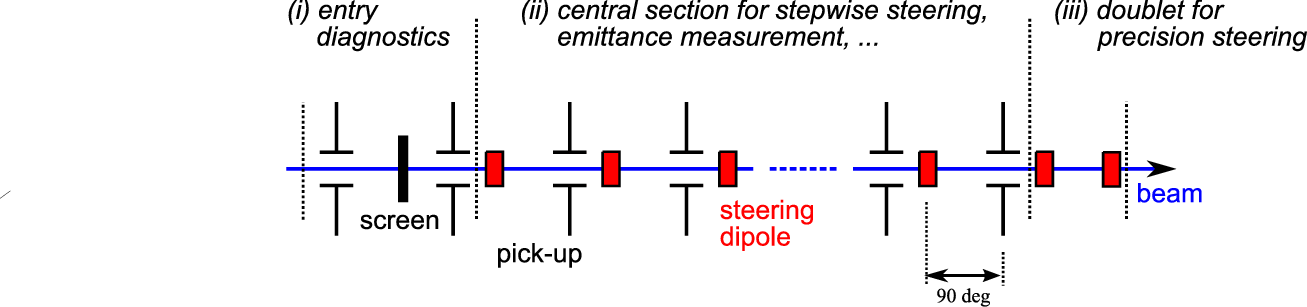}
\caption{Basic layout of transfer line diagnostics and correction elements for steering}
\label{fig:S5TLine}
\end{center}
\end{figure}

The usual philosophy used for beam steering in a transfer line is illustrated in \Fref{fig:S5TLine} and will be explained according to \Bref{Bryant94}.
\begin{description}
  \item[(i)] At the entry of the transfer line it is useful to have information about angle and position of the extracted beam together with qualitative information about the beam shape in order to have a first insight into optics mismatch. Angle/position information are gathered with a pair of pick-ups. From a practical point of view the precision and reliability of such measurements are greatly improved by having only a drift space between them. For the beam shape determination a transverse profile monitor (e.g., luminescent screen) is used.
  \item[(ii)] In the central section of the transfer line, each steering magnet is paired with a pick-up so that the trajectory can be corrected stepwise along the line. The phase advance between steerer and pick-up should be about 90$^{\circ}$ so that it is possible to reconstruct the betatron oscillation. However, in practice it is common to have fewer pick-ups, especially if there are long straight sections.\\
  The beam emittance measurement is usually performed in the central part of the transfer line in a dispersion-free region, see also next section.
  \item[(iii)] At the line exit, the last two steerer magnets are used as doublet to adjust beam angle and position to the values required for the uniquely determined closed orbit solution of the subsequent synchrotron. For maximum sensitivity, the steering magnets should be approximately a quarter betatron oscillation length apart. With one quadrupole for each transverse plane the optics mismatch for the injection in the subsequent accelerator section can be corrected.
\end{description}
Furthermore, horizontal and vertical planes should be independent for correction elements. Care has to be taken in positioning the elements in view of best sensitivity for beam control and observation. The most sensitive points are the maxima of the beta function which are situated at the positions of the quadrupoles. Therefore, both monitor and magnet should in general be located close to the quadrupoles.

\subsubsection{Emittance measurement}\label{Sec:TLineEmitt}

Emittance diagnostics in circular machines and transfer lines are based on transverse profile measurements, cf. Eq.~(\ref{Eq:SyncEmitt}). However, in a circular machine the betatron function $\beta$ is unambiguous and the emittance can be determined by a single profile measurement. In a transfer line the Twiss parameters ($\alpha, \beta, \gamma$) are not known a priori and have to be determined together with the emittance $\varepsilon$. With the constraint $\gamma = \frac{1 + \alpha^2}{\beta}$ there are three independent parameters, therefore an unambiguous determination can be achieved with at least three profile measurements.

There exist two common schemes for emittance diagnostics, (i) either the beam profile is varied by changing the focussing strength of a quadrupole upstream of a profile monitor, and the size is measured for each quadrupole setting, or (ii) the beam size is measured with different profile monitors for a fixed setting of the beam optics. The situation becomes even more complicated if the dispersion contribution to the beam size in Eq.~(\ref{Eq:SyncEmitt}) is taken into account. In order to avoid this additional complication, profile measurements should therefore be located in dispersion-free sections. More information about emittance diagnostics can be found e.g., in Ref.\cite{Bravin18}.

\subsubsection{Final remarks on instrumentation for the injector chain}

So far, diagnostic needs for accelerator subsections in a hadron injector chain have been discussed. Different monitors and diagnostic concepts were presented with examples mainly from the DESY proton injector complex.

\begin{figure}[!h]
	\begin{center}
		\includegraphics[scale=0.65,clip]{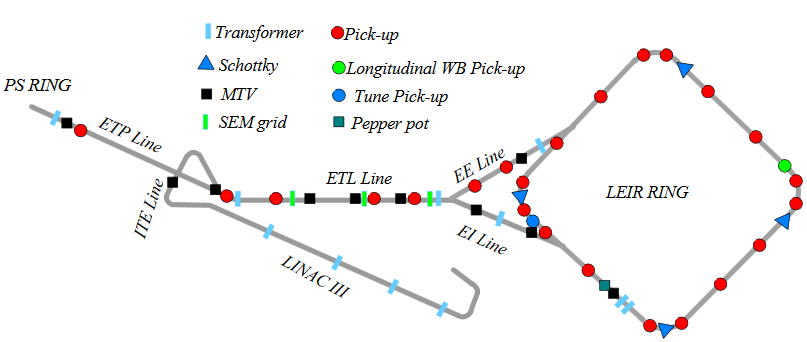}
		\caption{Layout of the LEIR complex together with the beam instrumentation \cite{Bal05b,Bal06}}
		\label{fig:S5syncLeir}
	\end{center}
\end{figure}
In addition, \Fref{fig:S5syncLeir} shows an example for instrumentation in a hadron injector chain (including linac, transfer line, and circular accelerator): the Low Energy Ion Ring (LEIR) at CERN is used to transform a series of long low intensity pulses from Linac 3 into short high density pulses, which are further accelerated in the PS and SPS ring before injected to LHC. The injected ion pulses will be stacked and phase space cooled via electron cooling before they are accelerated to the ejection energy of \Unit{72}{MeV/u}. As can be seen the diagnostics needs and the instrumentation for the LEIR complex are similar to those described above. More information about the monitors in use at LEIR can be found in Refs.~\cite{Bal05b,Bal06}. Moreover, \Bref{Strehl99} gives an overview of diagnostics for hadron machines with special emphasis on heavy ion machines.

After the beam is ejected from the first synchrotron it is transported via a transfer line to the next accelerator. This step may be repeated several times, see \Fref{fig:S5desy} for the HERA accelerator complex at DESY. From the diagnostics point of view the requirements for beam monitors in the subsequent injector sections are the same, and the requirements for the final storage ring/collider will be considered in the following.

\subsection{Storage ring}

As above the description of hadron storage ring diagnostics will be based mainly on examples from the HERA proton ring at DESY. However, diagnostic overviews for other machines can be found e.g., in Refs.~\cite{Jones07,Jones09,Trad17} for the LHC at CERN, in Refs.~\cite{Schwickert09,Reiter18} for the FAIR project at GSI, in Refs.~\cite{Shiltsev06,Moore09} for the Tevatron at Fermilab, and in Refs.~\cite{Bai03,Minty11} for the RHIC at Brookhaven National Laboratory. Reference \cite{Jones18} shows diagnostics examples from high energy colliders in general. For all these high energy hadron machines it is common to use superconducting magnets to achieve the required magnetic dipole fields.

The following list summarizes the demands for parameters and monitor systems which are needed to operate the storage ring/collider and which will be discussed in the subsequent sections:
\begin{itemize}
  \item intensity (bunch and mean current);
  \item beam orbit;
  \item tune, chromaticity, coupling;
  \item beam distribution, emittance (longitudinal and transverse planes);
  \item luminosity;
  \item beam energy;
  \item machine protection to avoid quenches of superconducting magnets.
\end{itemize}
Besides the requirement of a very sensitive magnet quench protection system (as the painful LHC accident from September 2008 made aware, see e.g., \Bref{Myers12}), the use of superconducting magnets has additional impact on the collider diagnostics as will be discussed in the following section.

\subsubsection{Remarks on superconductivity}

In order to obtain superconductivity the magnets have to be cooled down to liquid helium temperatures. At HERA the operational temperature was 4.4 K, the LHC is operated at 1.8 K. The effects on beam diagnostics to operate the machine in a cold environment are listed below:
\begin{itemize}
  \item Together with the magnets, a part of the beam diagnostics instrumentation has to be operated in the cold environment.
  \item For the design of cold instrumentation, care has to be taken in order to minimize heat transfer from the monitor to the cold environment (e.g., by higher order mode heating).
  \item No interceptive diagnostics can be used in or close to the cold sections: a particle shower may lead immediately to a magnet quench.
  \item Beam intercepting monitors must be protected against possible misuse, i.e., they have to be integrated in an interlock system.
\end{itemize}
In order to minimize the difficulties which arise because of the cold environment, the common strategy is to concentrate most of the beam instrumentation in warm sections, namely the straight insertions without need for particle bending (dipole magnets). Apart from the beam position and loss monitors which have to be installed all around the ring, the remaining monitors are located in warm sections.

\subsubsection{Monitor systems}

\subparagraph{Intensity}
Intensity related parameters are bunch charge resp. fill pattern and mean current. The bunch charge is monitored with an AC current transformer, the mean current with a DC or parametric current transformer.

Furthermore, from the current determination additional parameters can be deduced. DC current measurements performed in short time intervals result in a beam lifetime $\tau$ according to
\[
    \frac{1}{\tau(t)} = -\frac{1}{N}\,\frac{{\rm d}N}{{\rm d}t} \; .
\]
With knowledge of DC and bunch current, the coasting beam contribution $I_{cb}$ (i.e., the number of beam particles leaking out of the RF buckets occupied by bunches) can be determined by subtracting the sum of all bunch currents from the DC current:
\[
     I_{cb} = I_{dc} - \sum_{i}^{bunches} I_{ac,i} \; .
\]
For a reliable coasting beam indication careful monitor calibrations are mandatory.

\subparagraph{Beam orbit}

Position information is required for the determination of beam orbit related parameters. This includes closed orbit measurements from which lattice parameters are deduced and offers the possibility to compare the real machine with its design values, see e.g.,  Refs.~\cite{Wenninger08a,Wenninger18}. Additionally single turn information is required, especially for optimization of the injection. By minimizing injection orbit oscillations it is possible to eliminate e.g., mismatches of other parameters and related emittance blow-up.

The beam position is measured with BPMs located in cold and warm environments. In order to reach maximum sensitivity for a measurement of the orbit deviation they are usually installed close to the quadrupoles where the $\beta$-function has its maximum.

The choice of pick-up type depends on linearity, dynamic range, and required resolution. While HERA was equipped with stripline monitors \cite{Schutte87}, in the LHC cold button-type pick-ups are installed in the arcs and directional stripline couplers
in  all  interaction  regions because they are capable of distinguishing between counter propagating beams in the same beam pipe \cite{Jones07}.

\subparagraph{Tune, chromaticity, and coupling}

For a circular accelerator the tune defines the working point of the accelerator. As for the injector synchrotron, the basic principle of a fractional tune measurement relies on transverse beam excitations in combination with an FFT from a turn-by-turn position determined with a pick-up \cite{Jones18b,Steinhagen08}. In order to minimize the emittance blow-up, only small excitations are allowed. Therefore, a high sensitivity of the pick-up detectors together with minimum disturbing excitation schemes are required. Examples for widely used schemes are
\begin{itemize}
	\item Tune kicker\\
	This is based on the traditional kick method. The principle is simple and robust and used typically for the commissioning stage.
	\item Tune shaker\\
	The tune shaker relies on a continuous beam excitation. Therefore it can be used for continuous tune monitoring, and the signal can be integrated in a feedback loop for tune and chromaticity \cite{Steinhagen07}. There exist a variety of excitation schemes like (i) single frequency excitation, (ii) single frequency locked on the tune (PLL mode), (iii) single or dual frequency with adiabatic rise and fall-off (AC dipole mode), (iv) band-limited excitation, or (v) repetitive chirp excitation \cite{LHC_Tune}.
\end{itemize}
At HERA the signals for tune measurements were produced with a repetitive chirp excitation and measured via a resonant 'Schottky type' pick-up \cite{Herb91}. At LHC a \emph{direct diode detection} (3D) technique was developed to reach a sensitivity which allows us to observe beam betatron oscillations with amplitudes far below a micrometre. It is based on an increase of the betatron frequency content in the base band which relaxes the demands on the pick-up sensitivity \cite{Gasior12}. The so-called \emph{base band tune} (BBQ) system allows passive tune monitoring, tracking, and feedback, proving to be sufficiently reliable for controlling tune variations during almost every LHC ramp and squeeze \cite{Steinhagen11}. Furthermore, \Bref{Tan05} compares the use of different pick-up types and measurement schemes at the Tevatron for protons and antiprotons.

Apart from the methods described above, Schottky diagnostics can be used as passive method (i.e., without external excitation) to measure the (incoherent) tune and chromaticity.

\begin{figure}[!t]
	\begin{center}
		\includegraphics[scale=1.]{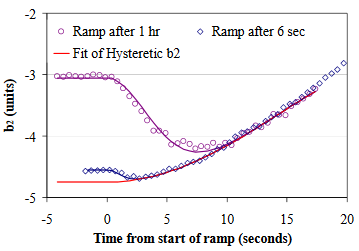}
		\caption{Chromaticity measurement as a function of time on the energy ramp for two different waiting times \cite{Martens05}}
		\label{fig:S5snapback}
	\end{center}
\end{figure}
However, the use of superconducting magnets has a strong impact on tune and chromaticity in the storage ring because of dynamic effects: superconducting eddy currents or persistent currents affect the multipole components of the dipole magnets especially at injection energy, and with it also the storage ring performance \cite{Mess96,Holzer96,Aquilina12}. Furthermore, these persistent currents are not really persistent and decay with time. The most important contribution comes from the sextupole component which may generate huge chromaticity errors. As an example, \Fref{fig:S5snapback} shows a chromaticity measurement from Tevatron (FNAL, USA) on the energy ramp \cite{Martens05}. The persistent currents are reinduced to their full strength on the first steps of the ramp, approaching the original magnet hysteresis curve. This \emph{snap-back} effect together with the persistent current decay needs correction in order to operate the accelerator under controlled conditions, i.e., a reliable control during the ramp is mandatory. Besides online measurements of magnetic multipole components and correction tables for the magnetic fields, feedbacks on tune and chromaticity were proposed in Refs.~\cite{Steinhagen08,Steinhagen07}.

\subparagraph{Transverse beam distribution and emittance}

Due to the unique solution for the beta function in the storage ring the emittance diagnostics can be reduced to a single transverse profile measurement. Depending on the extent of beam perturbations there exist three classes of transverse beam monitors.

\emph{Single pass monitors} are simple and robust. They are used typically during the commissioning stage where beam operation is performed only with single or few bunches. Therefore single pass monitors need a high sensitivity, but only modest demands on accuracy. A widely used monitor for this purpose is the luminescent screen.

\emph{Few pass monitors} are used typically for the study of injection mismatch, i.e., for Twiss parameter and dispersion matching via observation of shape oscillations during the first turns. For this purpose turn-by-turn acquisition is required (typically 10--20 turns), and the demands on the accuracy are still moderate. However, only moderate beam blow-up is allowed and energy deposition in the screen material becomes a critical issue. Therefore optical transition radiation monitors are widely used because they require only thin foils for light generation. A description of such monitors can be found e.g., in Ref.\cite{Afanasyev06}. A further option for turn-by-turn instrumentation are residual gas monitors, see e.g., Refs.\cite{Forck03,Jansson06}.

\emph{Circulating beam monitors} are required to study the evolution of the rms beam size, for emittance diagnostics, and to determine the beam tilt due to coupling between both transverse planes. These types of monitors can produce only a minimum beam blow-up, i.e., they have to rely on minimum resp. non-intercepting methods. Furthermore the demands on accuracy are high. Monitor principles in use are
\begin{itemize}
  \item Residual gas (luminescence) monitors.
  \item Flying wires\\
  Because of the partial beam intercepting signal generation, this type of monitor is sometimes used only for calibration of other transverse profile monitors. A limitation is high beam intensity combined with small beam sizes which can destroy the wire due to high heat load. Therefore wire speeds of \Unit{1}{m/s} and even more are required.
  \item Synchrotron radiation monitors\\
  In high energy proton machines the intensity of synchrotron radiation in the visible spectral region produced in the fringe field of a dipole magnet or from an undulator is sufficient to find applications in beam diagnostics. The first profile monitor based on this principle was realized at the SPS (CERN) \cite{Bossart79}, while later the Tevatron \cite{Thurman06} and HERA \cite{Kube06} used monitors of this kind. For the LHC a Mitsuhashi-type interferometric measurement scheme is tested \cite{Trad16} which exploits the spatial coherence of the radiations source \cite{Mitsuhashi98}.
\end{itemize}

\subparagraph{Longitudinal beam distribution and time structure}

Longitudinal diagnostics serve mainly for the determination of the classical {\it longitudinal profile parameters}, i.e., bunch centre of gravity, rms bunch length, and core distribution. At HERA bunch lengths (about \Unit{1.6}{ns} 1$\sigma$ at 920\UGeV{}) were monitored with a wide band wall current monitor. For the LHC with shorter bunch lengths (down to about \Unit{0.26}{ns} 1$\sigma$) a similar system based on a wall current monitor is used \cite{Lefevre10}.

In addition \emph{abort gap monitoring} is essential for superconducting storage rings, i.e., a continuous monitoring that the rise time gap of the dump extraction kicker will be free of particles. Particles located in this gap would not receive a proper kick when the dump system is fired which could lead immediately to the damage of machine components. Furthermore the \emph{detection of ghost bunches} is of interest. They occur due to diffusion of untrapped particles into nominally empty RF buckets. These ghost bunches may disturb the BPM system read-out or physics data taking. For these purposes the LHC is equipped with a longitudinal density monitor which is a single-photon counting system using visible synchrotron light \cite{Jeff12}.
\begin{figure}[!h]
\begin{center}
\includegraphics[scale=.3]{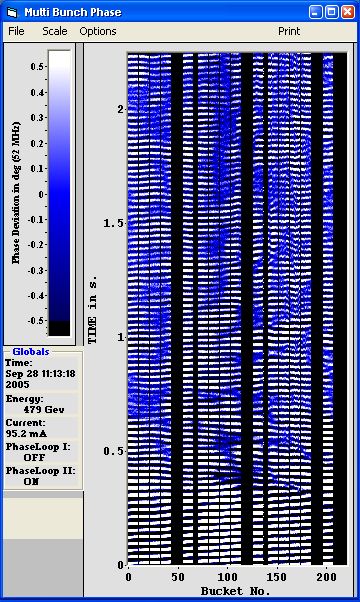}
\caption{Multi-bunch phase oscillation display in the HERA control room. The abscissa indicates the bucket number, the ordinate the time axis, and the colour code is a representation of the deviation from the reference phase. The appearance of a pattern indicates the presence of a longitudinal instability.}
\label{fig:S5HERAlong}
\end{center}
\end{figure}

Finally, the {\it observation of longitudinal instabilities} is of interest to optimize the machine performance. At HERA coherent oscillations of the proton beam during acceleration lead to an increase of the bunch length and a decrease of the luminosity. In order to investigate this effect a diagnostics system was developed to measure length and phase of every bunch together with the transient and accelerating RF voltages of all cavities \cite{Vogel00,Soloviev03}. Figure\ref{fig:S5HERAlong} shows a screen shot of the longitudinal diagnostics as seen in the control room, indicating the onset of a longitudinal instability. Based on this diagnostic a broadband longitudinal coupled bunch feedback system was developed to counteract the instability and preserve bunch length and luminosity \cite{Hoffmann06}.

\subparagraph{Luminosity}

Luminosity is the key parameter to determine the collider performance. The online luminosity diagnostic is important e.g., for the optimization of beam collisions at the interaction point. The principle of a luminosity determination is based on measuring a count rate: a reaction channel with known cross-section $\sigma_r$ is chosen and the corresponding event rate $\dot{N}_r$ is detected. According to Eq.~(\ref{eq:S5Lumi}) the luminosity is simply derived. Special care must be taken to suppress  background contributions to the measured event rate which falsify the luminosity value.

However, hadronic cross-sections are not precisely calculable because of the constituent particle nature. For the accelerator operation, reaction rates in a hadron collider do not serve for absolute luminosity monitoring, but are mainly used for optimization purposes. The absolute luminosity determination which is typically a complicated task is often the duty of the experiments. For this purpose it is important to measure DC and bunch currents with high precision.

\begin{figure}[!h]
	\begin{center}
		\includegraphics[scale=0.38,clip]{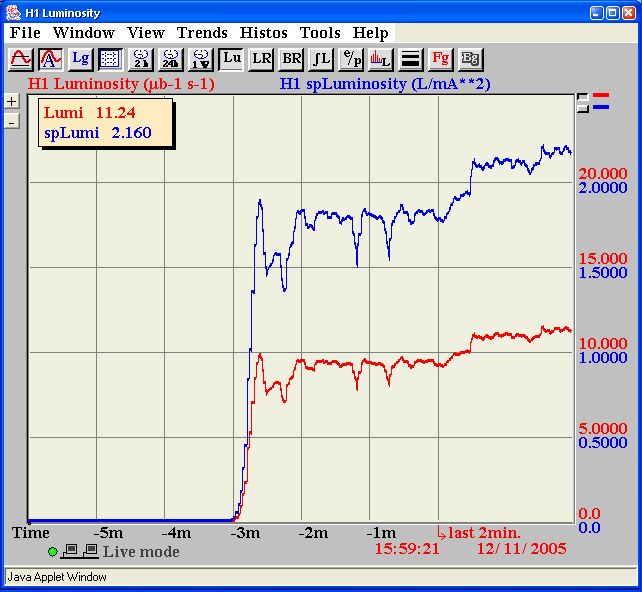}
		\caption{Online luminosity display for the H1 experiment at HERA, just after adjustment of beam collisions for a new run. The luminosity count rate is plotted as a function of time. The lower curve indicates the absolute luminosity, the upper one the specific luminosity (i.e., normalized to the beam current).}
		\label{fig:S5HERAlumi}
	\end{center}
\end{figure}
For the $\Pe\Pp$ collider HERA the determination of an absolute online luminosity was simpler because the reaction channel under investigation was the Bethe--Heitler process resp. Bremsstrahlung
\[
     \Pe \, \Pp \quad \longrightarrow \quad \Pgamma \, \Pe' \, \Pp'
\]
whose cross-section is well known. Figure \ref{fig:S5HERAlumi} shows the HERA luminosity display at the H1 experiment. Based on this information, the position and angle of both colliding beams were tuned for maximum collision rate at this interaction point.

More information about luminosity measurements can be found in Refs.~\cite{Johnson88,Potter94,Wenninger08}.

\subparagraph{Energy}

In a hadron collider, the absolute energy determination is of less importance. The hadrons have a constituent nature (quarks and gluons), and these constituents share the beam momentum. Therefore the total energy in a reaction is only loosely related to the beam energies. For the beam energy measurement the determination of the dipole magnet current is therefore sufficient. References~\cite{Johnson88,Muller08} contain additional information about energy diagnostics.

\subparagraph{Machine protection system and loss monitors}

\begin{figure}[!h]
	\begin{center}
		\includegraphics[scale=0.5,clip]{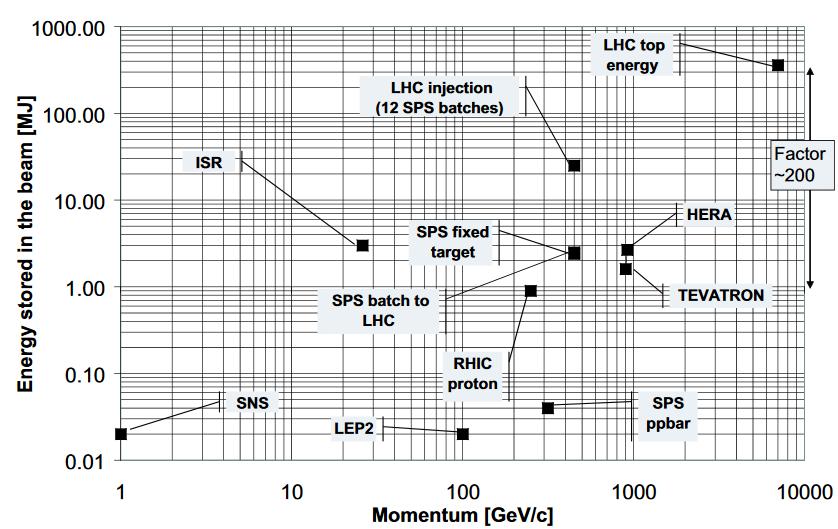}
		\caption{Comparison of the stored beam energy for various accelerators \cite{Schmidt08}}
		\label{fig:S5LHCloss}
	\end{center}
\end{figure}

A machine or quench protection system with integrated beam loss monitors is essential for the operation of superconducting accelerators. To emphasize this requirement \Fref{fig:S5LHCloss} shows the stored energy in the beam for different accelerators.

For HERA at 820 GeV beam energy, the He bath temperature was 4.4 K while the cable quench temperature was 5.2 K, i.e., a temperature rise of only 0.8 K in a magnet cable was sufficient to quench one of the superconducting magnets. For the LHC with about a factor of 200 more stored beam energy the situation is even more critical.

Therefore a very sensitive and reliable machine protection system is required which dumps the beam under controlled conditions to protect the equipment in case of component failure or non-tolerable background conditions, see also \Bref{Schmidt08}. One of the consequences of the LHC accident from September 2008 for example \cite{Myers12}, when during tests without beam several magnets and interconnects were damaged due to a resistive zone developed in the dipole busbar magnet interconnects, was a major upgrade of the machine (magnet) protection system \cite{Formenti10} which had proved too insensitive to protect the interconnect splices.

A very important part of the machine protection system is the beam loss monitors which have to detect irregular (uncontrolled, fast) losses. There are several considerations in selecting the appropriate type of beam loss monitor, see e.g., Ref.~\cite{Shafer02}. Monitors typically in use are gas ionization chambers, PIN diodes, photomultipliers with scintillators, and secondary emission tubes~\cite{Wittenburg08} In order to increase the sensitivity and discriminate against wanted signals, the loss monitor system can be optimized by locating the monitors as close as possible to the superconducting coils, i.e., inside the cold mass in a superfluid helium environment. For the LHC for example, a cryogenic loss monitor system was developed based on diamond detectors~\cite{Bartosik14}.

There is a special aspect for heavy ion synchrotrons operating with intermediate charge state beams. For these machines intensity dependent beam losses have been observed which are driven by a change of the charge state of beam ions owing to collisions with residual gas atoms. The resulting $m/q$ deviation (with $q$ and $m$ the ion charge and mass) from the reference beam ion leads to modified trajectories in dispersive elements, which finally results in beam loss. Beam ions hitting the beam pipe will release gas molecules which causes a local vacuum pressure increase. This pressure rise in turn will enhance the charge change and particle loss process,  finally causing significant beam loss within a very short time \cite{Omet08}. Especially for superconducting accelerators this loss mechanism may have severe impact on the accelerator performance.

In order to suppress and control the gas desorption process, dedicated scraper-based ion catcher systems are used. For the heavy ion synchrotron SIS100 (FAIR, GSI), a special cryogenic ion catcher system was developed \cite{Bozyk12} which in combination with a charge separator lattice \cite{Spiller04} shall provide 100\% efficiency to dump the ionized beam particles in a controlled manner.

\section{Additional examples for hadron accelerator instrumentation}

So far, the diagnostic needs for a hadron collider together with its injector chain were discussed. This section presents two additional examples for hadron accelerators and their requirements for beam instrumentation: (i) a spallation neutron source and (ii) a hadron therapy accelerator. However, the discussion will be less extensive because most of the instruments in use are based on the same monitor concepts described above. Therefore the description will be focussed on the peculiarities of these accelerator concepts and the consequences for beam diagnostics and instrumentation.

\subsection{Spallation neutron source}

In a spallation source neutrons are produced by the interaction of a high-energy proton beam with a heavy metal target. The neutrons are subsequently moderated to energies suitable for neutron scattering experiments (neV to eV). The number of neutrons produced by the protons depends on the primary beam energy in the range of 0.2--10\UGeV{}. At a beam energy of about 1\UGeV{} the neutron yield amounts to $\sim$30 neutrons per proton.

Following \Bref{Henderson13} there are four main configurations of spallation neutron sources: (a) short pulse sources where a long beam pulse (duration 0.1 – \Unit{1.0}{ms}) from a low energy ($\sim$100\UMeV{}) linear accelerator is accumulated in a rapid-cycling synchrotron (RCS) and accelerated to high energy, (b) short pulse sources where a long beam pulse from a full energy linear accelerator is accumulated in an accumulator or compressor ring, (c) long pulse sources
where a beam pulse of one or a few milliseconds delivered from a linear accelerator can be used without subsequent accumulation, and (d) continuous beam sources based on a cyclotron. Examples of machines of category (a) are JSNS (JPARC, Japan) \cite{JPARC} and CSNS (IHEP, China) which is recently under commissioning \cite{CSNS,CSNSb}, of type (b) the Spallation Neutron Source (SNS) at Oak Ridge (Tennessee, USA) \cite{Henderson07}, of type (c) the European Spallation Source ESS (Lund, Sweden) which is presently under construction and which will be the most powerful machine with 5 MW average beam power on the target \cite{ESS}, and for type (d) the SINQ at PSI, Switzerland \cite{SINQ}.

\begin{figure}[!b]
	\begin{center}
		\includegraphics[scale=.8,clip]{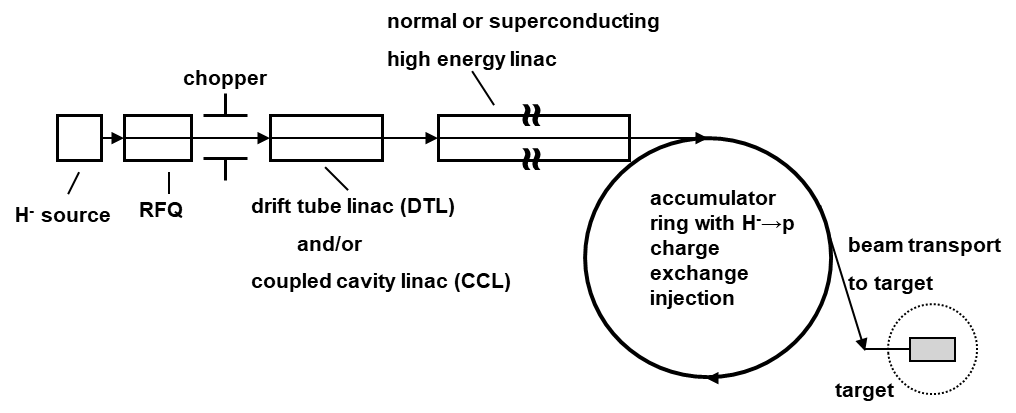}
		\caption{Schematic view of a short pulse neutron source}
		\label{fig:S6SNS}
	\end{center}
\end{figure}
Figure \ref{fig:S6SNS} shows a schematic view of a short pulsed neutron source of type (b) according to Ref.\cite{Chao02}. The principle set-up resembles that of a standard hadron linac and injector synchrotron, and with it also the concepts for beam instrumentation. However, the peculiarities of this accelerator type and the implications for beam instrumentation are especially the handling of high beam power. The following list gives an overview of the principle aspects:
\begin{enumerate}
  \item Achieving high beam power:\\
  diagnostics systems are required which help to understand the dynamics of intense beams. Examples are monitors to understand beam halo formation.
  \item Measuring high power beams:\\
  beam monitors have to measure the fundamental beam parameters during full power operation. Challenging in this context are especially transverse beam profile monitors because only non-interceptive principles can be used. Examples are ionization profile monitors or the laser profile monitor based on H$^-$ photo neutralization \cite{Blokland04}.
  \item Protecting the diagnostics:\\
  beam monitors that cannot survive the interaction with high power beams have to be protected, e.g., the machine protection system needs interfaces for intercepting devices.
  \item Protecting the facility:\\
  diagnostics systems are required that protect the facility from beam induced damage or activation. Examples are loss monitors or beam-on-target diagnostics.
\end{enumerate}
\begin{figure}[!h]
	\begin{center}
		\includegraphics[scale=.6,clip]{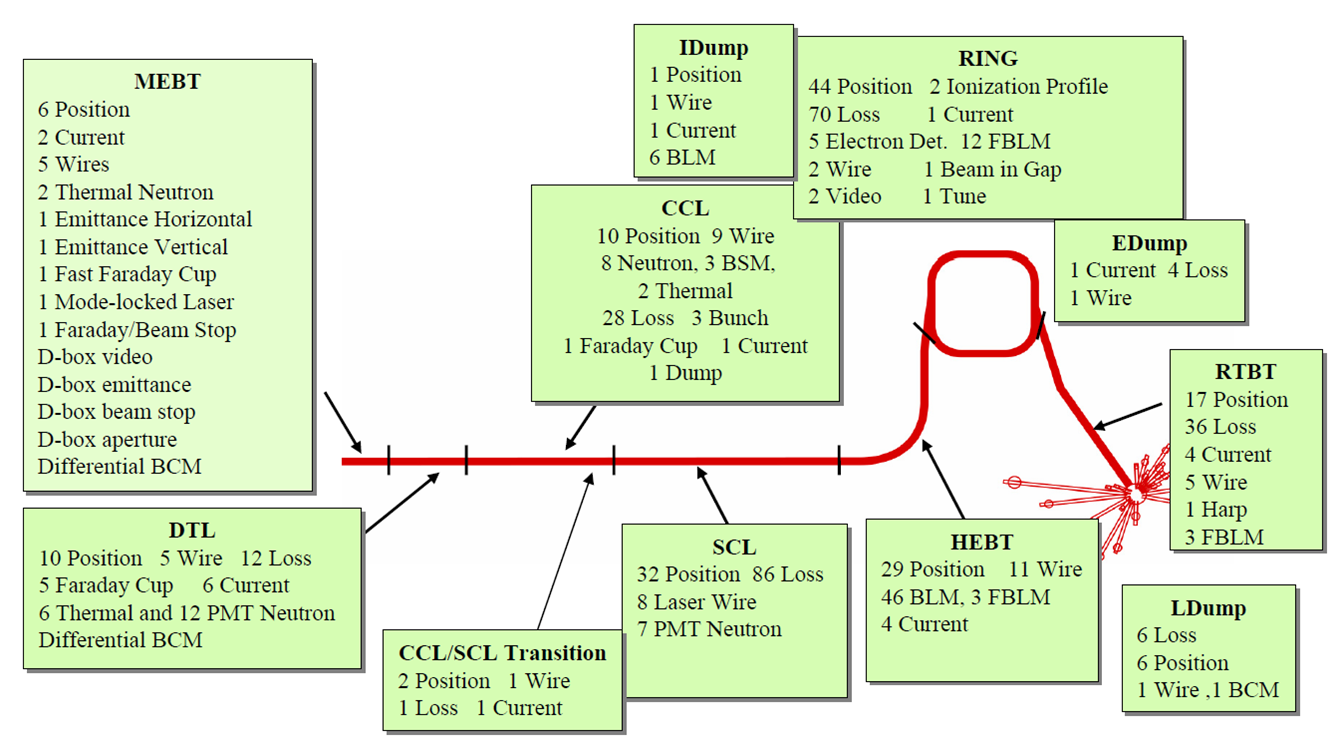}
		\caption{Layout of the SNS diagnostics systems \cite{Cameron03}}
		\label{fig:S6SNSdiagnose}
	\end{center}
\end{figure}
Figure \ref{fig:S6SNSdiagnose} shows the layout of the SNS diagnostic systems. Further information about the monitors can be found in Refs.\cite{Shea00,Shea01,Cameron03}.

\subsection{Hadron therapy accelerator}

Nowadays different types of hadron and electron accelerators are used for medical application \cite{Eickhoff08}. According to \Bref{Uesaka16}, conventional hadron accelerator schemes in use are (a) linacs for cancer therapy, (b) synchrotrons for proton and carbon cancer therapy, and (c) cyclotrons for proton cancer and boron neutron capture therapy and for rare isotope production for positron emission tomography. In the following synchrotron based facilities will be considered for the medical treatment of tumour patients using proton or light-ion beams. Examples are the National Centre for Oncological Hadrontherapy (CNAO) in Pavia (Italy) \cite{CNAO,CNAOb}, the MedAustron in Wiener Neustadt (Austria) \cite{Medaustron,Medaustronb}, and the Heidelberg Ion Therapy centre (HIT) in Heidelberg (Germany) \cite{HIT,HITb} with its branch in Marburg (Germany), the Marburg Ion Therapy centre (MIT) \cite{MIT}.

\begin{figure}[!h]
	\begin{center}
		\includegraphics[scale=.62,clip]{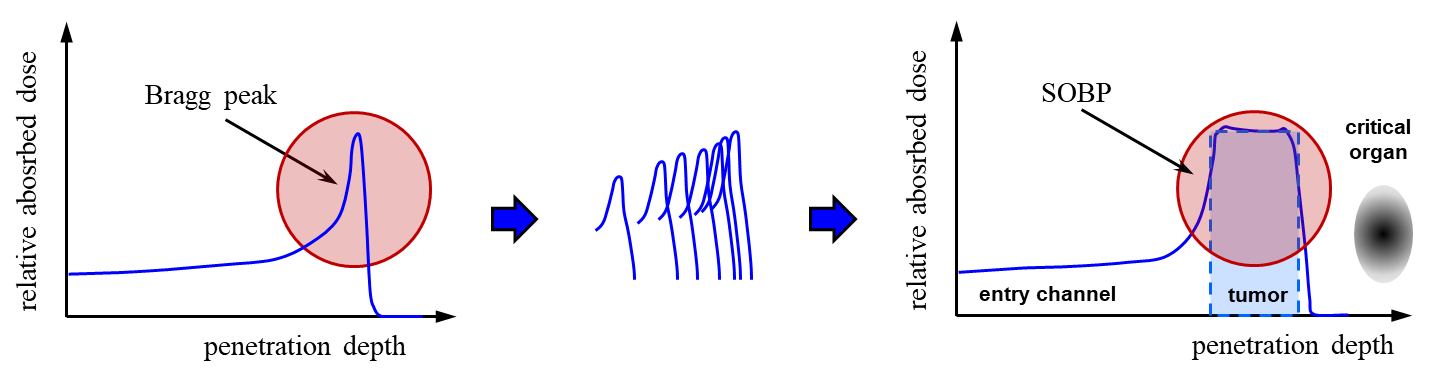}
		\caption{Dose profile and Bragg peak behaviour (left). In clinical applications, the Bragg peak must be extended to homogeneously cover the tumour region. This can be done by overlapping beams at different energies and intensities. The dose plateau (right) which results from this overlapping is called the spread-out Bragg peak (SOBP).}
		\label{fig:S6Tumor}
	\end{center}
\end{figure}
The radio-biological motivation for hadron therapy is to apply the energy of particles for the DNA destruction inside the nucleus of a tumour cell. The main requirements for this therapeutic method are illustrated in \Fref{fig:S6Tumor}. These are (i) a constant and high dose profile at the location of the tumour, and (ii) a low dose profile at critical organs. The basic advantage of hadron therapy is that charged hadrons deposit the maximum energy density at the very end of their range (the so-called Bragg peak), therefore the position where cells are damaged can be well localized in depth so that critical organs behind the tumour are safe. Furthermore, the use of carbon ions is advantageous because of the relative low radiation dose in the entry channel compared to e.g., protons. More information about hadron therapy and their technological developments can be found in Refs.~\cite{Bryant02,Suit09,Durante16} and the references therein.

In order to scan the particle beam over the whole tumour region, scanning mechanisms are required for both transverse planes and the longitudinal one.
As an example, the GSI pilot project for HIT developed an intensity-controlled raster-scan method which has been successfully applied in patient treatment \cite{Kraft98}.

By this method the tumour is painted with a pencil-ion beam using an active variation of the beam properties. In transverse directions a two-dimensional scanner magnet excitation is applied in order to vary the beam position, i.e., a precise knowledge of beam position and beam size is mandatory. In the longitudinal direction the fact is exploited that the particle penetration depth is determined by the kinetic energy of the beam, i.e., the position of the Bragg peak can be controlled via the primary particle energy. In contrast to the hadron accelerators described so far, for the hadron therapy accelerator an accurate knowledge of the particle beam energy is therefore mandatory. Moreover, to achieve a constant dose profile over the tumour region the particle beam intensity has to be adjusted together with the beam energy. As a consequence, the operation of a hadron therapy accelerator for patient treatment requires precise diagnostics for the determination of beam size, beam position, energy, and intensity.

The principle set-up of a hadron therapy accelerator resembles that of a standard hadron injector linac together with injector synchrotron and transfer lines. Therefore the instrumentation in use is similar to the one discussed before. However, there are some peculiarities which are to be considered in view of beam instrumentation. According to these specialities the diagnostic systems can be classified in three categories \cite{Peters00}:
\begin{description}
  \item[(i)] Non-destructive diagnostics systems that work online during patient treatment and in all other cases. These systems have to work reliably and must be easy to operate. Monitors of this type must be located in all important areas so that the operators can see at a glance if the accelerator is working correctly.
  \item[(ii)] Destructive devices that are used for daily checks of the machine performance and the beam stability, and in addition to solve simpler machine problems. This kind of diagnostic is positioned more densely along the machine. Handling and data interpretation of these devices may be more complicated and require complex algorithms controlled by software.
  \item[(iii)] Special devices that will be necessary during the machine commissioning and in the case of serious machine problems. Equipment of this kind will be used only by specialists and therefore can have a more difficult user interface.
\end{description}
Detailed information about specific instrumentation for medical hadron accelerators can be found e.g., in Refs.~\cite{Peters00,Schwickert04,Schwickert07} and in the report about medical applications in these proceedings \cite{Peters18b}. 
\section{Instrumentation for lepton colliders}

So far diagnostics needs for hadron accelerators have been considered. The present section is dedicated to the specific needs of lepton accelerators in general, and especially for high-energy physics colliders and storage rings. A comprehensive overview of $\Pem\Pep$ colliders can be found in \Bref{Oide14}. The description of the required instrumentation systems is mainly based on examples from the 27.5\UGeV{} \Pep(\Pem) ring of the electron-proton (ep) collider HERA at DESY. Descriptions of beam instrumentation for other machines can be found e.g., in \Bref{Biagini94} for the  0.7\UGeV{} collider DA$\Phi$NE (INFN Frascati, Italy), in \Bref{Cao08} for the 1--2.1\UGeV{} collider BEPCII (IHEP Beijing, China), in \Bref{Smaluk04} for the 6\UGeV{} collider VEPP-4 (Budker Institute Novosibirsk, Russia), in \Bref{Fisher98} for the 9\UGeV{} \Pem/3.1\UGeV{} \Pep collider PEP-II (SLAC, USA), in Refs.\cite{Bovet86,Borer91} for the 100\UGeV{} collider LEP (CERN), and in Refs.~\cite{Arinaga12,Fukuma16} for the 7\UGeV{} \Pem/4\UGeV{} \Pep collider SuperKEKB (KEK Tsukuba, Japan) where the first collisions were observed in April 2018. However, before starting with the description of the required instrumentation, the following subsection recalls the differences between lepton and hadron beams and their consequences for diagnostics.

\subsection{Lepton properties and the consequences for diagnostics}

The main differences between leptons and hadrons are that (i) leptons are simple point objects and have no constituent nature, and that (ii) their rest mass is much smaller than that of hadrons. Especially the second aspect has strong impact on accelerator physics as already indicated before:
\begin{itemize}
  \item Leptons are fully relativistic at a few MeV beam energy which is typically achieved early in the accelerator chain, behind the first accelerating sections.
  \item According to Eq.~(\ref{eq:S2CoulField}) they produce a strongly squeezed electromagnetic field. With increasing particle energy the electric field is increasingly `flattened' towards the plane perpendicular to the direction of motion. Characteristic parameter for the description of the field compression is the Lorentz factor $\gamma = E / m_0 c^2$.
\end{itemize}
The $\gamma$ scaling implies that non-propagating fields have a long transverse range, and especially that synchrotron radiation (SR) is emitted for bent particle motion. The emission of synchrotron radiation has strong influence on both particle dynamics and beam diagnostics.

The SR emitted power can be estimated according to
\[
    P_{\gamma} = 8.85 \times 10^{-2} \, \frac{E^4 [\mathrm{GeV}^4]}{\rho [\mathrm{m}]} \, I [\mathrm{A}]
\]
with $E$ the beam energy, $I$ the mean beam current, and $\rho$ the mean bending radius. For HERA with $E$ = 27.5\UGeV{}, $I$ = \Unit{50}{mA}, and $\rho$ = \Unit{550}{m} the emitted power amounts to $P_{\gamma}$ = \Unit{4.6}{MW}. This means that a huge level of radiation power is emitted from the beam which is deposited in the accelerator environment. Such a level may easily lead to the damage of components, i.e., sensitive accelerator components have to be protected from direct SR illumination and cooling is required in places where the power is deposited.

Furthermore, the extracted beam power results in an energy loss per turn which is estimated as
\[
    \Delta E_{\gamma} = 8.85 \, \frac{E^4 [\mathrm{GeV}^4]}{\rho [\mathrm{m}]} \, .
\]
For the HERA parameters the energy loss per turn amounts to $\Delta E_{\gamma}$ = 92\UMeV{}. Without restoration of this energy loss on every turn, a stable accelerator operation would be impossible. This is the duty of the RF system, i.e., cavities are required which have to provide sufficient voltage to bring the particles back to their nominal energy. Because of the large energy loss, typically a large number of cavities is necessary. HERA for example used 98 cavities, grouped in 8 section with 8 transmitter stations, each with \Unit{1.4}{MW} nominal power and fed by 2 klystrons.

In the following list the consequences of SR emission are summarized together with their impact on beam dynamics and beam diagnostics.
\begin{description}
  \item[(i)] A large number of cavities is required in order to restore the energy loss due to the radiation emission. However, each cavity represents a high impedance for the beam. A higher impedance increases the possibility to excite (multi-bunch) instabilities which have to be  damped. Therefore circular lepton accelerators usually require a multi-bunch feedback system for stable operation.
  \item[(ii)] The heat load on accelerator components due to the high SR power becomes critical. Therefore the accelerator together with the instrumentation has to be protected, and cooling is required in locations where direct SR illumination cannot be avoided.
  \item[(iii)] A high total cavity voltage $V_r$ is required in order to compensate the energy losses and to guarantee sufficient lifetime. However, the cavity voltage is connected with the rms bunch length $\sigma_t$. Above transition energy the following relation holds
  \begin{equation}\label{eq:S7BunchLength}
    \sigma_t = \frac{\alpha_c - 1/\gamma^2}{2 \pi f_s} \, \sigma_{\delta} \quad \propto \quad 1/\sqrt{V_r} \, ,
  \end{equation}
  with $\alpha_c$ the momentum compaction factor, $f_s$ the synchrotron frequency, and $\sigma_{\delta}$ the relative energy spread. According to Eq.~(\ref{eq:S7BunchLength}) a high cavity voltage implies smaller bunch lengths, i.e., the lepton beam spectrum contains much higher frequencies than the hadron spectrum.
  \item[(iv)] In a lepton accelerator SR emission leads to the formation of an equilibrium emittance in the horizontal plane owing to the concurring processes of radiation damping and quantum excitation. The vertical emittance is defined by the coupling between both transverse planes and typically very small. Therefore the beam emittance in the final storage ring is determined by the ring itself and not by the injector chain, i.e., emittance blow-up is not critical and the injector chain has relaxed requirements.
\end{description}
Similar to the hadron collider, in the subsequent sections a `walk' along the injector complex will be done. The properties of the various accelerators in the chain together with their beam diagnostics are briefly discussed.

\subsection{Injector complex instrumentation}

\begin{figure}[!t]
	\begin{center}
		\includegraphics[scale=0.8,clip]{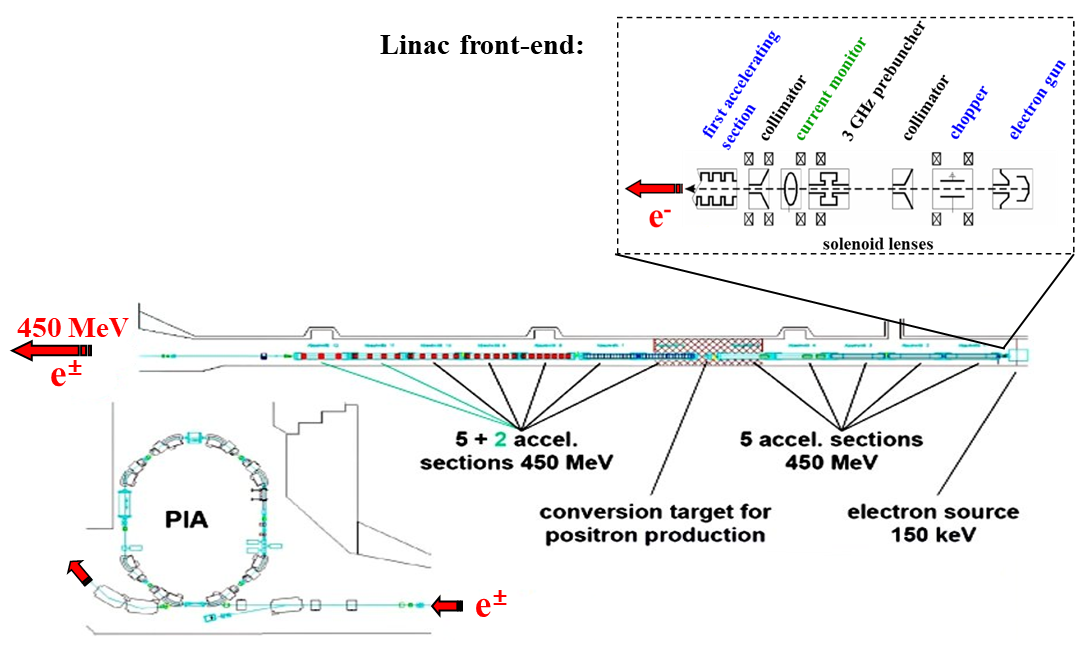}
		\caption{The 450\UMeV{} \Pep/\Pem injector linac for HERA together with the positron intensity accumulator ring PIA}
		\label{fig:S7Linac2}
	\end{center}
\end{figure}
Figure \ref{fig:S7Linac2} gives an overview of the \Pep/\Pem injector complex of HERA. It consists of the following subsections:
\begin{itemize}
  \item Thermionic gun (150\UkeV{}) which produces \Unit{3}{$\mu$s} long pulses at \Unit{50}{Hz} repetition rate.
  \item Chopper and collimator section for shortening of the long gun pulses (\Unit{60/20}{ns} for \Pep/\Pem).
  \item Single cell \Unit{3}{GHz} pre-buncher cavity for matching to the linac RF.
  \item 450\UMeV{} linac sections. These are \Unit{3}{GHz} (S-band) travelling wave structures which operate at a repetition rate of \Unit{50}{Hz}. After the linac the relative energy width is $\Delta E / E \approx$ 0.27 \% for electrons and $\approx$ 1 \% for positrons.
  \item Converter for \Pep production. The target consists of a \Unit{7}{mm} (2 radiation lengths) thick tungsten target which is located in a \Unit{1.8}{T} solenoidal  field to enhance the collection efficiency \cite{Stange79}.
  \item Positron intensity accumulator ring, PIA. In addition to intensity accumulation, PIA is utilized for the re-formation of the time structure for the subsequent synchrotron which operates at \Unit{500}{MHz} \cite{Mulhaupt76}. It has two RF systems operating at \Unit{10.4}{MHz} and at \Unit{125}{MHz}.
\end{itemize}
A similar set-up was used e.g., for the LEP pre-injector (LPI), consisting of the LEP injector linac (LIL) and the electron positron accumulator (EPA) \cite{Madsen87,Madsen88}.

As in the case for the hadron injector linac, the lepton injector complex needs beam instrumentation (i) for the adjustment of the beam transport through the individual accelerator sections, (ii) for the setting of the RF system, and (iii) in order to indicate the operating status during standard operation. The following list gives an overview of the standard instrumentation and their tasks:
\begin{itemize}
  \item Intensity:\\
  an intensity diagnostic is required for the determination of the transfer efficiency. Usually current transformers and/or wall current monitors are used.
  \item Beam position:\\
  the determination of the beam position is necessary for the beam steering through various accelerator sections. Common devices for position determination are BPMs with sufficient sensitivity for the long linac bunch trains. However, luminescent screens are also used. Their usage is less critical than for hadron beams because of the lower energy deposition in matter.
  \item Transverse beam profiles:\\
  the beam profile is an important prerequisite for the emittance determination and beam optics matching. Typical beam monitors are fluorescent or optical transition radiation (OTR) screens in straight sections and a synchrotron light monitor in the accumulator ring.
  \item Transverse emittance:\\
  a knowledge of the transverse emittance is required for the matching of different accelerator sections. It can be determined either via the multi-screen method or k-modulation of quadrupoles as in the case for transfer lines, or it is measured with synchrotron radiation diagnostics in the accumulator ring.
  \item Longitudinal plane:\\
  magnet spectrometers are typically used for energy (-spread) measurements. Similar to hadron beam diagnostics the spectrometer magnet is normally the entrance magnet of a diagnostics beamline behind the linac. The time structure can be measured with an RF deflector, a wall current monitor, or via coherent radiation diagnostics.
\end{itemize}
The beam instrumentation required for a lepton injector complex is in principle the same as that for hadron machines. Therefore a detailed description of each monitor is not needed. An overview of the LEP injector linac instrumentation can be found e.g., in \Bref{Battisti87}. Further examples for synchrotron light source injector linac diagnostics are \Bref{Lumpkin93} for the APS (Argonne, USA) or \Bref{Schlott99} for the SLS (Villingen, Switzerland).

So far the lepton injector complex together with its instrumentation has been presented. Nevertheless, there are some peculiarities in the positron production which have an impact on the instrumentation. They will be discussed in the following subsection.

\subsubsection{Comment on positron production}

\begin{figure}[!h]
\begin{center}
\includegraphics[scale=.64,clip]{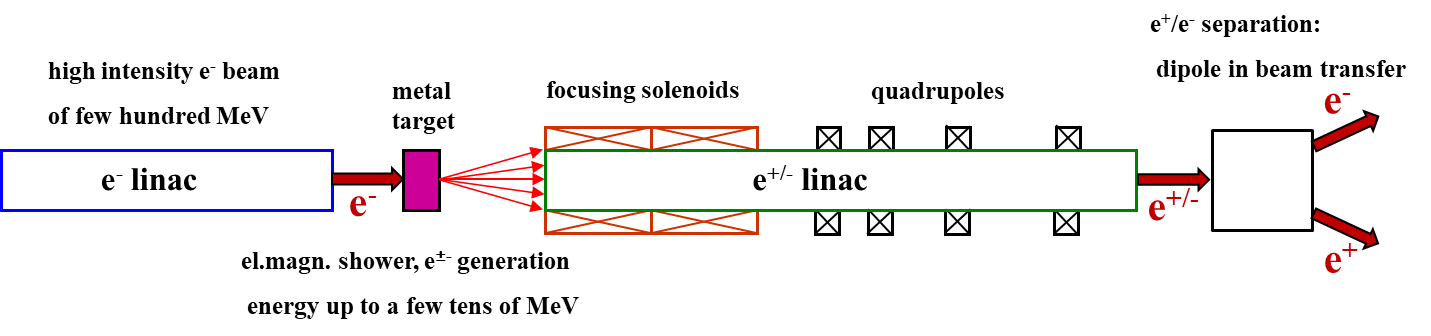}
\caption{Principle of positron generation}
\label{fig:S7Positron}
\end{center}
\end{figure}

The principle of positron production is illustrated in \Fref{fig:S7Positron} and explained according to \Bref{Hubner88}. A high-intensity electron beam of a few hundred MeV strikes a metal target. The resulting electro-magnetic showers generate a mixture of secondary positrons and electrons with energies up to a few tens of MeV. A focussing system behind the target maximizes the collection efficiency of secondaries and guides them to the subsequent accelerator sections. Solenoids on the first accelerating sections together with quadrupoles on the sections further downstream provide focusing. At the end of the beam transfer, \Pep and \Pem are separated by a dipole.

The positron production mechanism described above implies that the conversion target is located in a harmful radiation environment, and only radiation resistant instrumentation can be applied close to the target.

Furthermore, for a high positron yield a focus at the converter target is required. Therefore, a transverse profile monitor close to the conversion target is desirable for beam size optimization. Luminescent screens cannot be used in such radiation hard environments because of material degradation effects. As an example, for LIL two possible types of secondary emission monitors were in discussion. More information about this can be found in Ref.\cite{LEP83}.

Finally, the matching to the accumulator ring energy acceptance $\Delta E/E$ is of importance. As an example, for the LEP injector chain the EPA acceptance was $\pm$ 1 \%, and the energy spread was determined by three factors: the accepted momentum spread from the production target, the microbunch length, and beam loading \cite{Godot91}. In order to keep these effects under control, for a proper matching precise measurements of energy spread and bunch length are mandatory.

After injector linac and accumulator ring, the next elements in the injector chain are transfer line and booster synchrotron. However, there are no fundamental differences in comparison to hadron machines, and no fundamental differences in the instrumentation between an electron linac and a storage ring. Therefore the explicit description of diagnostic needs of these injector accelerators is not necessary and the discussion is continued with the description of the final storage ring instrumentation.

\subsection{Storage ring diagnostics}

The subsequent list gives an overview of typical lepton storage ring instrumentation together with some comments concerning their usage.
\begin{itemize}
  \item Current monitors (AC and DC):\\
  current monitors are required for the determination of bunch charge and stored DC current. The monitor concepts in use are the same as those presented in the previous sections.
  \item BPMs:\\
  BPMs are necessary for the determination of beam orbit and associated parameters.
  \item Tune measurement:\\
  the tune defines the working point of the machine, its determination is therefore mandatory for machine operation.
  \item Feedback systems:\\
  feedback systems are required in order to damp beam instabilities and to allow for stable operation with high beam current.
  \item Synchrotron light diagnostics:\\
  this kind of diagnostic is used to measure longitudinal as well as transverse beam profiles and emittances, and to study dynamical effects.
  \item Energy measurement:\\
  the beam energy is an important parameter for the experiments because it defines the centre-of-mass energy for particle production.
  \item Luminosity monitor:\\
  the luminosity is a key parameter for colliders and important for event rate optimization in the experiments. Leptons are simple point objects and their reaction cross-section into the final state is precisely calculable. Reference channel for luminosity determination in an \Pepm collider is usually Bhabha scattering. The cross-section is well known, therefore absolute luminosity determination is even possible without precise knowledge of beam and bunch currents. Nevertheless they are of importance for the specific luminosity.
  \item Beam loss monitors:\\
  beam loss monitors are required for loss control and machine optimization. In a normal conducting lepton machine the potential for damage of accelerator components in case of beam loss is less critical. Therefore loss monitors can be used not only for machine protection, but also for machine physics studies.
  \item Machine protection system:\\
  the machine protection system is necessary to protect sensitive and critical accelerator components especially against excessive heat load. Therefore a reliable temperature control is essential for safe machine operation.
\end{itemize}
In the following, selected instruments listed above will be presented and discussed and compared with those in hadron machines.

\subparagraph{Beam position monitors}

The bunch length in a lepton storage ring is much smaller than that in a hadron ring, typically of the order of \Unit{10-100}{ps}. Therefore button-type pick-ups are used. Owing to their high-pass characteristics they are suitable for small bunch lengths, but are also much simpler in construction and cheaper than a stripline monitor, for example.

\begin{figure}[!h]
	\begin{center}
		\includegraphics[scale=.9]{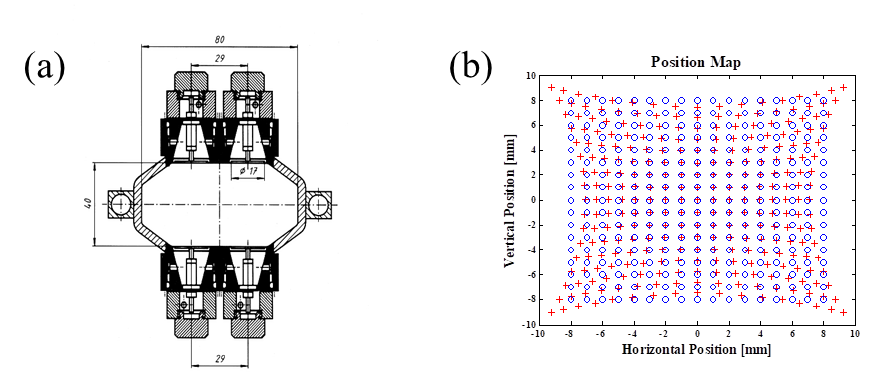}
		\caption{(a) Cross-section of a HERA BPM. The button pick-ups are mounted on top and bottom of the chamber. (b) BPM position map: blue dots correspond to the beam position inside the chamber, red crosses are the reconstructed positions according to the $\Delta/\Sigma$ algorithm.}
		\label{fig:S7BPMlepton}
	\end{center}
\end{figure}

To avoid direct synchrotron radiation illumination and the generation of additional charges due to the photo effect causing misleading position information, the buttons are mounted out of the orbit plane. Furthermore, in a lepton ring the vacuum chamber is not rotationally symmetric, see \Fref{fig:S7BPMlepton} (a). This has two reasons, (i) the horizontal emittance is much larger than vertical one because of synchrotron radiation emission in the horizontal plane, and (ii) injection oscillations in this plane due to off-axis injection which allows intensity accumulation.

However, the symmetry distortion implies non-linear monitor position characteristics. For illustration \Fref{fig:S7BPMlepton}(b) shows a position map calculation according to \Bref{Kube07c} for a similar BPM type as shown in \Fref{fig:S7BPMlepton}(a). As can be seen, with increasing distance from the monitor centre the difference between actual and reconstructed beam position also increases. Usually this non-linearity is corrected applying higher order polynomials \cite{Nosych14}.

\subparagraph{Tune and feedback}

Due to radiation damping caused by synchrotron radiation emission, beam blow up is not a critical issue. Therefore a small permanent beam excitation can be applied and an online tune control is possible.

In a high-energy lepton storage ring the particle's electromagnetic field has a long range according to Eq.~(\ref{eq:S2CoulField}) and can act back on the beam itself via the environment. Due to the short bunch lengths, the beam spectrum has broadband characteristics and excitations of instabilities are possible over a wide frequency range. As consequence, lepton storage rings typically have to be operated with multi-bunch feedback systems to damp coupled-bunch instabilities.

\begin{figure}[!t]
\begin{center}
\includegraphics[scale=.6,clip]{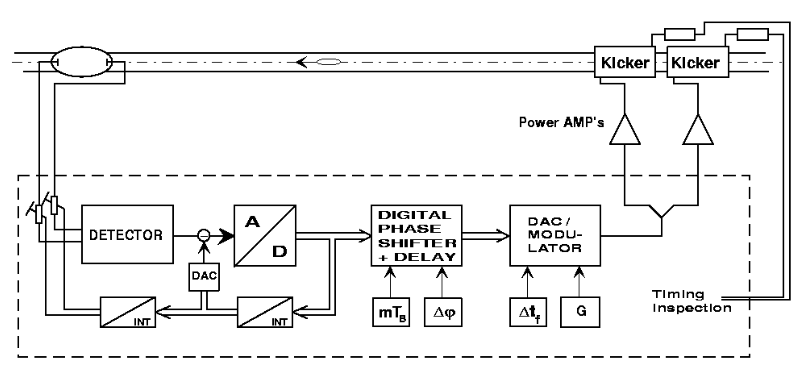}
\caption{Schematic view of the HERA transverse electron feedback system \cite{Balewski98}}
\label{fig:S7HERAfeedback}
\end{center}
\end{figure}
\Fref{fig:S7HERAfeedback} illustrates the components required for a multi-bunch feedback system. These are (i) a detection system to measure (longitudinal or transverse) beam oscillations which is normally a fast BPM, (ii) a signal processing unit to derive correction signals, and (iii) a broad band amplifier and beam deflector to act back on the beam. More information about feedback systems can be found e.g., in Refs.~\cite{Balewski98,Lonza08} and in the report in these proceedings \cite{Schmickler18}.

\subparagraph{Transverse profile and emittance}

For transverse profile or emittance measurements, imaging with synchrotron radiation is widely used as a non-destructive profile diagnostic. The imaging resolution of optical synchrotron radiation is usually sufficient. To give an example, the HERA horizontal and vertical beam sizes were about $\sigma_h$ = 1200 $\mu$m and $\sigma_v$ = 250 $\mu$m.

\begin{figure}[!h]
	\begin{center}
		\includegraphics[scale=.75,clip]{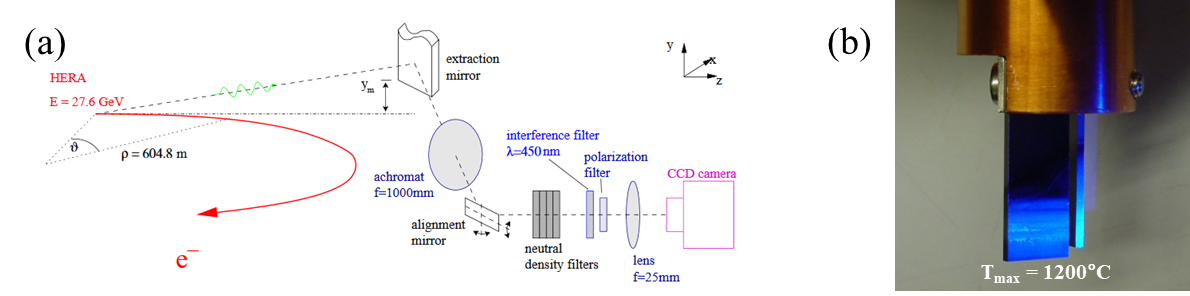}
		\caption{(a) Schematic view of the HERA profile monitor set-up \cite{Kube05}. (b) Photo of the HERA extraction mirror. Close to the lower mirror edge a surface deformation is clearly visible caused by the heat load.}
		\label{fig:HERAeSyLI}
	\end{center}
\end{figure}

Figure \ref{fig:HERAeSyLI}(a) shows a schematic view of the HERA monitor set-up \cite{Kube05}. The light is extracted out of the vacuum system by a mirror and the beam is imaged by a lens system onto the chip of a CCD camera. An interference together with a polarization filter serves for resolution improvement. Additional information about synchrotron radiation diagnostics in general can be found e.g., in Refs.~\cite{Hofmann05,Wilke95} and the references therein.

A problem which often arises is the heat load on the first extraction mirror which leads to image distortion and resolution deterioration. This is caused by the X-ray part of the synchrotron radiation absorbed in the mirror material. To keep the absorption to a minimum, materials with low absorption coefficients like beryllium are often used, and the mirror is additionally water cooled. However, this is not sufficient to prevent image distortion and even destruction of the monitor components. Figure \ref{fig:HERAeSyLI}(b) shows a photo of the HERA extraction mirror which was in use for several years. Close to the lower mirror edge a deformation of the surface can be seen caused by the heat load. Finite element calculations demonstrate that the mirror could heat up to a temperature of 1200$^{\circ}$ C in the region where the X-ray part of the synchrotron radiation fan hits the mirror.

There are different ways to overcome this problem, a few possibilities are presented in the following. At HERA out of plane observation was used, i.e., the extracting mirror bottom edge was placed above the orbit plane (cf. \Fref{fig:HERAeSyLI}(a)) so that the X-ray part of synchrotron radiation could not hit the mirror, and only optical synchrotron radiation emitted under larger angles was reflected out of the vacuum system. A drawback of this method was an increased diffraction limited resolution contribution \cite{Kube05}. The LEP 2 synchrotron light telescopes used adaptive optics for correction, i.e., cylindrically deformable mirrors to compensate the cylindrical extraction mirror deformation together with moveable detectors to compensate spherical deformation \cite{Burtin99}. At PEP-II a slotted mirror design together with an X-ray absorber was applied \cite{Fisher96}, but also with the drawback of increased diffraction limited resolution. At SLS (Villingen, Switzerland) a thin absorber is inserted in front of the extraction mirror which blocks the radiation, and the measurement is performed with synchrotron radiation in $\pi$ polarization \cite{Andersson06}. However, not only direct heating due to synchrotron radiation absorption may lead to a degradation of the extraction mirror performance. At LHC, heating of the light extraction system caused by electromagnetic coupling with the beam was observed which necessitated a refurbishment of the entire system \cite{Andreazza13}.

\subparagraph{Longitudinal profile}

For the investigation of longitudinal profiles of the order of picoseconds and larger, again visible synchrotron radiation based instrumentation is applied. In case of HERA with 27.5\UGeV{} beam energy, the duration of a synchrotron radiation pulse from a single electron (intrinsic time resolution) is in the order of \Unit{10^{-2}}{as}. Owing to this excellent intrinsic resolution, the time structure of the synchrotron radiation pulse emitted from the whole bunch is a replica of the bunch itself and therefore well suited to resolve longitudinal beam profiles of even sub-femtoseconds. An advantage of using the radiation pulse replica instead of the particle bunch itself is that  advanced optical measurement techniques can be applied which were originally developed for laser physics.

\begin{figure}[!h]
	\begin{center}
		\includegraphics[scale=.6,clip]{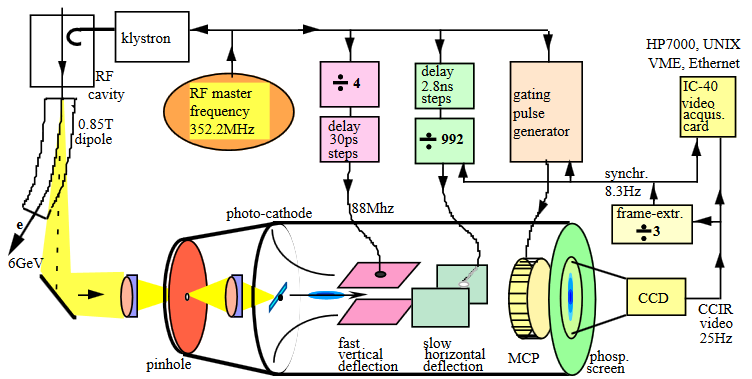}
		\caption{Schematic view of a dual sweep streak camera together with synchronization \cite{Scheidt96}}
		\label{fig:S7Streak}
	\end{center}
\end{figure}
The measurement device which is conventionally used at lepton storage rings is the streak camera. Their set-up together with the synchronization scheme is illustrated in \Fref{fig:S7Streak}. Apart from the report in these proceedings \cite{Gillespie18}, more  information about streak cameras can be found e.g., in Refs.\cite{Scheidt00,Yang06}.

\subparagraph{Beam energy}

Leptons are point-like objects, and in a lepton collider the total energy is of tremendous importance as it is a constraint on the final state detected in the experiment. Therefore the determination via the dipole current is not sufficient and more precise energy measurements are required.

A measurement technique widely applied at storage ring based lepton machines (colliders and light sources) is the resonant spin depolarization technique which is based on the destruction of the self-polarization arising from the Sokolov--Ternov effect \cite{Sokolov64}. The machine energy is varied until a ($g-2$) resonance is excited which causes the beam to depolarize. Assuming the resonance is uniquely identified, the spin tune
\[
    \nu_s = \gamma (g-2) / 2
\]
is determined, and with it the beam energy $E = \gamma m_0c^2$. The $g$ factor of the electron is known from precise quantum electrodynamical calculations, and the beam polarization is monitored by scattering a laser beam off the circulating beam and measuring the spin-dependent part of the Compton scattering by recording the angular distribution of the back--scattered $\gamma$ rays. This method was successfully applied e.g., at LEP \cite{Arnaudon95}.

Another scheme in use is the Compton backscattering technique which was originally implemented at the BESSY-I and BESSY-II storage rings \cite{Klein97}, and also used at the VEPP-4M collider (BINP, Novosibirsk) \cite{Bgk05}. More information about beam energy measurements can be found in \Bref{Muller08}.
\section{Instrumentation for light sources}

So far diagnostics needs for lepton accelerators with special emphasis on high-energy physics colliders were presented. This chapter is dedicated to a different class of electron accelerators, the light sources. Synchrotron light sources have become indispensable scientific instruments for basic and applied frontier research in various fields of science and technology.
Reference \cite{Schneider10} illustrates how progress in accelerator science and technology stimulated advancement in photon science. In particular, the development of light sources of the so-called third generation which are based on a low emittance electron storage ring boosted the development dramatically \cite{Zhao10}. However, currently operating sources have intrinsic limitations that restrict some of the experiments. New schemes are discussed in the community and upgrade plans have been discussed at several facilities or are already in a concrete planning, approval and even construction stage. In addition, totally new concepts like energy recovery linac sources \cite{Hajima10} are being discussed at various places, and X-ray free-electron lasers \cite{Schreiber10} have started operation. Reference \cite{Weckert15} discusses the present synchrotron light sources and their limitations and gives an overview of the next-generation sources such as diffraction-limited light sources, energy-recovery linacs, and free-electron lasers.

In the following storage ring based light sources are discussed, while the second part of this section is dedicated to the diagnostic needs for a linac (single pass) based fourth generation light source (FEL).

\subsection{Storage ring based light sources}

\begin{figure}[!t]
\begin{center}
\includegraphics[scale=.22,clip]{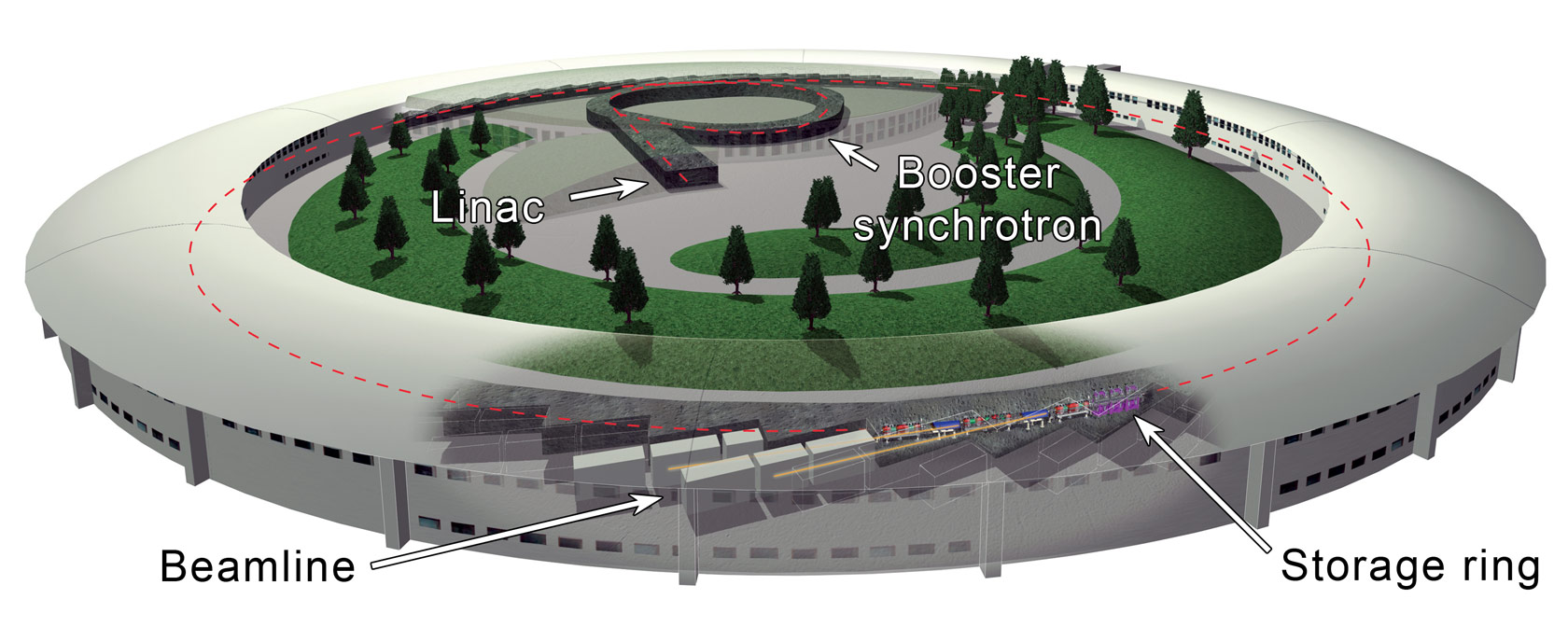}
\caption{Layout of a third generation storage ring based light source \cite{ESRF}. The storage ring is densely equipped with user beamlines for transportation of the radiation from the source (bending magnet, wiggler, or undulator) to the experimental hutches where the user experiments are located. The injector chain is short and consists of the electron gun, the subsequent linac, a booster synchrotron, and a short transfer line.}
\label{fig:S8esrf}
\end{center}
\end{figure}

A storage ring based light source resembles a lepton collider with the difference that only one species of particles is circulating in the machine. Figure \ref{fig:S8esrf} shows a typical light source layout \cite{ESRF}.
The storage ring typically has an energy in the range  1--8\UGeV{} and a circumference of the order of 100--\Unit{2000}{m}. Most machines operate at fixed energy, e.g., a special diagnostic to control the acceleration process is not required. The insertion devices (wigglers or undulators) are an integrated part of the machine located in straight sections, and user experiments are situated at the end of long photon beamlines which are typically 50--\Unit{100}{m} away from the source point.

The injector chain of a light source is much shorter than the one of a collider, it consists of a linac, a booster synchrotron to cover the energy gap between linac and storage ring, and a short transfer line to the injection in the storage ring. The instrumentation required for operation and diagnostics in the injector chain consists of standard monitors already described above. In the subsequent section some general remarks concerning light source requirements are summarized.

\subsubsection{Remarks on light sources}

Reference \cite{Weckert15} condenses the basics of synchrotron radiation generation and illustrates the interplay between accelerator parameters and radiation properties. A key parameter of a light source is the spectral brilliance $B$ which is defined by
\begin{equation}\label{Eq:Brilliance}
    B = \frac{{\rm Number~of~photons}}{[{\rm s}] \, [{\rm mm}^2] [{\rm mrad}^2] [{\rm 0.1\% bandwidth}]} \; ,
\end{equation}
i.e., it is a measure of the phase space density of the photon flux. The user requirement is to have a high brilliance which means as many as possible monochromatic photons emitted from a minimum source region on the sample. The brilliance can be rewritten as a function of the accelerator parameters in the following form:
\begin{equation}\label{Eq:Brilliance2}
    B \propto \frac{N_{\gamma}}{\sigma_x \, \sigma_x' \, \sigma_y \, \sigma_y'} \propto \frac{I}{\varepsilon_x \, \varepsilon_y} \, ,
\end{equation}
with $I$ the beam current and $\varepsilon_{x,y}$ the horizontal resp. vertical beam emittance. According to Eq.(\ref{Eq:Brilliance2}) there are two requirements for the accelerator to achieve a high brilliance:
\begin{description}
  \item[(i)] High beam current. This has the following implications for beam diagnostics:
  \begin{itemize}
   \item [(a)] To achieve the high currents.\\
   Diagnostics and instrumentation are required to detect and damp instabilities.
   \item [(b)] To cope with high heat load.\\
   Heat load changes may lead to a position drift of accelerator and beam line components which the user will see immediately as a change in the photon flux on the sample
  \end{itemize}
  \item[(ii)] Small beam emittances. The implications are:
  \begin{itemize}
    \item [(a)] To achieve small emittances.\\
    The choice of the proper magnet lattice defines the minimum achievable emittance. Lattices which are presently in use are e.g., the double bend achromat (DBA) or the triple bend achromat (TBA). However, the new class of diffraction limited storage rings is based on multi-bend achromats (MBAs), typically with 7--9 bending magnets per cell, see e.g., Refs.~\cite{Jiao18,Streun19}. Nevertheless the magnet lattice is a task for the lattice designer \cite{Ropert98,Streun05,Wolski14} and will not be covered in the following.
    \item [(b)] To measure small emittances.\\
    If the accelerator is designed to have a small emittance it is mandatory to have monitors which are capable of measuring this beam parameter. The measurement of a small transverse emittance in a storage ring means to be able to resolve small transverse beam sizes.
    \item [(c)] To preserve the emittance.\\
    There are various effects resulting in emittance growth and with it a deterioration of the beam quality. Examples are fast beam orbit motions as well as short term and medium term component position drifts, i.e., stability is again a critical issue.
  \end{itemize}
\end{description}
As can be seen from this list stability is crucial for the operation of a light source. Therefore in the following section stability issues and their implications on beam diagnostics will be addressed. Detailed discussions about this topic can be found e.g., in Refs.~\cite{Bocchetta98,Farvacque98,Boege04,Decker05,Nadolski08}. Further information about performance and trends of 3$^{rd}$ generation light sources in general can be found in \Bref{Bartolini08} and the references therein. Diffraction-limited storage ring based light source design and plans are discussed e.g., in Refs.~\cite{Eriksson14,Hettel14,Liu17}.

\subsubsection{Stability}

Although stability issues are mutually dependent, they can roughly be classified into three categories.
\begin{description}
  \item[(i)] Energy stability and suppression of energy broadening effects:\\
  these effects are mainly caused by (longitudinal) multi-bunch instabilities. They can result in a shift of the radiation harmonics from an undulator which the user sees as intensity fluctuations and line broadening. Examples for these effects can be found e.g., in Refs.\cite{Furman94,Kim95}. To combat against such instabilities multi-bunch feedback systems are required.
  \item[(ii)] Intensity stability:\\
  a change in background conditions or thermal load on beamline and machine components due to intensity variations may affect position stability. Therefore it is desirable to keep the intensity at a constant level. Intensity fluctuations may be caused by transverse multi-bunch instabilities which have to be damped by a feedback system. However, a change in intensity due to the natural beam lifetime is also not tolerable for many applications. Therefore the trend is going towards operating light sources in top-up mode, i.e., the natural losses are compensated by refilling small amounts of charge in short time intervals \cite{Ohkuma08,Aiba18}. This implies a vast dynamic range which is required for the beam monitors, starting from the injector chain.
  \item[(iii)] Position stability:\\
  instabilities in the beam position result in emittance growth and intensity fluctuations, i.e., in a reduction of the brilliance of the light source. In order to keep the orbit stable to a high level of precision, orbit feedback systems are required which include high resolution electron BPMs in the storage ring together with photon BPMs in the user beamlines. Reference \cite{Bulfone08} gives an overview of state of the art fast beam position feedback systems.
\end{description}

\subsubsection{Instrumentation examples}

In this section instrumentation is presented which is specific for the operation of light sources. These are beam position monitors and devices for emittance diagnostics. A general overview of diagnostics for third generation light sources and recent developments can be found e.g., in Refs.~\cite{Kuske94,Rehm08}. 
Besides the report about diagnostics examples from light sources in these proceedings \cite{Wittenburg18b}, additional information about beam instrumentation can be found in overview articles for light sources. Examples are \Bref{Lumpkin92} for the APS (Argonne, USA), \Bref{Rehm06} for Diamond (Oxfordshire, UK), \Bref{Nadji07} for Soleil (Gif--sur--Yvette, France), and Refs.~\cite{Balewski08,Kube08c,Balewski09} for PETRA III (DESY, Germany).

\subparagraph{Beam position monitoring}

The typical stability requirement for a light source is to keep the intensity fluctuations constant to a level of 0.1\%. This can be translated into a maximum allowed emittance growth of 20\% owing to short term orbit fluctuations, which is equivalent to an orbit stability requirement of 10\% of the (1$\,\sigma$) beam size resp. of the beam divergence \cite{Galayda94}. For PETRA III at DESY, the beam sizes close to insertion devices are $\sigma_h$ = \Unit{20}{$\mu$m} and $\sigma_v$ = \Unit{3}{$\mu$m}, i.e., the corresponding stability requirements are \Unit{2}{$\mu$m} and \Unit{0.3}{$\mu$m}.

\begin{figure}[!h]
\begin{center}
\includegraphics[scale=.4,clip]{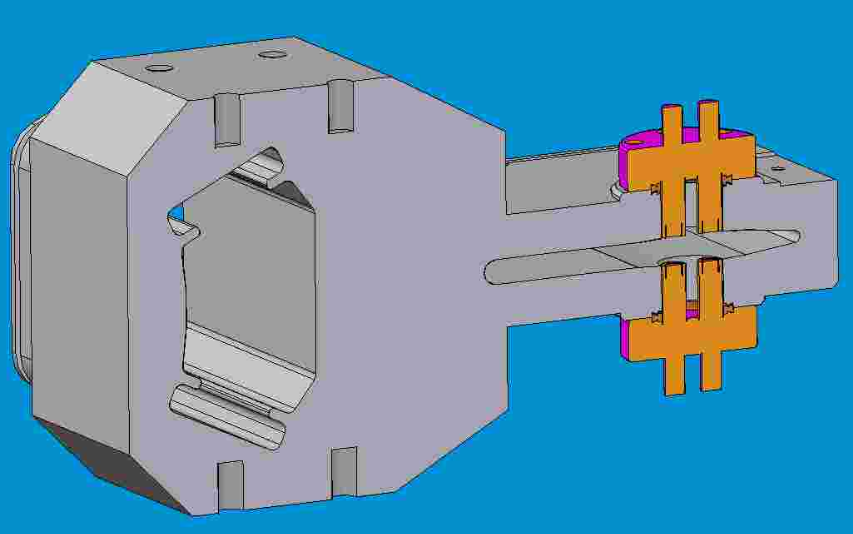}
\caption{PETRA III vacuum chamber between canted undulators with flanged BPM (cut). The channel on the left side is used for the mounting of NEG-strip pumps (courtesy A. Delfs, DESY).}
\label{fig:S8PetraBPM}
\end{center}
\end{figure}

In order to monitor the electron beam orbit, high-resolution button-type pick-ups are used. As an example, \Fref{fig:S8PetraBPM} shows a drawing of a PETRA III--BPM which is installed in sections between two canted undulators. Because of the small undulator gaps the chamber height is only 7 mm, and the chamber width is 83.5 mm and asymmetric in order to avoid heat load due to direct synchrotron radiation illumination. Both the strong ellipticity and the asymmetry of the vacuum chamber profile result in strong non-linearities in the beam position determination which has to be corrected.

Besides electron BPMs, additional X-ray BPMs are installed in photon beamlines. Usually two XBPMs are used per beamline in order to measure photon beam angle and position. The objective is to include the XBPM data in the orbit feedback system in order to stabilize the photon beam position at the location of the experiment. A monitor type that is widely used is the gapped photoemission blade monitor. It probes the outer fringes of the photon beam with metal electrodes, permitting the central core of the beam to pass through to a downstream experiment \cite{Johnson89,Loyer93}. However, gapped monitors suffer from the fact that they sample only the lower energy tails of the photon beam. Any asymmetry in illumination of the electrodes due to occlusion from upstream apertures, low energy scattered or reflected photons, or in the case of undulator radiation beamlines from nearby bending magnet radiation, leads to an error in determining the centroid beam position \cite{Kuske94}. In order to overcome this drawback, different concepts are discussed as residual gas X-ray BPMs \cite{Ilinski06}, vibrating wire sensors \cite{Arutunian08,Decker08}, and chemical vapour deposition (CVD) diamond based detectors \cite{Bergonzo99,Bergonzo06,Griesmayer16}. More information about XBPMs and beamline instrumentation can be found e.g., in Refs.~\cite{Kuske94,HOttini08}. 

\subparagraph{Emittance diagnostics}

Third generation light sources shall achieve horizontal emittances of the order of 1 $\pi\,{\rm nm}\,{\rm rad}$ and even less with emittance coupling of about 1\%. Therefore a precise measurement especially of the small vertical emittance is a challenge. As usual in a storage ring, synchrotron based diagnostics is used to measure the photon spot which contains information about the emittance.

Common way is to image the beam with an appropriate light optics, i.e., the photon spot is a measure of the transverse beam profile. However, in the imaging process there exists a principle limitation which is given by Heisenberg's uncertainty relation. In this specific case it can be reformulated as
\begin{equation}\label{Eq:S8Heisenberg}
    \Delta\sigma \approx \frac{\lambda}{2 \Delta\Psi}\, ,
\end{equation}
with $\Delta\sigma$ the resolution broadening due to diffraction, $\lambda$ the wavelength of observation, and $\Delta\Psi$ the opening angle of the emitted photon as measured from the orbit plane. While the horizontal emission angle is large due to the particle motion on a curved trajectory, the vertical one is small and thus imposes the fundamental resolution limit. For a typical optical wavelength of observation $\lambda$ = 500 nm and an opening angle $\Delta\Psi$ = 1 mrad, the resolution would amount to $\Delta\sigma$ = 250 $\mu$m. Considering that the vertical beam size in a modern light source is of the order of a few tens of microns and even less, such a monitor would have a totally diffraction limited resolution. In order to overcome this limit there exist different concepts which will be briefly addressed in the following. More information can be found in Ref.\cite{Kube07b} and the references therein.

The most straightforward way to overcome this limitation is imaging at smaller wavelengths in the ultraviolet (VUV), soft or even hard X-ray region, cf. Eq.~(\ref{Eq:S8Heisenberg}). In this case the discussion about a monitor concept is reduced to the question about the appropriate imaging optics. This can be a focusing optic like a reflective one (Kirkpatrick--Baez mirrors), a diffractive one (Fresnel zone plate or Bragg--Fresnel lens), or a refractive one (compound refractive lens). Beside these, focus-free imaging with X-ray pinhole cameras is used in many accelerator laboratories \cite{Elleaume95,Thomas13}. Furthermore, the wave optics features of the emitted radiation can be exploited \cite{Hansson12,Breunlin15}. In addition, coded-aperture imaging is sometimes applied \cite{Bloomer14,Alexander14} which is a technique well-developed among X-ray astronomers. However, in any case a necessary prerequisite for X-ray imaging as emittance diagnostics is to have a dedicated diagnostics beamline, at least with synchrotron radiation from a bending magnet acting as source.

\begin{figure}[!h]
	\begin{center}
		\includegraphics[scale=.7,clip]{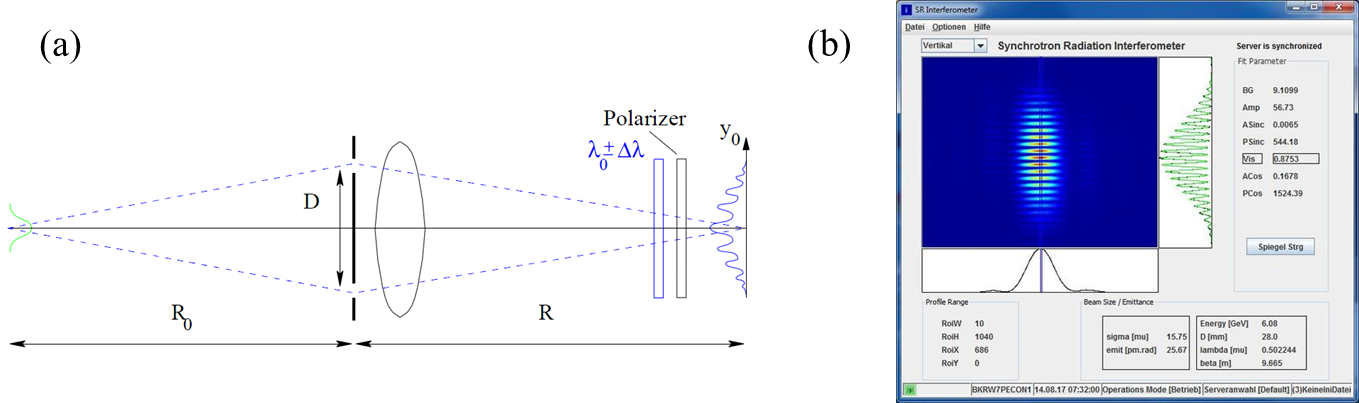}
		\caption{(a) Principle set-up for interferometric beam size measurements. (b) Screen shot of the PETRA III interferometer measuring the vertical emittance.}
		\label{fig:S8Interfero}
	\end{center}
\end{figure}
Another measurement concept is an interferometric approach using visible synchrotron radiation \cite{Mitsuhashi98} which is adapted from Michelson's stellar interferometry and applied for the determination of the extent of stars. It is based on the investigation of the spatial coherence properties of the radiation by measuring the blurring of the interferogram which depends on the particle beam size in a double-slit interferometric set-up, cf. \Fref{fig:S8Interfero}. The fundamental limit of this monitor principle is again Heisenberg's uncertainty relation which can be reformulated for an interferometric measurement as $\Delta n \Delta \Phi \sim 1$, with $\Delta \Phi$ the relative phase difference between the wave trains passing the two individual slits and $\Delta n$ the number of required photons \cite{Lipson95}. As a consequence, in order to measure the phase difference with high accuracy the intensity must be sufficient. For the determination of the beam size in both transverse directions in principle each plane requires a dedicated interferometric setup. As an alternative a two-dimensional interferometer can be operated as it is realized e.g., at PETRA III (DESY, Hamburg) \cite{Novoks17}.

A rarely used concept is the projection method which exploits the angular divergence instead of the beam size \cite{Scheidt06,Mueller06}. However, due to the horizontal fan of bending magnet radiation only the vertical emittance can be determined. The principle of this method relies on the fact that only a tiny fraction of very hard X-rays can fully penetrate the dipole crotch absorber and enter in the free air space behind. These X-rays are detected by a simple, compact and low-cost device, consisting of a CdWO$_4$ scintillator and a standard CCD camera system. With knowledge of the measured photon spot size $\sigma_{\gamma,y}$, the mean square photon emission angle $\langle \vartheta^2_{\gamma}\rangle$, the distance between source and image plane $L$, and the accelerator Twiss parameters at the emission point the emittance can be derived as
\begin{equation} \label{Eq:Projection}
  \varepsilon_y = \frac{\sigma_{\gamma,y}^2 - \langle \vartheta^2_{\gamma}\rangle L^2}{\beta_y + 2 \alpha_y L + \gamma_y L^2} \; .
\end{equation}
So far the peculiarities of storage ring based synchrotron light sources have been pointed out. The next section is devoted to the diagnostics needs of fourth generation light sources, i.e., linac (single pass) based FEL.

\subsection{Free electron lasers}

In the last couple of years, free electron lasers (FELs) have been a remarkable success as fourth generation light sources all over the world. Operating in the high-gain regime in the self-amplified--spontaneous-emission (SASE) mode, they produce laser-like radiation in a broad wavelength range and generate multi-gigawatt and femtosecond coherent X-ray pulses.
\begin{figure}[!b]
\begin{center}
\includegraphics[scale=1.,clip]{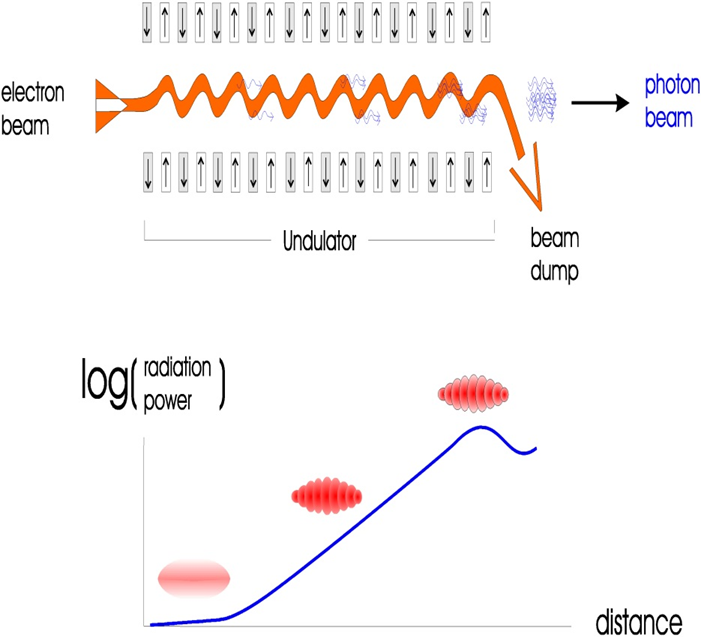}
\caption{Working principle of a single-pass free electron laser (FEL) operating in the self-amplified--spontaneous-emission (SASE) mode. The lower plot shows the bunch density modulation (micro-bunching), developing in parallel to the radiated power \cite{TESLA_TDR}.}
\label{fig:S8_SASEFEL}
\end{center}
\end{figure}

Figure \ref{fig:S8_SASEFEL} illustrates the working principle of a high-gain SASE FEL. It makes use of the fact that a high quality electron beam, passing a long undulator magnet, exponentially amplifies an initially existing radiation field. A prerequisite for this effect is that the photon wavelength $\lambda_r$ matches the resonance condition
\begin{equation} \label{Eq:FEL1}
  \lambda_r = \frac{\lambda_u}{2 \gamma^2} (1 + K^2)\, ,
\end{equation}
which is determined by the Lorentz factor $\gamma$, the undulator period $\lambda_u$, and the undulator parameter $K$ which is a measure for the magnetic field. The intensity amplification results from the modulation of the electron bunch with the undulator radiation field emitted by the bunch itself. This leads to a density modulation (micro-bunching) in the bunch so that more and more electrons radiate in phase until saturation is reached. Further information about the theory of X-ray FELs can be found e.g., in textbooks \cite{SSY00,SDRB14} or in review articles \cite{Schreiber10,Huang10,Pellegrini16}.

\begin{figure}[!h]
	\begin{center}
		\includegraphics[scale=0.24,clip]{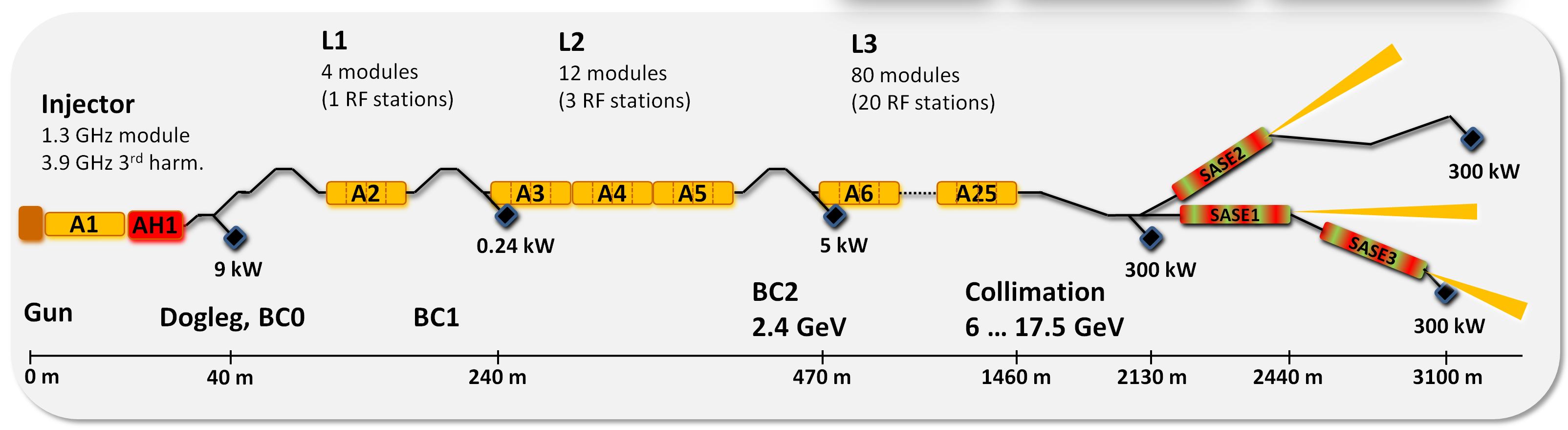}
		\caption{Block diagram of the European XFEL}
		\label{fig:S8xfel}
	\end{center}
\end{figure}

There exist a number of soft and hard X-ray SASE FELs in different accelerator laboratories worldwide. Examples are the  Linac Coherent Light Source (LCLS) at SLAC (USA) \cite{LCLS1} and its planned upgrade LCLS-II \cite{LCLS2}, the SACLA facility at SPring8 (Japan) \cite{SACLA1,SACLA2}, the SwissFEL at PSI (Switzerland) \cite{SwissFEL}, the PAL--XFEL at the  Pohang Accelerator Laboratory (South--Korea) \cite{PAL}, and the machines which are built and operated by DESY (Hamburg, Germany): the FLASH facility \cite{FLASH1,FLASH2} and the European XFEL \cite{XFEL1,XFEL2,XFEL3} which is shown schematically in \Fref{fig:S8xfel}. In addition, an example for a seeded instead of a SASE FEL is the FERMI facility at Elettra (Italy) \cite{FERMI}.

In the next subsection, requirements for FEL operation are addressed and their impact on beam diagnostics and instrumentation is pointed out. The discussion relies especially on experience with the machines built and operated by DESY, the FLASH and the XFEL facility. However, from  diagnostics viewpoint the FEL type is not of such importance.

\subsubsection{FEL requirements}

A high current density is mandatory in order to have sufficient energy transfer from the electron beam to the radiation field. For a rough current estimation the number of electrons per wavelength
\begin{equation} \label{Eq:FELcurrent}
  N_{e,\lambda} = \frac{I \lambda_r}{e c}
\end{equation}
is considered as natural scale. According to Eq.(\ref{Eq:FELcurrent}) the current in a slice of a micro-bunched beam containing a single electron is calculated as
\[
N_{e,\lambda} = 1
\quad \Rightarrow \quad
I = \;
\begin{cases}
\; 0.5~\mu{\rm A} \quad & \text{for}~~\lambda_r = \text{100}~\mu{\rm m}, \\
\; 0.5~{\rm A} \quad & \text{for}~~\lambda_r = \text{0.1}~{\rm nm}.
\end{cases}
\]
However, the energy transfer from a single electron to the radiation field is not sufficient and very high peak currents are required depending on the wavelength. At FLASH for example the peak current of the uncompressed bunch amounts about \Unit{70}{A}, but a peak current exceeding \Unit{1}{kA} is required for laser operation. In order to achieve these currents additional longitudinal bunch compression is required for this increase of the current density. For this purpose a correlation between particle energy deviation and longitudinal position is introduced with an RF system. In the following dispersive beam line section particles with different energies have different path lengths. With the proper parameter settings, the bunch tail has a shorter path and can catch up to the head so that the bunch is effectively compressed. At FLASH the bunch compression using magnetic chicanes is done in two steps at different energies, and the resulting bunch lengths are in the order of much less than \Unit{100}{fs} which is about three orders of magnitude smaller than in a lepton storage ring.

Together with the requirements of extremely short bunches, a very good electron beam quality is essential to sustain the lasing. In order to have resonant energy exchange from the beam to the radiation field the demands on the beam energy spread are
\[
\frac{\sigma_E}{E} \approx 10^{-4} \, .
\]
The radiation field and electron beam must have additionally a good overlap which defines constraints on the transverse beam emittance:
\[
\varepsilon \le \frac{\lambda}{4 \pi} \quad {\rm with} \quad \varepsilon = \varepsilon_n / \beta\gamma \; ,
\]
i.e., a high beam energy is helpful to achieve this condition.

\begin{figure}[!h]
\begin{center}
\includegraphics[scale=.65,clip]{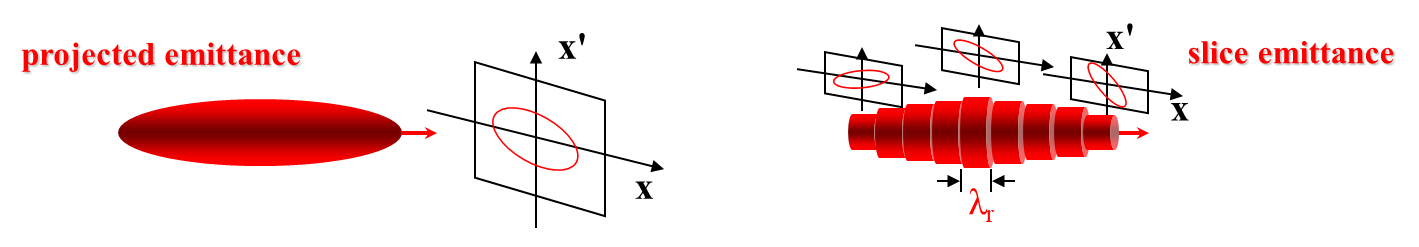}
\caption{Difference between projected (left) and slice (right) emittance of a bunch}
\label{fig:S8ProSlEmit}
\end{center}
\end{figure}

The FEL working principle relies on the fact that the electrons slip back in phase with respect to the photons by one radiation wavelength $\lambda_r$ each undulator period, i.e., the FEL integrates over the slippage length. As consequence it is the slice emittance and not the projected one which is of primary concern, see also \Fref{fig:S8ProSlEmit}.

To conclude, the operation of a SASE FEL requires very high demands on the full six-dimensional phase space. But differently from a circular accelerator, in a linac based accelerator there exists no radiation damping and the beam quality is determined already from the gun. This implies the necessity of careful diagnostics and control of the relevant beam parameters starting from the gun along the whole linac.

Besides the demands on the phase space, stability is a critical issue. This includes (i) the energy stability which translates directly into a wavelength stability as
\[
\frac{\Delta\lambda}{\lambda} = - 2 \, \frac{\Delta E}{E}
\]
according to Eq.~(\ref{Eq:FEL1}), i.e., energy jitter causes wavelength jitter. Due to the use of magnetic chicanes the energy stability defines also the stability of the photon arrival time which is of importance for the users, especially in pump-probe experiments.

(ii) Position stability is crucial because the overlap between electron and photon beam in the undulators has to be maintained. To give an  example, the length of the undulator sections for the European XFEL is of the order of 100 -- \Unit{150}{m} with a transverse rms beam position stability requirement of 10\% of the $\sim$ \Unit{30}{$\mu$m} beam size (1 $\sigma$). In order to be sensitive to small orbit variations in the micron region, high resolution BPMs are required. In the case of the European XFEL, cavity BPMs with a (single bunch) resolution of better than \Unit{1}{$\mu$m} are used for the intersections between the undulator segments \cite{Lipka10,Lipka17}. For linac based FELs which can produce long bunch trains (which is usually the case for FELs using superconducting accelerating structures such as the European XFEL with up to 3250 bunches) it is even possible to stabilize the beam orbit with an intra-bunch-train feedback system which damps harmful beam position perturbations \cite{Keil15}.

To achieve a high average power in a SASE FEL, some of the accelerators use superconducting accelerating structures. Examples are FLASH and the European XFEL with their nine-cell standing wave structures of about \Unit{1}{m} length and fundamental TM mode at \Unit{1.3}{GHz}. These cavities are identical to the so--called TESLA cavity \cite{TESLA}, made from solid niobium, and bath cooled by superfluid helium at \Unit{2}{K}. The use of superconducting accelerator structures requires again that parts of the instrumentation are operated in cold environment, especially the pick-ups. These components must additionally be suitable for an assembly in a particle-free environment.

\subsubsection{Comments on FEL instrumentation}

This section gives an overview of general instrumentation which is essential for FEL operation and some peculiarities which will be addressed in the following. The discussion is mainly based on the experience with the operating European XFEL facility at DESY. A general overview of beam instrumentation and diagnostics strategies can be found in \Bref{Feldhaus04}.

Due to the nonlinear and stochastic FEL process the accelerator, the electron beam parameters along the FEL undulator, and the user beamlines with experiments are strongly coupled. To operate the FEL under controlled conditions diagnostic systems are required for all sections. The stochastic nature of the radiation requires additionally an event oriented data acquisition system for machine operation and user experiments to be able to correlate parameters. According to \Bref{Feldhaus04} the required instrumentation can be arranged in three categories:
\begin{description}
  \item[(i)] Standard electron beam instrumentation to operate the linac.\\
  Instrumentation is required to measure the necessary beam parameters. These are the electron beam orbit, the bunch charge, the beam size, and the beam phase (with respect to the RF).\\
  Fast protection systems with response time in the $\mu$s range are mandatory to shut off the beam in case of high losses in order to prevent damage to the undulators (demagnetization) and to the vacuum system (leakage). Such systems are essential for the operation of superconducting linacs due to the large amount of transported energy.
  \item[(ii)] Instrumentation required to control and optimize the FEL.\\
  For this purpose the phase space of both the electrons and the photons have to be measured and controlled. This includes the determination of the transverse emittance, bunch length and bunch shape (i.e., compression), and energy as well as energy spread.\\
  For the determination of beam size and transverse emittance the transverse beam profile has to be measured at different positions along the machine with a resolution of \Unit{10}{$\mu$m} and better. Instruments in use for this purpose are screen monitors and/or wire scanners.\\
  Bunch lengths smaller than \Unit{100}{fs} have to be measured. If no 3$^{rd}$ harmonic cavity is used, the lasing process is supported by only a small fraction of the charge in a narrow spike produced by the strong bunch compressor, so that it is not sufficient to determine only the first moment of the charge distribution. As consequence, a measurement is required which allows the full reconstruction of the bunch shape.\\
  As described above it is the slice emittance resp. the slice energy spread that determines the performance of the FEL, and a measurement of these parameters is required with a longitudinal resolution less than a radiation pulse length for complete characterization of the electron phase space distribution.\\
  An online signal is needed for the optimization of the SASE process that can be used to determine the optimal phase for bunch compression with a precision of about 0.1$^{\circ}$.
  \item[(iii)] Instrumentation required for user experiments.\\
  The FEL is a pulsed radiation source, and user experiments will be pulse resolved. Due to the fact that the SASE process starts from noise, each radiation pulse is different, and all relevant photon beam parameters have to be measured with single pulse resolution. The characterization of each radiation pulse requires the determination of photon energy and spectral distribution.\\
  Time-resolved experiments need information on duration and temporal structure of the radiation pulse. They require a precise determination of the arrival time which is especially important in pump-probe experiments, i.e., beam synchronous timing is crucial.
\end{description}
\Tref{tab:XFEL} gives an overview of the diagnostic devices which are in operation at the European XFEL. In the following some of the monitor concepts will shortly be presented which are specific for control and optimization of the FEL. This includes BPMs, screen monitors, and longitudinal diagnostic systems resp. slice emittance diagnostics.
\begin{table}[!h]
	\begin{center}
		\caption{Diagnostic devices at the European XFEL \cite{Noelle18}}
		\label{tab:XFEL}
		\begin{tabular}{lc}
			\hline\hline
			\textbf{System}  & \textbf{Number}\\
			\hline
			\emph{BPM system} & \emph{453}\\
			\hspace*{0.3cm} Button BPMs & 303\\
			\hspace*{0.3cm} Cavity BPMs & 126\\
			\hspace*{0.3cm} Re-entrant cavity BPMs & 24\\
			\emph{Charge monitors} & \emph{51}\\
			\hspace*{0.3cm} Faraday cups & 4\\
			\hspace*{0.3cm} Dark current monitors & 10\\
			\hspace*{0.3cm} Toroids & 37\\
			\emph{Beam size measurements} & \emph{79}\\
			\hspace*{0.3cm} Screens & 52\\
			\hspace*{0.3cm} Off-axis screens & 12\\
			\hspace*{0.3cm} Dump window camera & 3\\
			\hspace*{0.3cm} Wire scanners & 12\\
			\emph{Loss monitors and dosimetry} & \emph{1104}\\
			\hspace*{0.3cm} Beam loss monitors & 470\\
			\hspace*{0.3cm} Beam halo monitors & 4\\
			\hspace*{0.3cm} RadFet dosimeters & 630\\
			\emph{Longitudinal diagnostic systems} & \emph{16}\\
			\hspace*{0.3cm} Bunch compression monitors & 4\\
			\hspace*{0.3cm} Bunch arrival-time monitors & 7\\
			\hspace*{0.3cm} Electro-optical diagnostics & 3\\
			\hspace*{0.3cm} Transverse deflecting structures & 2\\
			\hline\hline
		\end{tabular}
	\end{center}
\end{table}

While standard instrumentation for linac operation was already covered in the preceding sections, 
information about diagnostic methods required for user experiments as e.g., streaking spectroscopy can be found in Refs.~\cite{Feldhaus04,Helml14,Duesterer14,ArdanaLamas16} and the references therein.

The important topic of beam synchronous timing is discussed e.g., in Refs.~\cite{Wilcox05,Winter06,Loehl10,Schulz11,Xin17,Kaertner17} together with the report in these proceedings \cite{Gallo18}. As an example, \Bref{Schulz15} reports about the first femtosecond all-optical synchronization of an X-ray FEL.

\subsubsection{Instrumentation examples}

Information about diagnostics and instrumentation in general required for the FEL operation can be found e.g., in \Bref{Jordan07} and in the report about diagnostics examples from lepton linacs and FELs in these proceedings \cite{Cianchi18a}. Additional information about beam instrumentation can be found in overview articles for FELs. Examples are \Bref{Loos14} for LCLS and LCLS-II (SLAC, USA), \Bref{Maesaka12} for SACLA (SPring8, Japan), \Bref{Schlott15} for the SwissFEL (PSI, Switzerland), \Bref{Baboi14} for FLASH2, and Refs.~\cite{Lipka17,Noelle18} for the European XFEL, the latter two machines  built and operated by DESY (Hamburg, Germany).

\subparagraph{Beam position monitors}

Reference \cite{Keil10} presents an overview of the European XFEL beam position monitor system, a short list of the BPM specifications for the different accelerator sections is summarized in 
\Tref{tab:BPM}. 
\begin{table}[!h]
	\begin{center}
		\caption{Short version of the BPM specifications for different sections of the European XFEL. Besides the beam pipe diameter, single bunch and train averaged resolution (rms) are quoted. The specified bunch charge range is 0.1 – \Unit{1}{nC}. Courtesy: D. N\"{o}lle (DESY).}
		\label{tab:BPM}
		\begin{tabular}{lllll}
			\hline\hline
			\textbf{Section} & \textbf{Beam pipe}  & \textbf{Type} & \textbf{Single bunch rms}  & \textbf{Train averaged rms} \\
			\hline
			Standard & \Unit{40.5}{mm} & button & \Unit{50}{$\mu$m} & \Unit{10}{$\mu$m}\\
			Cold section & \Unit{78}{mm} & button / & \Unit{50}{$\mu$m} & \Unit{10}{$\mu$m}\\
			& & re-entrant & & \\
			Transfer line & \Unit{40.5}{mm} & cavity & \Unit{10}{$\mu$m} & \Unit{1}{$\mu$m}\\
			Undulator section & \Unit{10}{mm} & cavity & \Unit{1}{$\mu$m} & \Unit{0.1}{$\mu$m}\\
			\hline\hline
		\end{tabular}
	\end{center}
\end{table}
As can be seen from this comparison, the variation in specification is large such that they cannot be fulfilled by a single BPM type. Therefore different BPM systems are required, in case of the European XFEL these are button-type BPMs \cite{Treyer13,Treyer17}, re-entrant type cavity BPMs \cite{Simon16}, and cavity BPMs \cite{Lipka10,Stadler14}. Another example is \Bref{Choi13} for the BPM systems of the PAL--XFEL.

\subparagraph{Transverse profile}

Transverse beam profile diagnostics in electron linacs is widely based on optical transition radiation (OTR) as a standard technique which is generated when a charged particle beam crosses the boundary between two media with different dielectric properties. Unfortunately, microbunching instabilities in high–brightness electron beams of modern linac–driven FELs can lead to coherence effects in the emission of OTR, thus rendering it impossible to obtain a direct image of the particle beam. The observation of coherent OTR (COTR) has been reported by several facilities (see e.g., \Bref{Wesch11}), and in the meantime the effect of the microbunching instability is well understood \cite{Stupakov14}.

In order to allow beam profile measurements in the presence of the instability, transition radiation based imaging in the EUV spectral region was successfully tested \cite{Sukhikh14,Sukhikh15}. Another possibility which was demonstrated to work properly is to eliminate the COTR influence by using a fast gated camera in combination with a scintillator screen (so-called temporal separation) \cite{Yan11,Behrens12}. This method relies on the fact that light emission in a scintillator is a multistage stochastic process from many atoms and therefore completely insensitive to the longitudinal bunch structure, while OTR light emission is an instantaneous process. However, both monitor concepts are either difficult to realize or rather expensive.

For the European XFEL it was therefore decided to use scintillation screen monitors and observe the screen under an angle such that COTR can be spatially separated from the scintillation light. In a series of test measurements performed in the past few years, the applicability of inorganic scintillators for high resolution electron beam profile measurements was investigated \cite{Kube10,Kube12}. Most notably, the dependency of the resolution on the scintillator material and on the observation geometry was studied with respect to resolve beam profiles of the order of several tens of micrometres, and it was concluded that LYSO (Lu$_{2(1-x)}$Y$_{2x}$SiO$_5$:Ce) is a suitable material because it gives the best spatial resolution. Based on these measurements, screen monitor stations were designed for the European XFEL using \Unit{200}{$\mu$m} thick LYSO screens \cite{Wiebers13}. In a high resolution beam profile measurement using an XFEL-type screen it was demonstrated that it is possible to resolve a vertical beam size of $\sigma_y$ = \Unit{1.44}{$\mu$m} \cite{Kube15}.

\begin{figure}[!h]
	\begin{center}
		\includegraphics[scale=.8,clip]{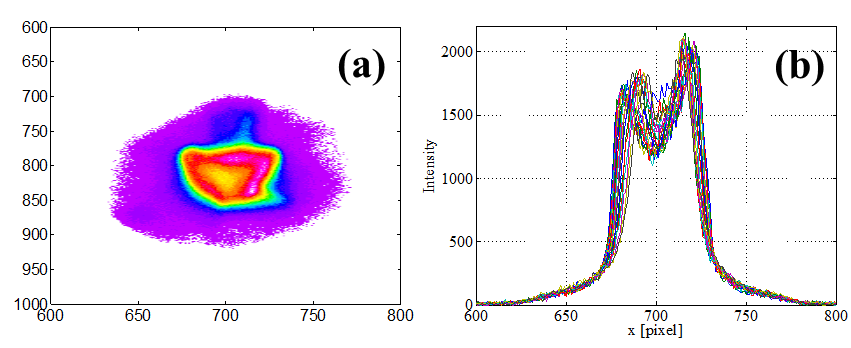}
		\caption{(a) Typical smoke ring shaped beam profile as measured with a European XFEL screen monitor based on a \Unit{200}{$\mu$m} thick LYSO screen. (b) Various horizontal cuts through the 2D profile demonstrate the intensity drop in the central part of the beam spot.}
		\label{fig:S8SmokeRing}
	\end{center}
\end{figure}
However, the experience during the commissioning of the XFEL showed that the measured emittance values were significantly larger than the expected ones \cite{Beutner17,Noelle18}. In addition, measured beam profiles showed a smoke ring shaped structure, see \Fref{fig:S8SmokeRing}. While the contribution of COTR emission from the scintillator surface, beam dynamical influence, and camera effects could be excluded to explain this observation, it is assumed that the beam profile distortions are caused by effects from the scintillator material. In \Bref{Kube18} a simple model was developed which takes into account quenching effects of exciton carriers inside a scintillator in a heuristic way. Based on this model, the observed beam profiles could be understood qualitatively. Up to now the question for suitable scintillator materials for beam profile diagnostic applications remains open, detailed studies are still ongoing.

\subparagraph{Bunch length and bunch profile diagnostics}

The resolution limit of a streak camera is typically of the order of about a picosecond. In order to resolve time distributions in the femtosecond region new diagnostic concepts are required. New schemes have been applied for the measurements of sub-ps longitudinal charge distributions. These are coherent radiation diagnostics (CRD), electro-optical sampling (EOS), and the application of a transverse deflecting structure (TDS). They will be briefly described in the following, more information can be found in the report in these proceedings \cite{Gillespie18}.

Radiation is emitted coherently if the wavelength is in the order of the bunch length, i.e., information about bunch length and shape is encoded in the emission spectrum which is exploited in CRD \cite{Grimm07,Schmidt18}. In case of coherent emission the spectral intensity is strongly amplified which can be expressed in the following form:
\begin{equation} \label{Eq:CRD}
  \frac{{\rm d}U}{{\rm d}\lambda} = \left(\frac{{\rm d}U}{{\rm d}\lambda}\right)_1
  \left(
  N + N (N - 1) |F(\lambda)|^2
  \right) \; .
\end{equation}
Here $({\rm d}U/{\rm d}\lambda)_1$ is the single particle emission spectrum, $N$ the number of particles in the bunch, and $F(\lambda)$ the bunch form factor. It is related to the normalized bunch profile $S(z)$ via a Fourier transform:
\begin{equation} \label{Eq:CRD2}
  F(\lambda) = \int^{+\infty}_{+\infty} {\rm d}z \; S(z) \; e^{-2 \pi i z / \lambda} \; .
\end{equation}
According to Eq.~(\ref{Eq:CRD}), from a measurement of the spectral intensity and with knowledge of the single electron spectrum together with the bunch charge, the form factor can be determined. Inverting the Fourier transform Eq.~(\ref{Eq:CRD2}) results in the reconstructed bunch profile $S(z)$. However, the situation is more complicate because it is the magnitude $|F(\lambda)|$ of the form factor which is determined rather than the complex form factor itself. Reconstruction is possible only if both amplitude and phase are available. Although a strict solution of this phase-reconstruction problem is not possible, a so-called minimal phase can be constructed with the Kramers--Kronig relation which gives a handle to solve this problem satisfactory. A detailed treatment of this problem and further discussions can be found e.g., in Refs.~\cite{Schmidt18,Grimm06,BTaheri16,Wu18}.

\begin{figure}[!h]
	\begin{center}
		\includegraphics[scale=1.1,clip]{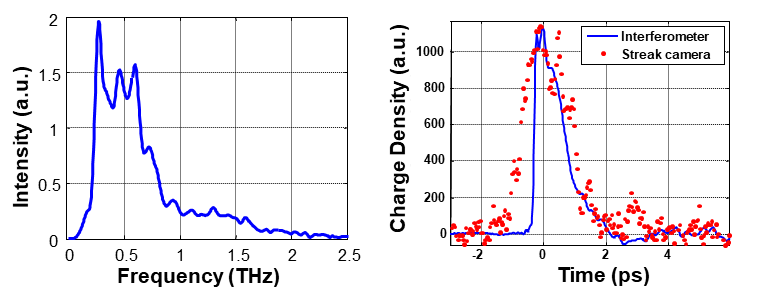}
		\caption{Example of a bunch shape reconstruction with coherent radiation diagnostics at FLASH. Left: measured synchrotron radiation spectrum. Right: reconstructed charge distribution together with a streak camera measurement, performed at the same location using visible light for an uncompressed bunch (courtesy L. Fr\"{o}hlich, DESY).}
		\label{fig:S8crd}
	\end{center}
\end{figure}

In principle any kind of coherent radiation can be used as a radiation source. Measurements were performed with coherent synchrotron radiation, transition radiation, diffraction radiation, and Smith--Purcell radiation. \Fref{fig:S8crd} shows a measurement with coherent synchrotron radiation together with the reconstructed bunch shape. As can be seen from the comparison with an independent streak camera measurement, CRD allows to resolve fine details of the bunch shape.

Electro-Optical (EO) techniques can be applied for ultra-relativistic electrons where the particle Coulomb field is purely transversal, i.e., the time structure of the non-propagating particle field is a measure of the longitudinal bunch profile. If the bunch passes close to an electro-optical crystal (ZnTe or GaP), its Lorentz contracted Coulomb field induces a change in the crystal refractive index (so-called Pockels effect). The information about the longitudinal profile is therefore encoded in a refractive index change which can be converted into an intensity variation by means of a laser together with polarizers.

\begin{figure}[!t]
	\begin{center}
		\includegraphics[scale=.5,clip]{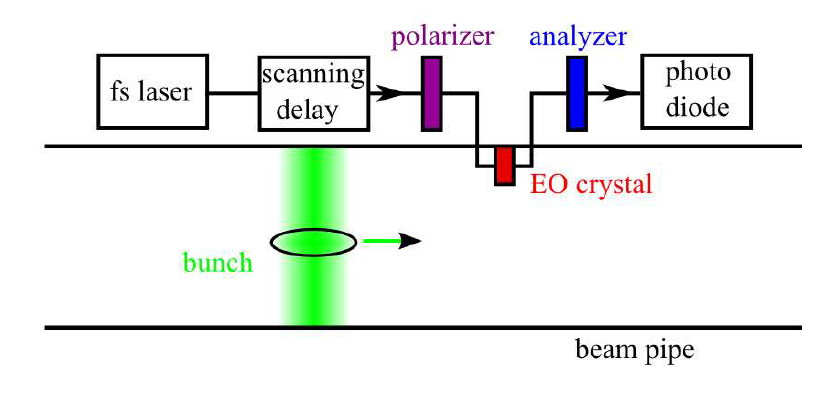}
		\caption{Schematic drawing of an EOS sampling set-up using crossed polarizers. The laser pulse passes through the polarizer and the electro-optical crystal in the beam pipe. In the presence of a transverse electrical field induced by the particle bunch the polarization becomes elliptical. The analyser turns the elliptical polarization into an intensity change which is measured by a photo diode. Changing the delay of the laser pulse, a different longitudinal position of the Coulomb field along the bunch is probed.}
		\label{fig:S8EOS}
	\end{center}
\end{figure}

Figure \ref{fig:S8EOS} illustrates a simple EOS set-up using a variable delay. In this scheme a polarized laser beam is scanned along the bunch, and the change in intensity is recorded as function of the time delay. There exist more sophisticated schemes with even the capability of single shot resolution like spectrally \cite{Wilke02}, temporally \cite{Berden04}, or spatially \cite{Cavalieri05} resolved detection.

Principle resolution limit of EO bunch length measurements is the existence of transverse optical phonon frequencies in the THz frequency range which show up as resonances in the dispersion relation, thus distorting the temporal profile of the fields as they propagate through the crystal. The resonance effect is immaterial when trying to measure electron bunches $\geq$ 120 fs for ZnTe and $\geq$ 60 fs for GaP \cite{Helle12}. Detailed information about EO techniques and measurements can be found e.g., in \Bref{Steffen09}.

\begin{figure}[!b]
	\begin{center}
		\includegraphics[scale=.65,clip]{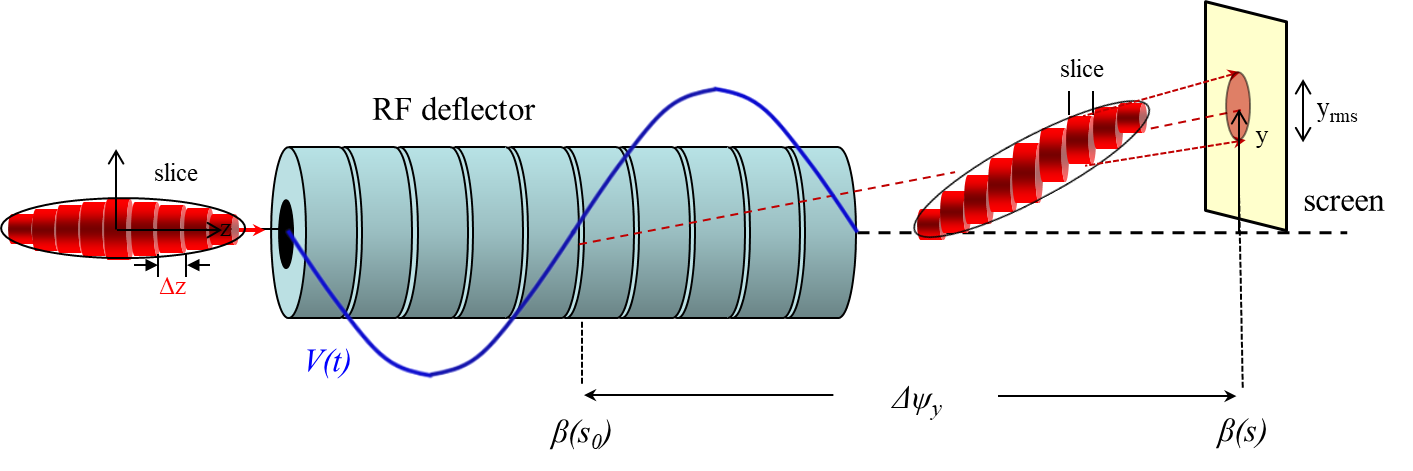}
		\caption{Working principle of a transverse deflecting structure (TDS). A bunch traversing the structure at RF zero crossing experiences a vertical ($y$) kick which depends linearly on time and vanishes in the bunch centre. Due to the vertical deflection, the vertical position of the electrons inside this bunch are linearly correlated to their longitudinal coordinates. A fast kicker (not shown) deflects the bunch onto an off-axis screen.}
		\label{fig:S8tds}
	\end{center}
\end{figure}

The most state-of-the-art instrument for bunch length measurements and even more is the transverse deflecting structure (TDS) \cite{Emma00,Roehrs09}. A TDS is an iris loaded RF waveguide structure designed to provide hybrid deflecting modes (HEM$_{11}$), a linear combination of TM$_{11}$ and TE$_{11}$ dipole modes resulting in a transverse force that acts on the synchronously moving relativistic particle bunch. The working principle resembles that of an intra beam streak camera: a single bunch inside the bunch train, traversing the structure at an appropriate RF phase, experiences a vertical kick which depends linearly on time and vanishes in the bunch centre. Due to the vertical deflection, the vertical position of the electrons inside this bunch are linearly correlated to their longitudinal coordinates. A fast kicker deflects the bunch onto an off-axis screen. Figure \ref{fig:S8tds} shows a scheme of the TDS set-up. Besides other parameters, the TDS resolution depends on the deflecting RF frequency and on the deflecting voltage. By using an X-band structure in combination with RF pulse compression via SLED technique, at the LCLS at SLAC (Stanford, USA) it was possible to achieve a TDS resolution of 0.5 fs \cite{Behrens14,Krejcik16}.

\begin{figure}[!h]
	\begin{center}
		\includegraphics[scale=1.,clip]{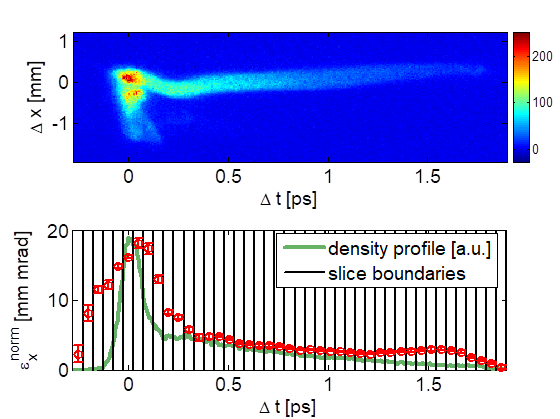}
		\caption{Top: longitudinal phase space distribution as measured at FLASH. The image is rotated by 90$^{\circ}$, the horizontal axis represents the time coordinate (vertical streak) and the vertical one the horizontal coordinate. The colour code indicates the particle density. Bottom: $1 \, \sigma$ slice emittance (dots) together with slice boundaries and density profile (line) \cite{Roehrs06}.}
		\label{fig:S8tds2}
	\end{center}
\end{figure}

Besides the excellent time resolution, a major advantage of a TDS is the access to the slice emittance. The beam image measured at the off-axis screen contains information about the particle distribution in the longitudinal--horizontal plane, see the measurement example in the upper plot of \Fref{fig:S8tds2}. If now the quadrupoles upstream of the structure are scanned and the horizontal beam size in each slice is determined as function of the quadrupole settings, this technique allows us to determine the horizontal slice emittance. In \Fref{fig:S8tds2} (bottom) the measured slice emittance together with slice boundaries and density profile is shown. In addition, with a screen located in a horizontally dispersive section behind the deflecting structure it is possible to measure directly the longitudinal phase space distribution in a single shot. More information about this technique can be found in \Bref{Roehrs06}.

By using a conventional TDS system it is possible to characterize the slice properties of an electron beam in the transverse direction perpendicular to the streaking direction. Therefore only one of the transverse slice envelopes can be measured. Presently the development of an advanced modular X-band TDS system is ongoing \cite{Marchetti17} which will have the new feature of providing variable polarization of the deflecting field \cite{Grudiev16}, thus allowing for 3D characterization of the phase space using tomographic methods \cite{Marx17}.

\section{Outlook}

So far diagnostic needs for different hadron and lepton accelerators were presented. The description covered accelerators in operation since longer time, machines which were recently commissioned, or even machines already shut down, such as LEP and HERA. However, from the beam diagnostics viewpoint each new accelerator represents a new challenge. Therefore, although the LHC presently stopped operation and is in process to undergo the high-luminosity upgrade \cite{LHC17}, the next generation of colliders for high-energy physics is already in discussion. 

In the preceding sections, injector linacs including \Pep production, lepton storage rings with the opportunity of radiation damping, and high brilliance electron linacs with stringent requirements for the whole 6-dimensional phase space were covered. Now one can combine these types of accelerators and end up with a linear collider.

\begin{figure}[!t]
	\begin{tabular}{lr}
		\vspace*{-0.4cm}
		\includegraphics[scale=.18,clip]{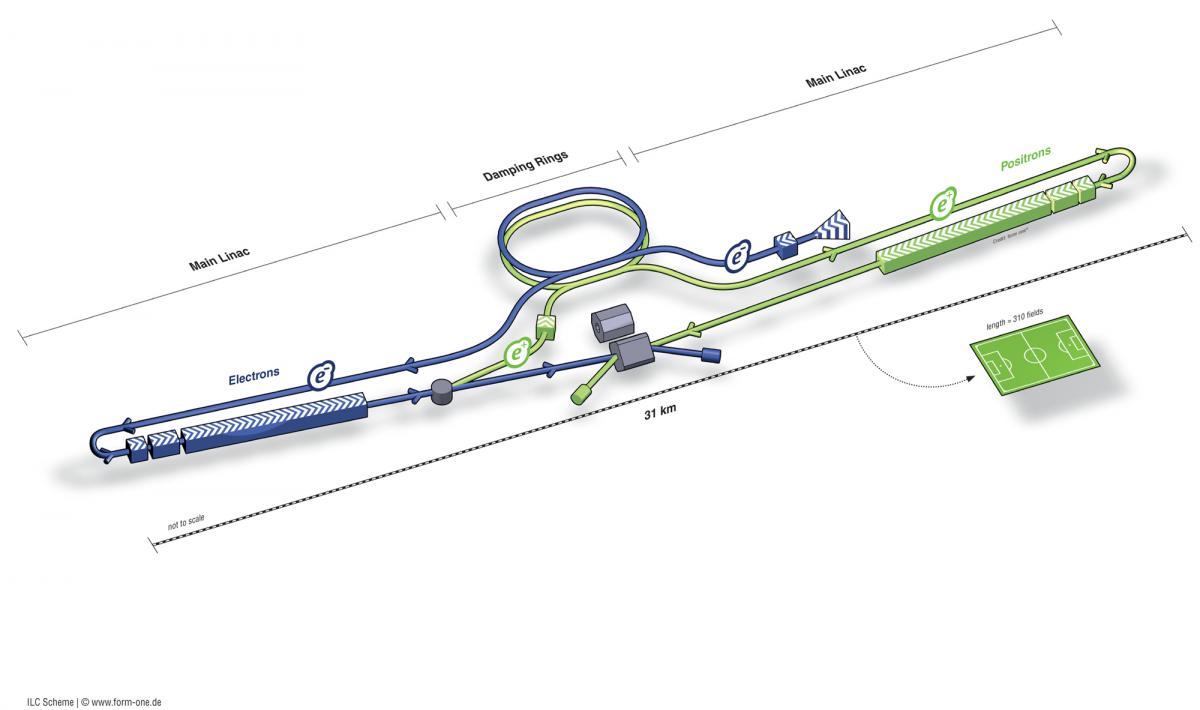}&
		\vspace*{0.5cm}
		\includegraphics[scale=.4,clip]{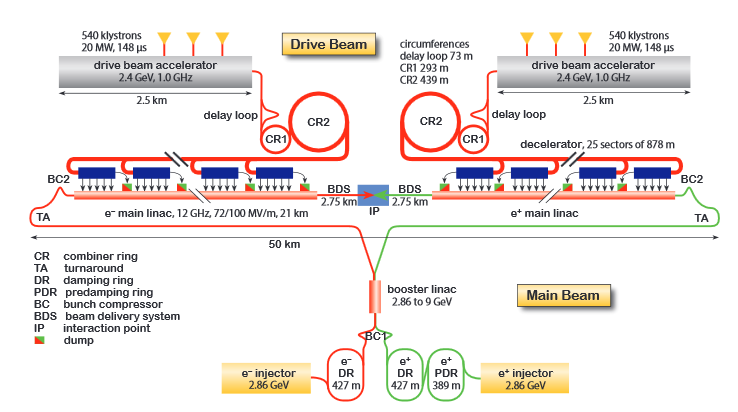}
	\end{tabular}
	\caption{Schematic layout of (right) the ILC for 500\UGeV{} centre-of-mass energy \cite{ILC1}, and (left) of CLIC for 3\UTeV{} \cite{CLIC}}
	\label{fig:S9LC}
\end{figure}

Figure \ref{fig:S9LC} shows the schematic layout of both linear collider projects which were subject of an animated debate during the past years: the International Linear Collider (ILC) \cite{ILCweb} and the Compact Linear Collider (CLIC) \cite{CLICweb}. The ILC baseline design stipulates a collision energy of 500\UGeV{} (250\UGeV{} per \Pep\Pem beam), based on superconducting RF L-band 1.3 GHz Tesla technology with a possible upgrade to TeV energies \cite{ILC}. In the final stage 3 CLIC will even operate at 3\UTeV{} centre-of-mass energy based on normal conducting RF X-band 12 GHz technology \cite{CLIC}.

\begin{table}[!h]
	\begin{center}
		\caption{Nominal values of ILC beam parameters at the interaction point \cite{ILC}}
		\label{TAB:ILC1}
		\begin{tabular}{p{6cm}p{2.5cm}l}
			\hline\hline
			Bunch population  & & 2 $\times$ 10$^{10}$ \\
			Number of bunches  & & 1312 \\
			Linac bunch interval & / ns & 554 \\
			Luminosity pulse repetition rate & / Hz & 5 \\
			Normalized emittance at IP $\gamma\varepsilon_{x,y}$ & / mm~mrad & 10 / 0.035 \\
			Beta function at IP $\beta^*_{x,y}$ & / mm & 11 / 0.48 \\
			Rms beam size at IP $\sigma_{x,y}$ & / nm & 474 / 5.9 \\
			Rms bunch length $\sigma_z$ & / $\mu$m & 300 \\
			Average beam power per beam  & / MW & 5.3 \\
			Energy of single pulse  & / MJ & 1.051 \\
			Luminosity $\mathfrak{L}$ & / cm$^{-2}$s$^{-1}$ & 1.8$\times$10$^{34}$ \\
			\hline\hline
		\end{tabular}
	\end{center}
\end{table}
The subsequent discussion is based on the ILC specifications. To get an impression about the beam diagnostic requirements for this machine, \Tref{TAB:ILC1} summarizes the nominal values of the key parameters at the interaction point \cite{ILC}. Besides the diagnostic and instrumentation requirements listed in this reference, \Bref{Tenebaum06} includes a discussion about ILC diagnostics aspects in greater depth. As can be concluded from this table, beam position stability and beam profile monitoring systems are particularly challenging.

Following \Bref{ILC}, maintaining the stability of the beam delivery system (BDS) just in front of the IP is an essential prerequisite to produce the required luminosity. Since the beams have vertical sizes of roughly \Unit{6}{nm} at the IP, vertical offsets of about \Unit{1}{nm} will noticeably reduce the luminosity. In addition, the beam--beam interaction is so strong that the luminosity is extremely sensitive to small variations in the longitudinal shape of the bunch caused by short-range wakefields. Finally, the size of the beam at the IP is sensitive to the orbit of the beam through the final-doublet quadrupoles and other strong optical elements of the BDS. Therefore, besides mechanical and temperature stabilization, beam-based orbit-feedback loops are required to maintain size and position stability of the beam at the IP.  All feedback loops shall use BPMs  with at least micron level, in some cases even sub-micron-level resolution. Two basic forms of feedbacks are foreseen: (i) train-by-train feedback, operating at the \Unit{5}{Hz} ILC repetition rate, and (ii) intra-bunch-train feedback which can apply a correction to the beam between bunches of a single train.

There are a variety of research and development activities for ILC instruments at different laboratories worldwide. Warm cavity BPMs studied at the KEK Accelerator Test Facility (ATF) in Japan have already achieved position resolutions of 8.7 nm for a bunch with 0.68 $\times$ 10$^{10}$ particles over a dynamic range of 5 $\mu$m \cite{Inoue08}. In the case of beam profile monitors non-invasive instrumentation has to be applied because otherwise the low emittance would destroy the monitor. A monitor concept in discussion is the laser wire scanner \cite{Agapov07} where a finely focused beam of laser light is scanned across the bunches. The resulting rate of Compton scattered photons is measured downstream in a detector as a function of the laser beam position. Prototype laser wire scanners have been tested at PETRA III \cite{Aumeyr10} and ATF \cite{Boogert10}.

Besides linear colliders, in recent years circular collider studies for high energy physics gained more and more interest. Two new projects are the main focus of attention. The Future Circular Collider (FCC) study, hosted by CERN, is designing a 100\UTeV{} collider to be installed inside a new \Unit{100}{km} tunnel in the Lake Geneva basin \cite{Benedikt18a,Benedikt18b}. The 100\UTeV{} Future Circular Collider (hadron version FCC-hh) would require \Unit{16}{T} dipole magnets based on Nb$_3$Sn superconductors.
\begin{figure}[!h]
\begin{center}
	\includegraphics[scale=1.2,clip]{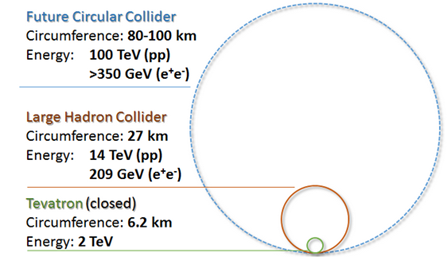}
	\caption{FCC accelerator parameters compared to the LHC and Tevatron (FNAL) \cite{FCC}}
	\label{fig:S9FCC}
\end{center}
\end{figure}
In order to get a better impression of the accelerator dimensions, \Fref{fig:S9FCC} shows a comparison of the FCC with the LHC and the Tevatron (FNAL). The FCC study scope also includes the design of a high-luminosity $\Pep\Pem$ collider (FCC-ee) operating at centre-of-mass energies of 90–365\UGeV{} as a possible first step, as well as a proton--electron collision option (FCC-he) at one interaction point where a 60\UGeV{}  electron beam from an energy recovery linac would be collided with one of the two 50\UTeV{} proton beams circulating in the FCC-hh.

Another international collaboration, centred at IHEP Beijing, launched a study of a similar  50-\Unit{100}{km} ring collider. It would first serve as an $\Pep\Pem$-collider for a Higgs factory and be named the Circular Electron--Positron Collider (CEPC). The facility would then be upgraded by adding a 70-100\UTeV{} Super Proton--Proton Collider(SppC) in the same tunnel with the magnet design realized with cables made from Fe-based high-temperature superconductors \cite{SppC,CEPC}.

Following the discussion in \Bref{CEPC}, the peculiarities of these colliders in view of beam instrumentation are directly connected to the large size of the accelerator. Considering the  long distances, due to cable attenuation the usage of copper cables for signal transmission from a beam monitor in the accelerator tunnel to an electronic cabinet is not to recommend. Therefore the analogue signals should directly be digitized inside the tunnel and transmitted via optical fibres to the local stations in an auxiliary tunnel. As consequence, sensitive electronics has to be installed in the tunnel which requires a careful shielding from radiation. Furthermore, the collider design supposes that both beams will circulate in the same beam pipe. Since both beams pass through the same monitors it must be ensured to distinguish the signals from the counterpropagating bunches. In case of BPMs the problem could be solved in principle by usage of stripline monitors as directional couplers. However, for CEPC it is planned to use button-type monitors for the 2324 BPMs. Positrons and electrons passing through the same monitor shall be distinguished by signal polarity \cite{CEPC}.

For accelerator based light sources, the progress towards next generation machines is well advanced. In the case of diffraction limited storage ring based light sources, with MAX IV (MAXLab, Sweden) the first accelerator using an MBA-type lattice is already in operation \cite{Eriksson16} and will be followed by Sirius (LNLS, Brazil) with commissioning expected soon \cite{Rodrigues16}. Diagnostic challenges of these new light sources are discussed in Refs.~\cite{Hettel14,Maesaka15} and will shortly be summarized below. Especially challenging is the high degree of transverse, longitudinal, and intensity stability which is required for maximum photon source performance. For user operation, pointing stability of the synchrotron radiation is the most important issue. The stability target of the optical axis of each photon beamline should be as good as 1/10 of the beam size and the beam divergence, corresponding to \Unit{0.5}{$\mu$m} for the source  position and \Unit{0.5}{$\mu$rad} for the angle. Therefore, the electron beam position should also be stabilized within 1/10 of the beam size. In addition, reliable X-ray photon BPMs have to be developed and included for the orbit feedback loop such that the optical axis can directly be stabilized. In this context, especially problematic are photon BPMs for soft
X-ray elliptically polarizing undulators whose transverse intensity distribution changes dramatically with the polarization setting. As general trend one can conclude that so far in 3$^{rd}$ generation light sources the state-of-the-art diagnostic techniques have been developed in view of \emph{electron-beam-oriented diagnostics}, based on the idea that the photon beam performance is guaranteed by the electron beam quality. For the new generation of light sources, \emph{photon-beam-oriented diagnostics} will be crucial to maximize the photon beam performance at the user experimental stations in the beamlines.

In case of linac based light sources (FELs) two general trends can be observed. Firstly the experience with the operation of FLASH and XFEL at DESY (Hamburg, Germany) shows that there is a high demand for attosecond photon pulses which requires to operate the accelerators with low bunch charges in order to minimize space charge broadening effects. For the beam instrumentation this implies a high sensitivity in order to cope with small bunch charges, but also a high dynamic range if the FEL is running at higher bunch charge for standard user operation. At the other hand there is strong interest to design and build X-ray FELs not for pulsed but for cw operation. An example is the LCLS-II high power upgrade to LCLS~\cite{LCLS2} which includes a superconducting \Unit{1.3}{GHz} cw linac with nominal \Unit{1}{MHz} bunch repetition rate, operating at 4.5\~UGeV{} and containing up to \Unit{250}{kW} average beam power. Following \Bref{Krejcik17} the key diagnostic challenges at pulsed X-ray FELs are high brightness and high peak current beams which require measurements of very small transverse beam sizes and ultra-short bunch lengths. For the new generation of FELs this would imply to add the requirements of superconducting cw linacs: high average beam power needs minimally invasive diagnostics, high repetition rates require high data acquisition rates, and diagnostics and instrumentation must be fully integrated in the timing and machine protection systems.

Finally the development of advanced accelerator concepts with the aim to increase the gradient of accelerators by orders of magnitude using new power sources (e.g., lasers and relativistic beams) and new materials (e.g., dielectrics, metamaterials, and plasmas) \cite{Colby16} requires in some cases completely different approaches for beam diagnostics and instrumentation. However, this issue goes far beyond the subject of this contribution and is discussed in the report about diagnostic needs for wakefield accelerator experiments in these proceedings \cite{Cianchi18b}.

\section*{Acknowledgements}
Many thanks to K. Wittenburg and N. Potylitsina--Kube (DESY) for discussions and help in the preparation of this lecture.

\end{document}